\documentclass[useAMS,usenatbib,letterpaper]{mn2e}
\usepackage[totalwidth=480pt,totalheight=680pt]{geometry}
\usepackage{graphicx,amssymb}                                                   \input psfig 

\title[Exoplanet Ejection from Dying Multiple Star Systems]
{The Great Escape II: Exoplanet Ejection from Dying Multiple Star Systems}
\author[Veras \& Tout]{Dimitri Veras$^{1}$\thanks{E-mail:veras@ast.cam.ac.uk}, Christopher A. Tout$^{1}$\\
$^{1}$Institute of Astronomy, University of Cambridge, Madingley Road, Cambridge CB3 0HA}

\begin{document}

\date{Accepted 2012 February 12.  Received 2012 February 5; in original form 2011 December 7}

\pagerange{\pageref{firstpage}--\pageref{lastpage}} \pubyear{XXXX} 

\maketitle

\label{firstpage}

\begin{abstract}
Extrasolar planets and belts of debris orbiting post-main-sequence single stars may
become unbound as the evolving star loses mass.  In multiple star systems, the presence
or co-evolution of the additional stars can significantly complicate the 
prospects for orbital excitation and escape. Here, we investigate the dynamical 
consequences of multi-phasic, nonlinear mass loss and
establish a criterion for a system of any stellar multiplicity
to retain a planet whose orbit surrounds all of the parent stars.  For single stars
which become white dwarfs, this criterion can be combined with the Chandrasekhar
Limit to establish the maximum allowable mass loss rate for planet retention.  We then apply the criterion to circumbinary planets in evolving binary systems 
over the entire stellar mass phase space.  Through about $10^5$ stellar evolutionary
track realizations, we characterize planetary ejection prospects
as a function of binary separation, stellar mass and metallicity.  
This investigation reveals that planets residing at just a few tens
of AU from a central concentration of stars are susceptible to
escape in a wide variety of multiple systems.  Further, planets
are significantly more susceptible to ejection from multiple star systems
than from single star systems for a given system mass.  For system
masses greater than about $2 M_{\odot}$, multiple star systems 
represent the greater source of free-floating planets.   
\end{abstract}


\section{Introduction}

Roughly one third of all stars in the Galactic disc
are components of multiple star systems \citep{lada2006}
and the majority of multiple star systems
are thought to be binary systems \citep{duqmay1991}.
Further, as of February 2012, several tens of extrasolar 
planets have now been detected or are suspected of existing
in binary systems.  Some planets have been
reported in systems with even higher 
stellar multiplicities \citep[e.g.][]{cocetal1997,ragetal2006,gueetal2009,desetal2011}.
\cite{mugneu2009} provide a helpful list of known exoplanets
in multiple star systems, as of 2009.  In binary systems, a planet
may orbit {\it one} of the stars in what is sometimes called an
S-type orbit \citep[see, e.g.][]{lowetal2002,baketal2006,eggetal2006,coretal2008}.  Alternatively the planet may orbit
{\it both} stars in a P-type orbit 
\citep[see, e.g.][]{sigetal2003,leeetal2009,beuetal2010,beuetal2011,kuzetal2011,potetal2011,qiaetal2011,qiaetal2012}.
The latter case describes circumbinary planets, whose
existence has been bolstered by the
recent transit-based discoveries of 
Kepler-16b \citep{doyetal2011}, and Kepler-34b 
and Kepler-35b \citep{weletal2012}.
Therefore, understanding the dynamics of planets in 
multiple star systems and, in particular, binary
systems is becoming increasingly relevant. This
understanding includes how such planets form, and how
they die.

One potential avenue for planetary death is dynamical
ejection as the star evolves beyond the main sequence
and loses mass \citep[][hereafter, Paper I]{veretal2011}.  Evidence for
free-floating planets \citep[in][]{lucroc2000,zapetal2000,zapetal2002,bihetal2009} provides observational motivation to investigate this physical mechanism.
In particular, \cite{sumetal2011} recently discovered 10 
wide-orbit or free-floating bodies and calculated that $1.8^{+1.7}_{-0.8}$
free floating planets exist per main-sequence
star.  Therefore, more planets may travel between
stars than orbit them, and this vast population of free-floaters
cannot be explained by planet-planet scattering immediately subsequent to system formation
alone \citep{verray2012}.   In Paper I, the authors analytically described
the conditions which can lead to planetary 
ejection from a single star.  They assumed isotropic mass loss 
and demonstrated that three key factors must be taken into account.
These are i) the mass loss timescale, ii) the planetary orbital
timescale and iii) the total mass of the system.
That study was limited to consideration of 
a single phase of stellar evolution for mass
loss which occurred in a linear manner.

Here, we perform a three-tiered extension to that work by 
considering i) multiple
stars, ii) multiple evolutionary phases and
iii) nonlinear mass loss. 
Our primary application is the determination of 
the prospects for circumbinary (P-type) 
planets to escape as the parent binary evolves.
Doing so provides us with the foundation to understand
systems of higher stellar multiplicities, systems which
often include tight binaries.

The physical evolution of a single star can, in principle, be
described by just the star's initial mass and metallicity.
However, the phase space of binary star evolution is
significantly broader.  In addition to each star's mass 
and metallicity, the separation and eccentricity of their
mutual orbit (possibly including orbital variations due
to massive planetary companions) are additional parameters which
must be taken into account.  The metallicities of both stars
are often assumed to be equivalent because the stars are
assumed to have formed from the same molecular cloud
and not to have subsequently accreted enough planetary
material to significantly alter their original [Fe/H] 
values.  Our understanding of the physics of stars is not yet
good enough to definitively link the above parameters
with the wind velocity, the accretion rate on to
the companion and the amount of 
spin angular momentum transferred between the stars.
Therefore, physical properties such
as these represent a further broadening of the 
potentially explorable phase space.

We focus our study by considering a single planet
which is far from the parent binary (which may be tight
or wide) and by analyzing the
amount of mass loss from the system as a function
of time.  In the absence of mass loss, the 
approximately elliptical orbit of the
planet will be negligibly perturbed
by the potentially complex interactions
between both members of the binary.
In addition, such complex details of this
star-star interaction need only be modeled to the extent
that they correctly yield the mass-loss rate from the 
entire system.  If the binary separation is large
enough so that both stellar components evolve independently, then
the planet's dynamical evolution reduces to the two-body case 
which was presented
in Paper I, with the difference that
the total system mass here is larger because of the
presence of the binary companion.

In Section 2, we consider theoretically how planetary ejection
is affected by multiple phases of nonlinear mass loss for 
systems of any multiplicity
and establish a criterion for retention we
use throughout the rest of the paper.  In Section 3,
we numerically simulate binary star evolution to identify 
which types of systems are susceptible
to planet ejection and quantify where planets must reside
in these systems in order to remain bound.  We 
compare the binary and single star cases, and discuss the 
implications for higher multiplicities, other types of orbits,
and extensions to this work in 
Section 4, and conclude in Section 5.

\section{Characterizing Multi-phasic Nonlinear Mass Loss}

This section provides the analytic background which
motivates the numerical simulations in Section 3 and
presents formulas which may be applied to other investigations.

\subsection{Phases, Regimes and Stages}

Here we define our nomenclature.  We adopt the same meaning of phases
which is commonly used in stellar evolution studies.  Stages are treated
as subsets of phases, and regimes refer the evolutionary environment
of a planet due to stellar mass loss.

\subsubsection{Phase Identification}

Realistic stellar evolution occurs across several phases.
A sequence of phases qualitatively characterizes a star's history.  
We adopt the definitions of phase in the SSE \citep{huretal2000}
and BSE \citep{huretal2002} stellar evolutionary codes:

0 = Low mass ($M < 0.7 M_{\odot}$) main-sequence star

1 = High mass ($M > 0.7 M_{\odot}$) main-sequence star

2 = Hertzsprung gap

3 = First giant branch

4 = Core helium burning

5 = Early asymptotic giant branch

6 = Thermally pulsing asymptotic giant branch

7 = Naked helium star main sequence

8 = Naked helium star Hertzsprung gap

9 = Naked helium star giant branch

10 = Helium white dwarf

11 = Carbon/oxygen white dwarf

12 = Oxygen/neon white dwarf

13 = Neutron star

14 = Black hole

15 = Massless remnant.

\noindent{We} denote phase numbers by $k$.
Within each phase, the mass-loss rate can vary.  Therefore, 
detailed modeling of a particular
system necessitates fitting mass-loss rates with a piecewise
nonlinear model.  However, doing so for a broad exploration 
of stellar evolution phase space is not feasible.

\subsubsection{Regime Identification}

Instead, we seek to construct nonlinear mass loss profiles
by a sequence of linear approximations.  A key benefit
of this approach is that we can utilize analytic results
from Paper~I and hence gain a better understanding of
what conditions must exist for a planet to be significantly
perturbed or to escape the system.
Assume the stellar mass-loss
rate is constant and equal to $-\alpha$, where
$\alpha>0$.  In order to characterize the
tendency of a system to eject a planet, 
we use the dimensionless mass loss index, $\Psi$ such that

\begin{eqnarray}
\Psi &\equiv& \frac{\rm mass \ loss \ timescale}{\rm orbital \ timescale}
= \frac{\alpha}{n\mu}
\nonumber
\\
&=& \frac{1}{2\pi} 
\left( \frac{\alpha}{1 M_{\odot} {\rm yr}^{-1}}\right)
\left( \frac{a}{1 {\rm AU}}\right)^{\frac{3}{2}}
\left( \frac{\mu}{1 M_{\odot}}\right)^{-\frac{3}{2}}
,
\label{mlindex}
\end{eqnarray}

\noindent{where} $a$ and $n$ represent the planetary semimajor
axis and mean motion, and $\mu = \sum_w M_w + M_p$, where $M_w$ represents
the masses of all of the stars in the system 
and $M_p$ represents the mass of the planet.  We use the 
term planet to loosely describe a bound body
of any mass that is not a star and does not perturb
the orbits of the stars.

A planet is said to be evolving in one of two regimes
depending on the value of $\Psi$.  In the adiabatic
regime, when $\Psi \ll 1$, $a$ increases but the planet's 
eccentricity, $e$, remains constant.
In the runaway regime, when $\Psi \gg 1$,
the planet's semi-major axis continues to increase
and $e$ can now vary (increase or decrease)
across all possible values.  Therefore, a planet
{\it may} escape only if $\Psi \gtrsim 1$.
This bifurcation point is not exact.  It
is a weak function of both $e$ and the planet's 
true anomaly, $f$ (see Paper I).  In fact, $\Psi$
may instead be defined as $\alpha T/\mu$, where $T$ represents
the planet's orbital period.  In this case, $\Psi$ does
not contain the factor of $1/(2\pi)$ that is present in Eq. (\ref{mlindex}).
We will henceforth refer to this factor as $\kappa$.

If the planet is evolving in the runaway regime,
the eccentricity evolution is a function of
the eccentricity at the start of that regime and
the planet's location along its orbit. 
For example, if the planet is close to pericentre on
an already highly eccentric orbit, then the planet
is ejected immediately.  Alternatively, if the planet
is close to apocentre on the same highly eccentric
orbit, then the planet is never ejected.
If the planet is on a circular orbit, regardless
of $f$, it is ejected when the
star has lost exactly half of its mass from its
value at the start of that regime.

\subsubsection{Stage Identification}

Now suppose that within a given phase, the system undergoes $N$ consecutive
stages of constant mass loss evolution with mass-loss
rates of $\alpha_1, \alpha_2, ..., \alpha_i, ... , \alpha_N$.  Then, 
based on the results from Paper I,
a planet experiencing $i$ consecutive stages of mass loss in the adiabatic regime 
will have a semimajor axis
and eccentricity at the end of stage $i$ of 

\begin{equation}
a_i = a_0 \left( \frac{\mu_i}{\mu_0} \right)^{-1} \equiv a_0 \beta_{i}^{-1}
,
\label{aadia}
\end{equation}

\noindent{and}

\begin{equation}
e_i = e_0
,
\label{eadia}
\end{equation}

\noindent{where} $\beta_i \equiv \mu_i/\mu_0$ represents the percent of
the original system mass that is retained by the end of stage
$i$.   Because eccentricity remains constant, a planet cannot be 
ejected in this regime (unless the semimajor axis
evolution carries the planet out of the system),
even if almost all of one or both stars' mass is lost. 
The mass loss index at the end of stage $i$ is:

\begin{equation}
\Psi_i = \kappa \alpha_i \left( \frac{a_0 \mu_0}{\mu_{i}^2} \right)^{\frac{3}{2}}
.
\label{psii}
\end{equation}

\noindent{where} $\kappa = 1$ or $2\pi$ depending on if the orbital timescale
is defined with respect to the orbital period or the mean motion.  
Equation (\ref{psii}) indicates 
that escapability of a planet
is independent of the details of the intermediate
stages of mass loss provided that both $\mu_{i}$
and $\alpha_{i}$ are known.


\subsection{Stage-based Retention Criterion}

Here, we describe the theory behind the 
critical system mass fraction lost 
for a given semimajor axis (Sections 2.2.1 - 2.2.4)
and the planet's critical semimajor axis for a given
system mass loss prescription
(Section 2.2.5).  Computation of the critical semimajor 
axis will be the focus of subsequent sections.

\subsubsection{The Critical Mass Fraction}

Equation (\ref{psii}) demonstrates that, at the end of 
the $i$th stage of evolution,
the minimum fraction of the original system mass which
must be retained to guarantee that a planet with a given $a_0$
remains bound is

\begin{eqnarray}
\beta_{{\rm crit}_i} &\equiv& \left( \frac{\mu_i}{\mu_0} \right)_{\rm crit}
\nonumber
\\
&=&\kappa^{\frac{1}{3}} 
\left( \frac{\alpha_i}{1 M_{\odot} {\rm yr}^{-1}}\right)^{\frac{1}{3}}
\left( \frac{a_0}{1 {\rm AU}}\right)^{\frac{1}{2}}
\left( \frac{\mu_0}{1 M_{\odot}}\right)^{-\frac{1}{2}}
.
\label{crit}
\end{eqnarray}

\noindent{Note} importantly how this criterion can
be written in a form which is independent of $\mu_i$
and $a_i$.  If $\beta_{{\rm crit}_i} > 1$, then the planet is
not guaranteed to remain bound, regardless of how much mass
is lost from the system.  Equation (\ref{crit}) demonstrates
that a single short but powerful ejection of mass
from a star at any time can unhinge a planet.  If this event occurs
late in the life of a star, when $\mu_i$ is low, the 
probability is increased.  
Equation (\ref{crit}) provides a useful way to quickly characterize
an ensemble of systems.  The equation holds for any stellar multiplicity.
For single stars which become white dwarfs, the Chandrasekhar Limit
can be inserted into $\mu_i$.  We may then determine the maximum possible $\alpha_i$
that can protect a planet.

\begin{figure}
\centerline{
\psfig{figure=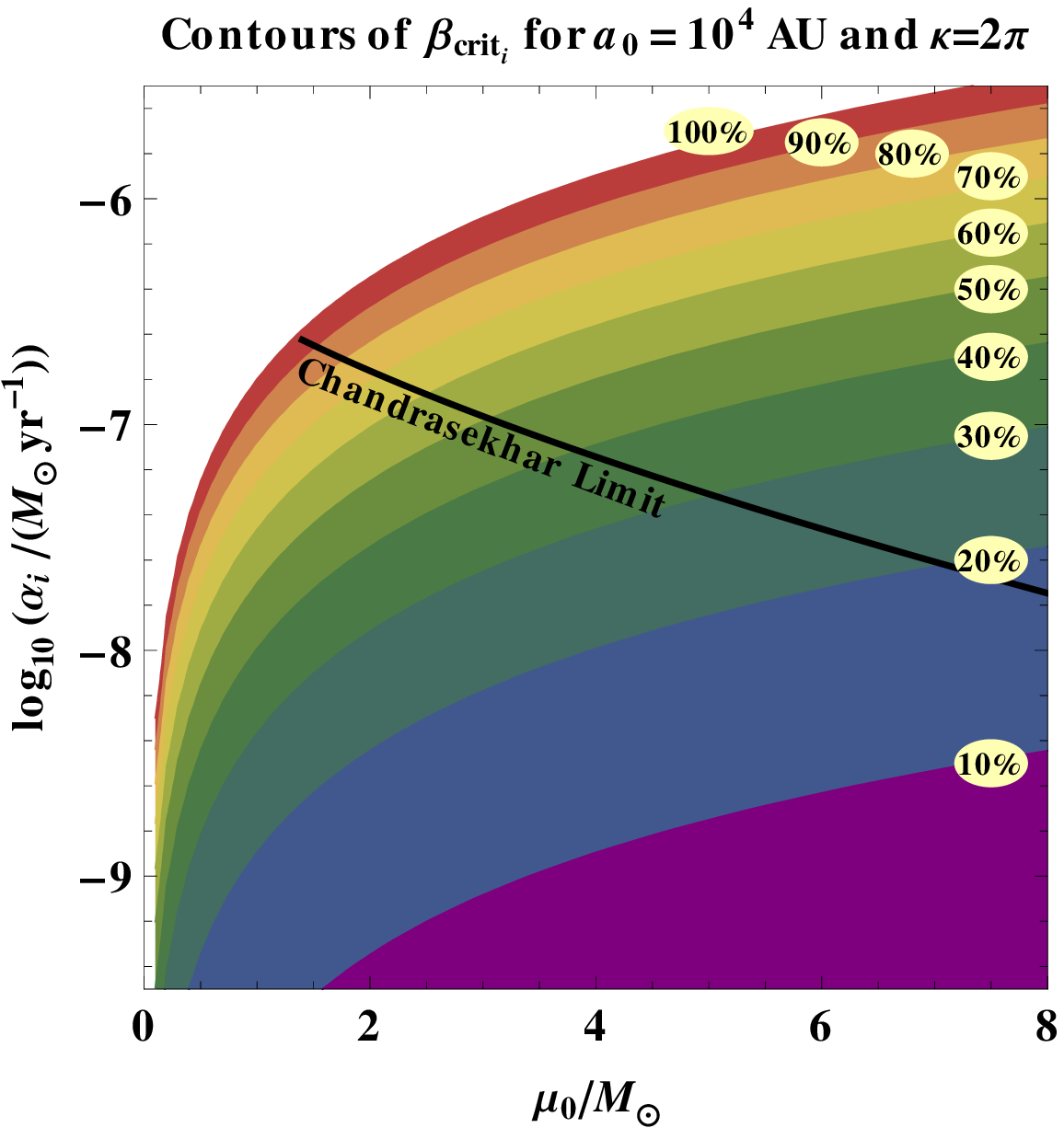,height=8.5cm,width=8.5cm} 
}
\centerline{}
\centerline{
\psfig{figure=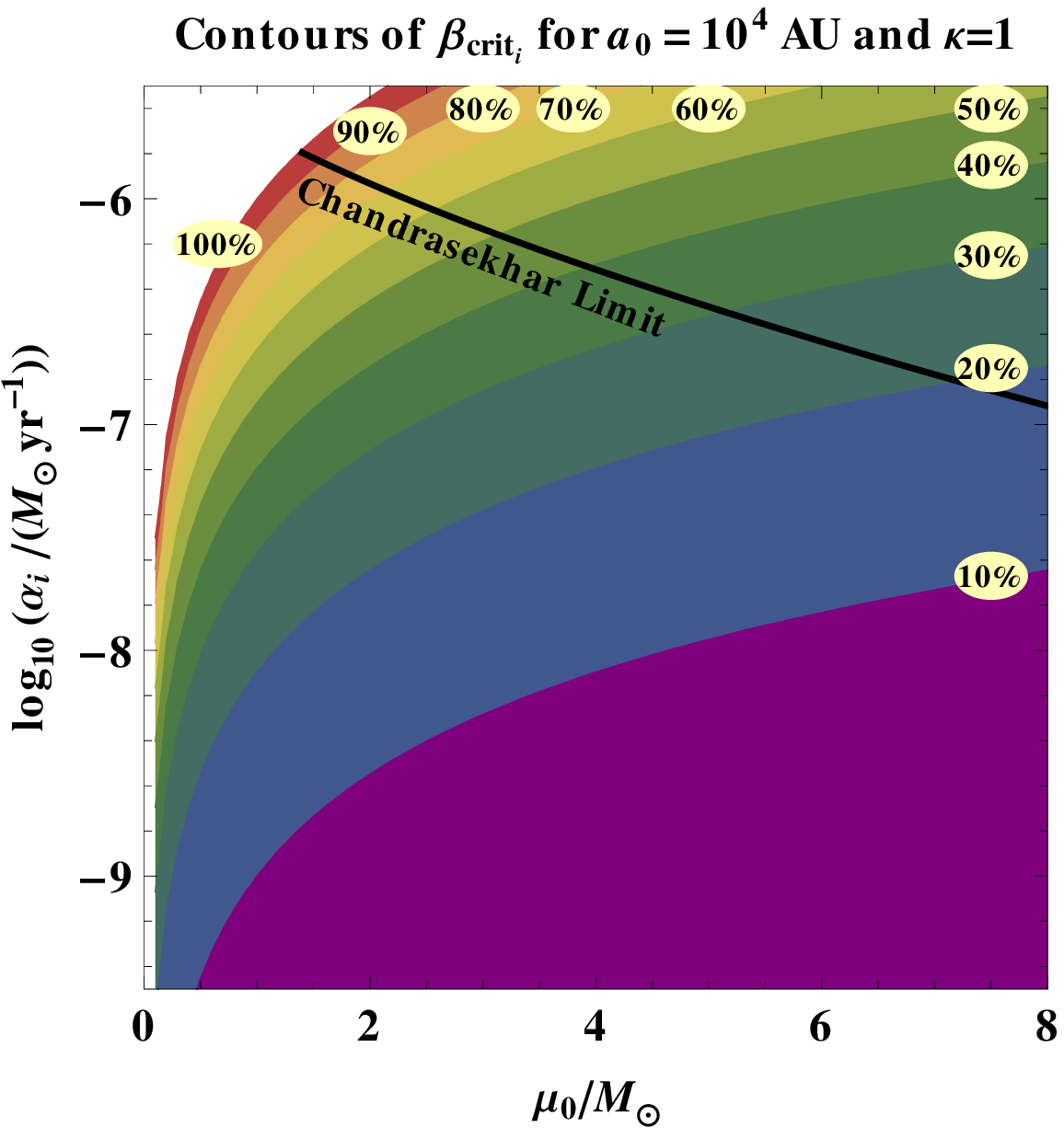,height=8.5cm,width=8.5cm} 
}
\caption{
Criteria to ensure planet retention.  
Each contour represents $\beta_{{\rm crit}_i}$, the minimum percent 
of the original system mass ($\mu_0$) which
must be retained by the end of an arbitrary evolutionary
stage $i$ in order to guarantee that a planet (at $a_0 = 10^4$ AU)
remains bound.  Different colours distinguish
regions between 10 per cent differences in contour lines. In the upper panel,
the mass loss index is defined with respect to the mean
motion (corresponding to $\kappa = 2\pi$; Eq. \ref{mlindex}).
In the lower panel, the mass loss
index is defined with respect to the orbital period ($\kappa = 1$). 
The $y$-axis, which has the same scale on both panels, represents the mass
loss rate from the system during the $i$th stage
of stellar evolution.
The thick black curve applies only to single stars
where the planetary mass is negligible compared to the stellar
mass and represents the maximum fraction
of mass retained by the star if it is to become a white dwarf.
Decreasing or increasing $a_0$ by one order of magnitude adds
or subtracts approximately $1.5$ units to the values on the vertical
axes but otherwise leave the plots unchanged.
}
\label{chandra}
\end{figure}

\subsubsection{Visualizing the Criterion}

We can visualize the limits on a planet's escapability
through Fig. \ref{chandra}, which 
can be applied to systems
of any stellar multiplicity in which a planet's orbit
surrounds all of the stars.  The upper and lower panels
respectively illustrate planet retention bounds when
one characterizes the mass loss index 
with respect to the mean motion (corresponding to $\kappa = 2\pi$; Eq. \ref{mlindex})
and the orbital period ($\kappa = 1$).
The contours are
values of $\beta_{{\rm crit}_i}$, and are a function
of $\alpha_i$ and $\mu_0$ for a planet at $a_0 = 10^4$ AU.
Colours denote regions separated by
adjacent contour levels in 10 per cent intervals in masses.  
The upper panel of the figure demonstrates, for example, that a single $2 M_{\odot}$ star 
which loses $60$ per cent of its mass by
the end of the $i$th stage of evolution is guaranteed to retain an orbiting
planet at $a_0 = 10^4$ AU if $\alpha_{i} \lesssim 10^{-7.6} M_{\odot}$ yr$^{-1}$.


Stars with masses up to $8 M_{\odot}$ may
become white dwarfs, whose maximum mass cannot exceed $1.4 M_{\odot}$. 
Therefore, in the single star limit,
the thick black curve denotes the Chandrasekhar Limit. 
Thus, in the upper panel, a $4 M_{\odot}$ star which becomes a white dwarf
has retained at most about $35$ per cent of that mass
during the transition.  Planetary retention in such a system
is guaranteed only if the mass loss rate during the transition is 
$\alpha_{i} \lesssim 10^{-7.2} M_{\odot}$ yr$^{-1}$.

For planetary semimajor axes other than $10^4$ AU, the plots are shifted
vertically but otherwise remain unaltered.  For every order
of magnitude that $a_0$ is decreased, the vertical axis values
are increased by $10^{1/6} \approx 1.5$ units.  Therefore,
for $a_0 = 10^5$ AU, which represents a typical Oort cloud distance,
$\alpha_{i}$ is rarely low enough to ever guarantee protection of
Oort cloud comets. 

\subsubsection{Stellar Multiplicity Comparison} \label{seclabel}

We can also extend the figure to 
selected multiple star situations.  Suppose that the primary
has just become a white dwarf but the secondary is still on
the main sequence and is losing a negligible amount of mass
such that $M_2 = M_{2_{0}} = \gamma M_{1_{0}}$.
In this case, the maximum value of $\alpha_{i}$ which can guarantee
planetary protection becomes

\begin{equation}
\alpha_i = \left[ \frac{1.4 M_{\odot} + \gamma M_{1_0}}{\sqrt{a_0 M_{1_0} \left(1 + \gamma\right)}}  \right]^3
.
\label{inhibit}
\end{equation}

\noindent{This} equation describes a limiting curve which can be 
drawn on Fig. \ref{chandra} for a given binary system.
Here, $\alpha_i$ takes on an absolute maximum when $\gamma M_{1_{0}} = 1.4 M_{\odot}$.
Hence, the maximum possible value of the ratio of the binary to single star values of 
$\alpha_i$ for a given
$M_{1_{0}}$ and $\gamma$ is about 
$107$
for the extreme bounds of $M_{1_{0}} = 8 M_{\odot}$ and $\gamma = 1$.
For the vast majority of realistic systems, however, this ratio is about
a few.  This exercise, which could be generalized to higher multiplicities,
demonstrates that the mass-loss rate for a main-sequence--white dwarf
binary system needs to be slightly higher than in the single star--white
dwarf case in order to eject a wide-orbit planet.

\subsubsection{Stellar Stage Comparison}

In order to deal with generic systems
at all evolutionary stages,  
reconsider Eq. (\ref{crit}).
Because $\beta_{{\rm crit}_i}$ is not 
explicitly dependent on the mass loss rates
during the earlier stages of evolution, we can
establish a critical value for the entire system
evolution of

\begin{equation}
\beta_{\rm crit} = {\rm Max}\left(\beta_{{\rm crit}_i}\right)
,
\label{bcrit}
\end{equation}

\noindent{where} $i=1...N$ represents the $i$th stage of evolution.

Comparing $\beta_i$ with $\beta_{\rm crit}$ 
determines the prospects for ejection.
If $\beta_i > \beta_{\rm crit}$ for all 
$\beta_i$, then
ejection {\it cannot occur}.   
If $\beta_i \lesssim \beta_{\rm crit}$
for at least one $\beta_i$,
then ejection {\it may occur}.
This is a general condition which may
be applied to systems with any
stellar multiplicity.  In this work,
we focus on binary systems, as they
often represent components of higher
multiplicity stellar systems and feature
a representative set of physical processes
that may be found in those systems.

\subsubsection{The Critical Semimajor Axis}

Now suppose that $a_0$ is not given.  Then we can determine the
critical value of $a_0$, $a_{\rm crit}$, for which the planet is guaranteed
to remain bound through Eq. (\ref{crit}).  We obtain

\begin{equation}
\frac{a_{\rm crit}}{1 {\rm AU}} 
\approx
\kappa^{-\frac{2}{3}}
\left( \frac{\mu_0}{1 M_{\odot}} \right)
{\rm min}
\left[
\beta_{{\rm crit},i}^2
\left( \frac{\alpha_i}{1 M_{\odot}/{\rm yr}}\right)^{-\frac{2}{3}}
\right]
.
\label{acrit}
\end{equation}

\noindent{where} the minimum is taken over all values of $i$.  Although 
\cite{verwya2012} apply this criterion for the specific case of the Sun, 
the criterion is generally applicable to multiple star systems.  We
use Eq. (\ref{acrit}) to determine $a_{\rm crit}$ throughout the
remainder of the paper.

\begin{figure*}
\centerline{
\psfig{figure=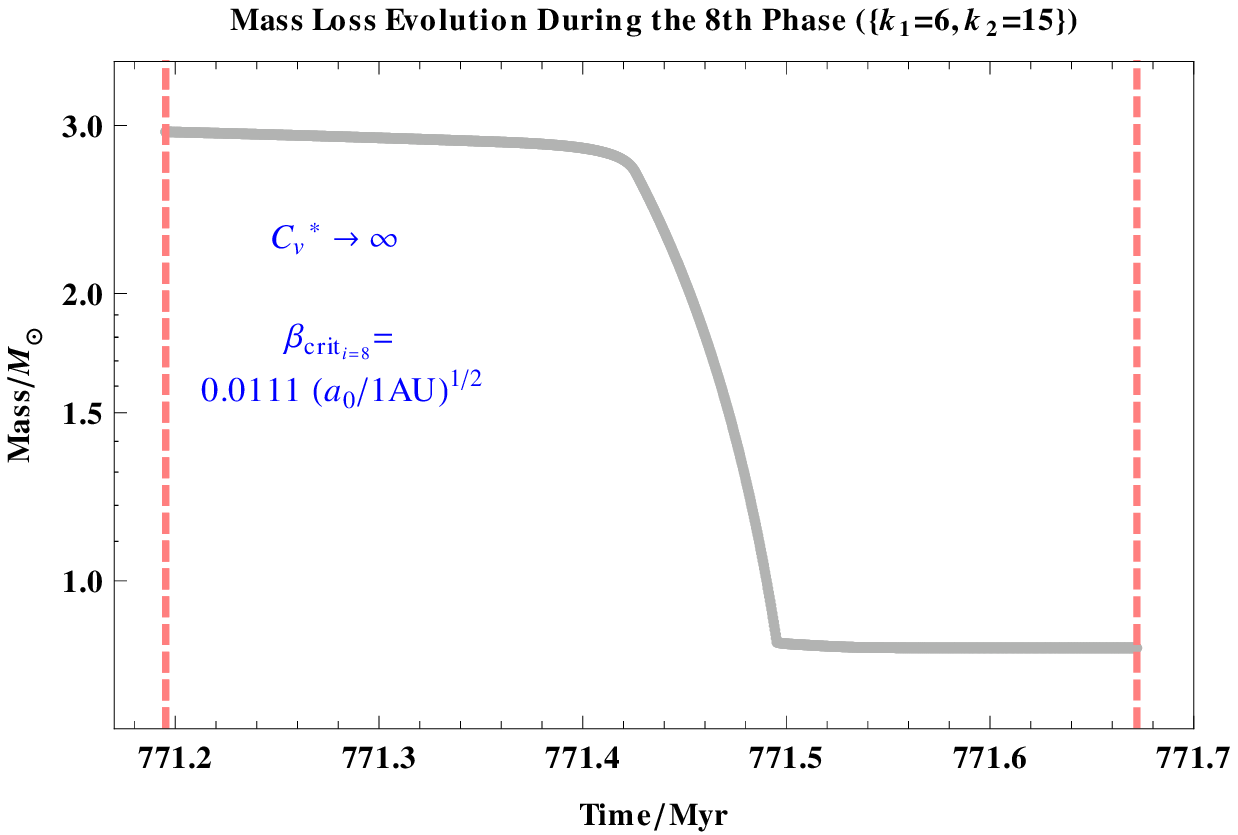,height=6.2cm,width=9.0cm} 
\psfig{figure=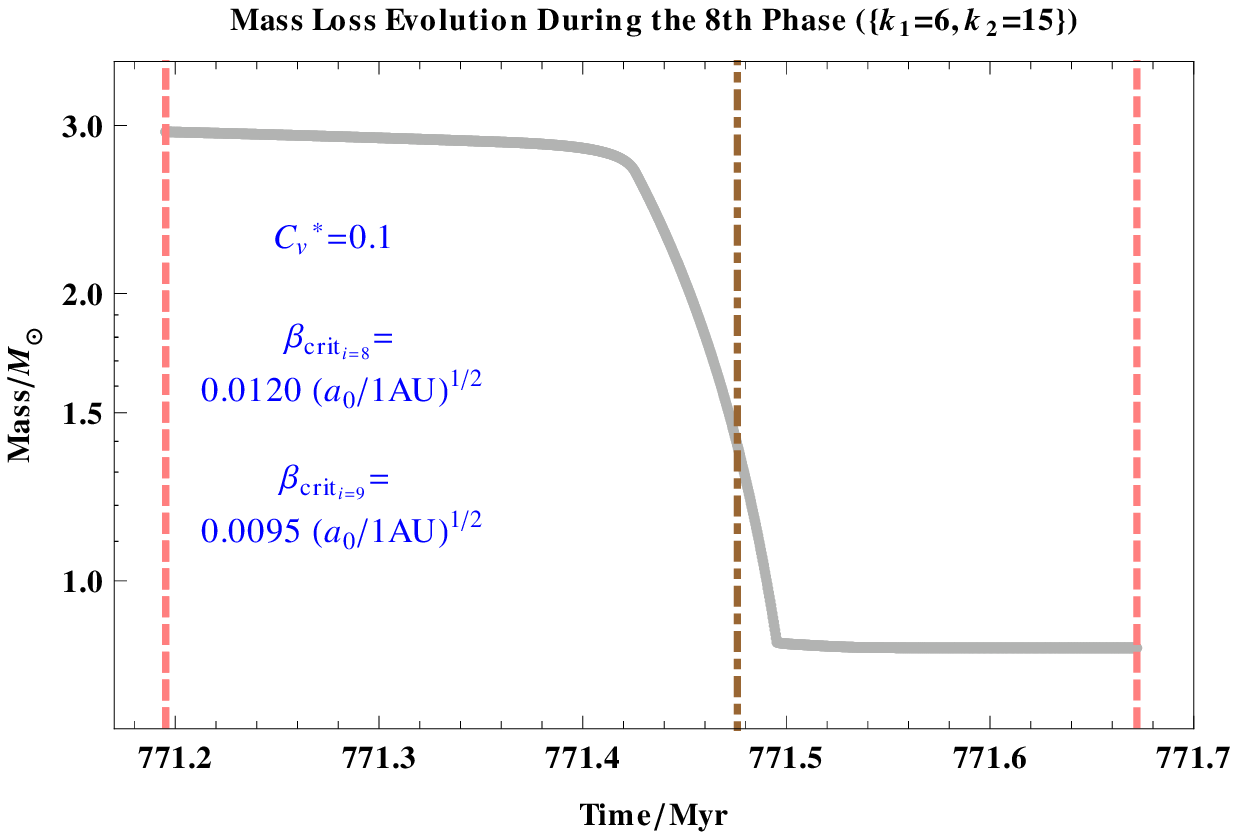,height=6.2cm,width=9.0cm}
}
\
\
\centerline{
\psfig{figure=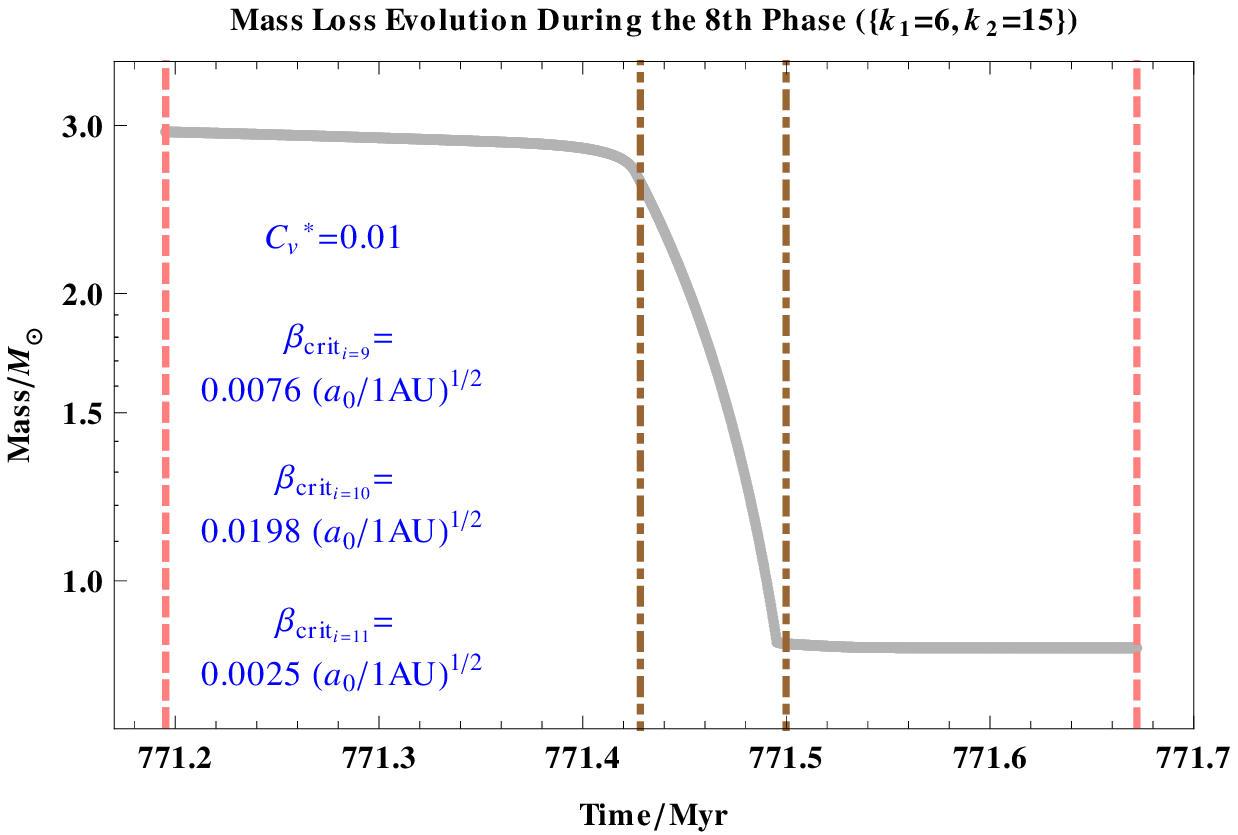,height=6.2cm,width=9.0cm} 
\psfig{figure=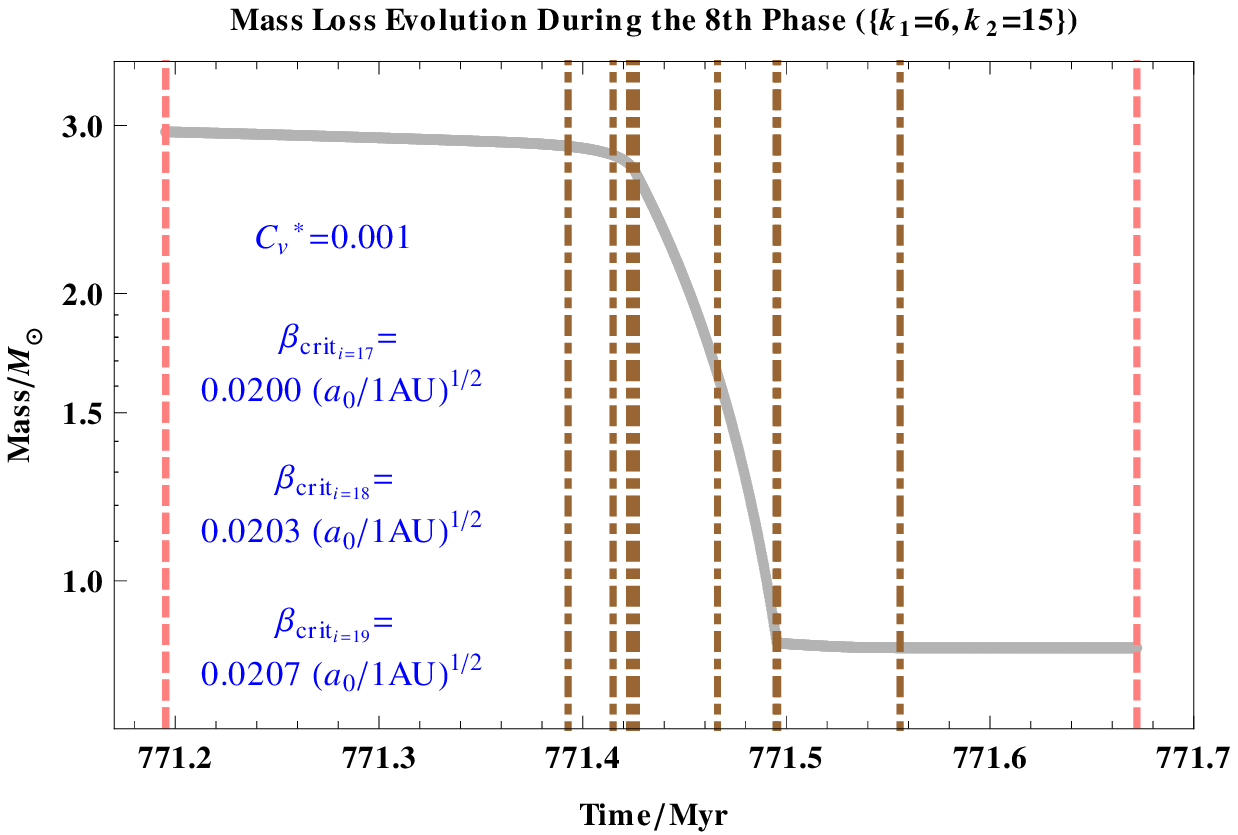,height=6.2cm,width=9.0cm}
}
\caption{
The representation of nonlinear mass loss as a sequence of
consecutive linear stages.  The coefficient of variation, $C_{v}$,
about a linear fit to the data determines
how to partition nonlinear mass loss within a phase.
Plotted are partitions for the 
8th stellar
evolutionary phase of a system with
$M_{1_0} =5.2 M_{\odot}$,
$M_{2_0} =2.2 M_{\odot}$, 
[Fe/H] $=$ [Fe/H]$_{\odot} = 0.02$,
$a_{B_0} = 100 R_{\odot}$ and $e_{B_0} = 0.0$,
when $M_1$ and $M_2$ have merged and are
undergoing AGB evolution.
The thick gray curve represents the 
actual mass loss profile.
The phase boundaries are given by the red dashed
lines, and the stage boundaries within this phase
are given by the brown dot-dashed lines.  
Values of $\beta_{{\rm crit}_i}$
are computed using Eq. (\ref{crit}).  The maximum
of these values (Eq. \ref{bcrit}) determines the potential for 
the system to retain a planet and must increase as the
number of stages increases.  This potential
is diminished as the profile is modeled more accurately.
}
\label{lintest}
\end{figure*}

\subsection{Treating Nonlinear Mass Loss}

Ideally, a nonlinear mass loss profile may be approximated
with the finest possible partition into linear segments.
\cite{verwya2012} achieved this accuracy by considering
the duration between every SSE timestep as a segment.
Although effective for a detailed study of one system,
this approach might not provide a consistent measure
or be computationally feasible for a population of systems.
Hence, we now consider alternatives.

\subsubsection{The Theory}

We compute $\alpha_k$ by dividing the change 
in mass with the change in time
over the entire phase $k$.  However, in cases
where the mass-loss rate changes nonlinearly within a phase,
we can model the mass loss evolution within the phase
by consecutive stages of approximately linear mass loss.
We select the boundaries for these stages based
on a coefficient of variation, $C_v$, of the data. 
If we fit a linear model to a given time series of data,
then $C_v$ represents the residual sum of squares about that fit
divided by the mean of the response variable (the mass).

An exact fit to the linear model yields $C_v = 0$.  If we insist that
each evolutionary stage satisfies $C_v \le C_{v}^{\ast}$,
where $C_{v}^{\ast}$ is a user-provided parameter,
then the quality of our analytical treatment from Section 2
relies on the value of $C_{v}^{\ast}$.  If $C_{v}^{\ast}$ is too low,
and a phase is split into too many stages, the resulting
analysis becomes computationally expensive.  If $C_{v}^{\ast}$ is too high,
and each phase is represented by a single stage, then the
analysis may represent a poor physical approximation.

Suppose a phase $k$ is split into $l$ stages, where $j=1,...l$.
Then Max$(\alpha_j) \ge \alpha_k$ because mass loss from 
the system is monotonic and at least one of the segments of a 
piecewise linear curve between both phase endpoints
is steeper than the original curve.
Therefore, the limiting value $C_{v}^{\ast} \rightarrow \infty$ does
not define any additional stages and hence provides a 
conservative estimate for planetary ejection.

\subsubsection{An Example}

Consider the binary system with $M_{1_0} = 5.2 M_{\odot}$,
$M_{2_0} = 2.2 M_{\odot}$, 
[Fe/H] $=$ [Fe/H]$_{\odot} = 0.02$,
$a_{B_0} = 100 R_{\odot}$, and $e_{B_0} = 0.0$, where the subscript
$B$ refers to the primary-secondary binary.  Suppose both
stars began life on the main sequence.  This system
undergoes 9 distinct phases of evolution such that 
$k_{1,2} = \lbrace(1,1), (2,1), (3,1), (7,1), (8,1), (11,1), (6,1), (6,15), (11,15)\rbrace$.
This sequence illustrates that $M_1$ evolves
off the main sequence first, on to the Hertzsprung gap and then the red 
giant branch.  The primary's envelope is then blown away, leaving
behind a naked helium star, which again begins a main sequence
then Hertzsprung gap phase.  The primary eventually becomes a white dwarf.
However, at this point, the secondary is roughly three times more massive
and is siphoning matter on to the primary.  The primary is revitalized
as a thermally pulsing AGB star, which soon after merges with 
the secondary.  The new star continues AGB evolution until it becomes
a white dwarf and dies out as such.

The mass loss for the $k_{1,2} = (6,15)$ phase is markedly nonlinear.
Fig. \ref{lintest} shows the mass loss evolution of this phase
for four different values of $C_{v}^{\ast}$.  The latter three split the
phase into stages.  Phase and stage boundaries are given by vertical
red dashed lines and brown dot-dashed lines, respectively.
The total number of stages 
throughout the system evolution that are generated for 
$C_{v}^{\ast} \rightarrow \lbrace\infty,0.1,0.01,0.001\rbrace$ 
are $\lbrace9,10,12,22\rbrace$.  The values of $\beta_{{\rm crit}_{i}}$ are
obtained from $\alpha_{i}$, which in turn are computed
from the intersections of the mass loss curve with the
stage boundaries.  Displayed in the lower-right
panel are the three highest $\beta_{{\rm crit}_{i}}$ values
of the many from that stage.
Note that, as $C_{v}^{\ast}$ decreases, 
$\beta_{{\rm crit}} \equiv {\rm Max}\left(\beta_{{\rm crit}_i}\right)$
increases.  This must be the case because at least one of the segments
of a piecewise linear curve between both phase endpoints is
steeper than the original curve.  The most accurate
$\beta_{{\rm crit}}$ value is a factor of $1.86$ greater
than the least accurate value and corresponds to a factor of
$\approx 3.5$ in $a_0$.

\section{Application to Realistic Systems}

In this section, we attempt to sample the
entire phase space of binary systems
in order to determine prospects for planetary
ejection.  We first argue that large regions of the
phase space may be treated in a similar manner.  For 
the remaining
regions, we utilize a proven binary stellar evolution
code that can generate tracks quickly, without
solving coupled differential equations.  
Finally, we present our results through a series of
contour plots that provide information about
features of the binary systems studied
as well as useful parameters such as the critical
semimajor axis at which a planet is guaranteed
to remain bound.

\subsection{Stellar Evolution Code}

We use a slightly modified form of the BSE stellar 
evolution code \citep{huretal2002}.  This code 
utilizes empirical, algebraic formulae
derived from observational sources and theoretical
frameworks in order to model evolving binary stars.
Although the code tracks a wide range of stellar parameters,
here we are concerned solely with the mass-loss
rate from the system and the physical reasons
for this mass loss.  We assume that the circumbinary 
orbit of the planet is always wider than
the apocentre and common envelope radius
of the binary.  As shown in Paper I, for monotonic
stellar mass loss, a planet's semimajor axis
{\it must} always be increasing, 
regardless of the shape of the orbit
or the planet's position along that orbit.

We have modified BSE with an updated form
of the mass loss for stellar wind from naked
helium stars. We have included a dependence on
metallicity of $\dot{M} \propto \sqrt{Z}$ according 
to Eq. (22) of \cite{nuglam2000}.  The mass-loss prescription we used 
is also incorporated in the NBODY6 N-body
code (which may be downloaded from the website,
http://www.ast.cam.ac.uk/$\sim$sverre/web/pages/nbody.htm,
as of December 2011).  We also modified BSE to withhold output data dumps
within evolutionary phases
until at least $0.01$ per cent of the entire system mass has been
lost.  This allows us both to use a mass-dependent 
uniform standard for analysis of all our simulations
and to enable us to cover a wide region of phase
space by limiting the number of outputs 
(particularly for main-sequence phases) in each simulation.
We always output the first instance of a new evolutionary phase.

\subsection{Phase Transitions}

The phase and stage formalism described
so far does not treat mass lost from a system 
during a phase transition.  The BSE
code computes the timescales and mass loss
for most phase transitions.  In these cases,
we consider the mass loss between consecutive
timesteps that encompass the phase transition
to be linear.  However, BSE treats some transitions 
as instantaneous.  Although this approximation
may be adequate for numerous stellar applications,
such as population
studies, here that approximation would cause
$\alpha \rightarrow \infty$ and hence
cannot be used.

In the single star
case, the only example of such an
instantaneous transition occurs
during a supernova. 
However, for binary stars, there are several
more examples.   In a comprehensive sampling
of phase space, each of these instances must
be accounted for and a timescale established for each.  
We found that all such
instances can be partitioned into four types
of violent phenomena, i) core-collapse supernovae,
ii) black hole core collapse, iii) thermonuclear supernovae
and iv) common envelope evolution.  In the following 
arguments, we assume that mass loss is isotropic and do 
not distinguish amongst neutrinos, ejecta and 
gravitational waves as separate sources of energy or 
mass liberation.  Accurate modeling of these phenomena 
would likely require removing these simplifications and 
is beyond the scope of this study.

\subsubsection{Core Collapse Supernovae}

Commonly referred to as a Type II, Type Ic or Type Id supernovae,
a core-collapse supernovae is a stellar explosion
of a massive star (of $8-20 M_{\odot}$), typically
while in a giant phase ($k \in \lbrace 3,5,6,9 \rbrace$).
The explosion leaves behind a neutron star ($k = 13$)
or a black hole ($k = 14$).  This event may occur 
in a system of any stellar multiplicity and the fraction
of the exploding star's original mass lost is typically
$50-95$ per cent.

If we assume the tightest-known exoplanet 
orbit around a main sequence 
star\footnote{See the Extrasolar Planets Encyclopaedia at http://exoplanet.eu/}$^{,}$\footnote{
See the Exoplanet Data Explorer at http://exoplanets.org/}
($a \approx 0.015$ AU) and the largest known
progenitor mass which results in a neutron star
($M \approx 20 M_{\odot}$), then no
planet at any semimajor axis is guaranteed
protection if $\alpha \gtrsim 0.9 M_{\odot}$ hr$^{-1}$.
This conservative rate is comparable to the
mass loss timescales that are predicted
from supernova ejection 
velocities \citep[e.g.][]{hampin2002,fesetal2007}.

Such planets are likely destroyed by
the progenitor during its giant phase.
For planets further away, with semimajor axes of $1$ AU, 
this critical mass loss
rate decreases by a factor of about $544$, and
would be enhanced by just a factor of a few in a binary
system, as argued in Section \ref{seclabel}.
Hence, realistic mass-loss rates from supernova would
well exceed the critical mass-loss rate for any planet
that could survive the progenitor's pre-supernova evolution.

Therefore, no planets orbiting stars of progenitor
masses of between about $8$ and $20 M_{\odot}$ can be guaranteed
protection under the criterion of Eq. (\ref{bcrit})
and we can neglect sampling this region of stellar
mass phase space in our simulations.  There are
some low-metallicity cases with
$6 < M/M_{\odot} < 8$ where the star
might undergo core collapse supernova.  In these cases, which correspond
to 

\begin{equation}
k_{w} \in \lbrace 0,1,2,3,4,5,6,7,8,9 \rbrace 
\rightarrow \lbrace 13,14 \rbrace
\end{equation}

\noindent{where} $w = 1$ or $w = 2$, we 
assume that no planet is guaranteed protection.

\subsubsection{Black Hole Core Collapse}

Considering black hole formation in this work is particularly
important because all known black holes are found in binary
systems \citep{beletal2011}.  Progenitor stars with masses greater
than about $20 M_{\odot}$
typically have extended layers of high density
which prevent mass from escaping during an explosion.
Nevertheless, the progenitor still forms a black
hole through core collapse.  The amount of mass loss 
which may accompany this core collapse is unconstrained
by observations and can take on any value.  A theoretical
upper-bound for the timescale for this process is $10$ sec
\citep{ocoott2011}.

This mass-loss timescale is orders of magnitude shorter than
the mass-loss timescale for core collapse 
supernova.  Further, no
planet at any semimajor axis is guaranteed
protection in a system with 
the largest theoretically
postulated progenitor mass which might
undergo core collapse and produce a black hole
($\mu \approx 300 M_{\odot}$, \citealt*{croetal2010})
if $\alpha \gtrsim 0.015 M_{\odot}$/s.  Therefore,
any massive progenitor undergoing supernova-less
core collapse need only lose a few tenths of a solar
mass to place any orbiting planet in danger of ejection.
In a binary system, this amount may be enhanced by a factor
of just a few.  Hence, we can claim conservatively that only in exceptional
circumstances is a planet in this class of systems
guaranteed to remain bound.

This claim allows us to neglect progenitor
masses of $M \gtrsim 20 M_{\odot}$ in our simulations.


\subsubsection{Thermonuclear Supernovae}

Commonly referred to as a Type Ia supernova, a thermonuclear supernova
is an explosion and probable disintegration 
of a white dwarf ($k \in \lbrace 10,11,12 \rbrace \rightarrow 15 $) 
owing to mass accretion from a companion.  Although this
companion is often a giant star, the standard model has subgiant donors.
Alternatively, the companion may represent a white dwarf
\citep{paketal2010}.  The velocity of the ejecta is
comparable to that from core collapse supernovae but
the total amount of mass ejected is typically less than
$1 M_{\odot}$ \citep{mazetal2007}.
Despite this relatively low velocity of ejecta mass, the mass-loss rate
is comparable to that from core collapse supernovae
to within an order of magnitude.  Therefore, because thermonuclear 
supernovae afford little more protection for orbiting planets,
we treat planets subjected to this type of mass
loss in the same way as with core-collapse supernovae
and assume that the planets cannot be guaranteed protection.

\subsubsection{Common Envelope Evolution}

Interactions between close binary stars resulting from
Roche Lobe overflow may precipitate the formation
and subsequent ejection of a common envelope.
Details of common envelope evolution are complex \citep{ivanova2011}
and progress for greater understanding is being made
through three-dimensional hydrodynamical simulations
\citep[e.g.][]{taaric2010,pasetal2011}.  Therefore, establishing a timescale
for common envelope evolution, $t_{\rm ce}$, is difficult.  However,
this timescale, $t_{\rm ce}$, is thought to be within a couple
orders of magnitude of the orbital period of the 
binary\footnote{This orbital period is commonly referred to as a dynamical timescale.}.
Even the lowest estimates for common envelope evolution
timescales guarantee that some planets remain bound.  Therefore,
this transition cannot be treated in the same manner as supernovae
or black hole core collapse.

In cases where BSE treats common envelope evolution
instantaneously, we assign a value to $t_{\rm ce}$.
A value of $t_{\rm ce} = 10^3$~yr represents an upper estimate for this timescale
and with this value we can establish a conservative estimate for escape.  
We also sample values of $t_{\rm ce} = 10, 100$ and $10^4$ yr
for relevant regions of phase space.

\begin{figure*}
\centerline{
\psfig{figure=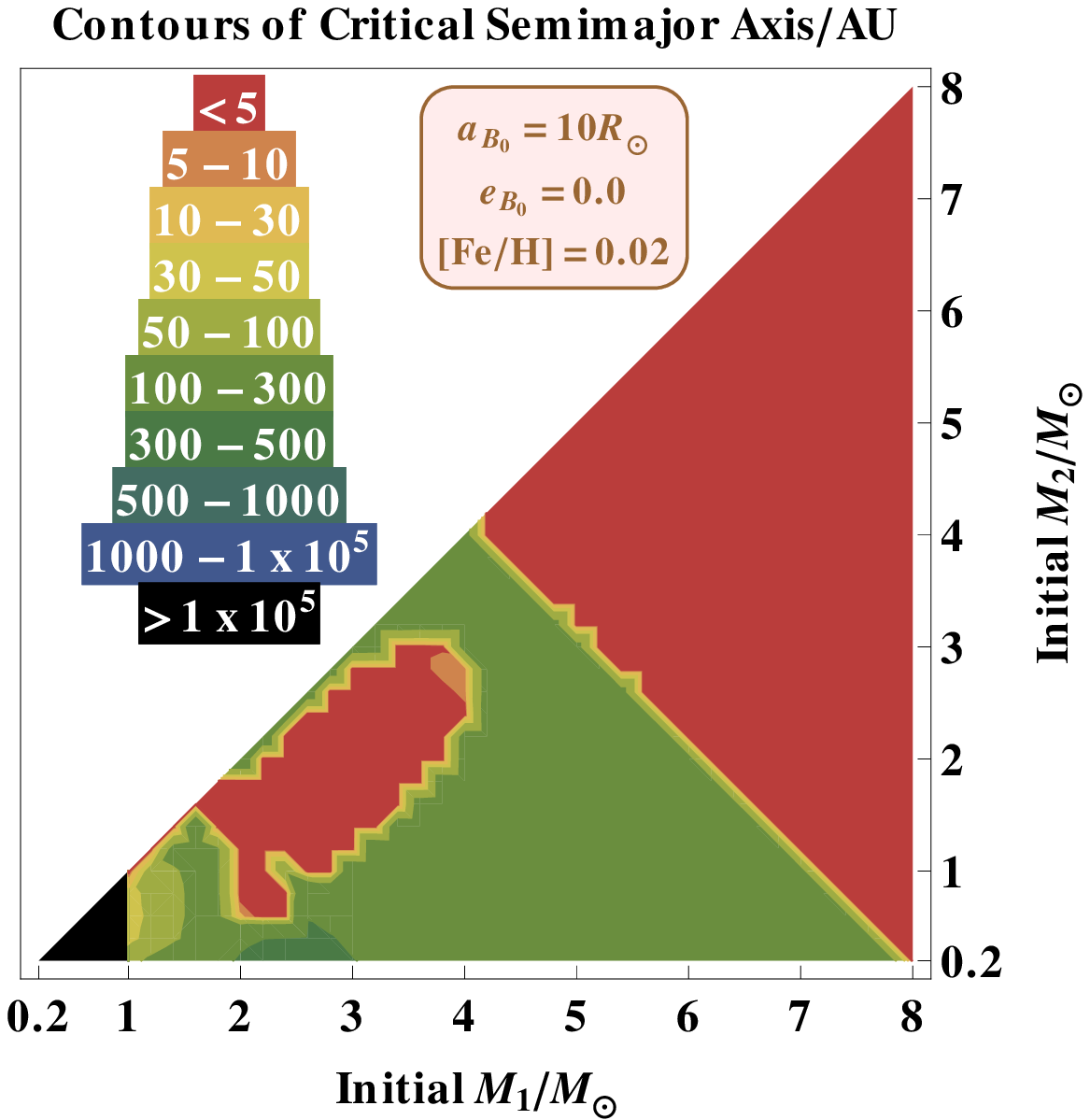,width=6cm} 
\psfig{figure=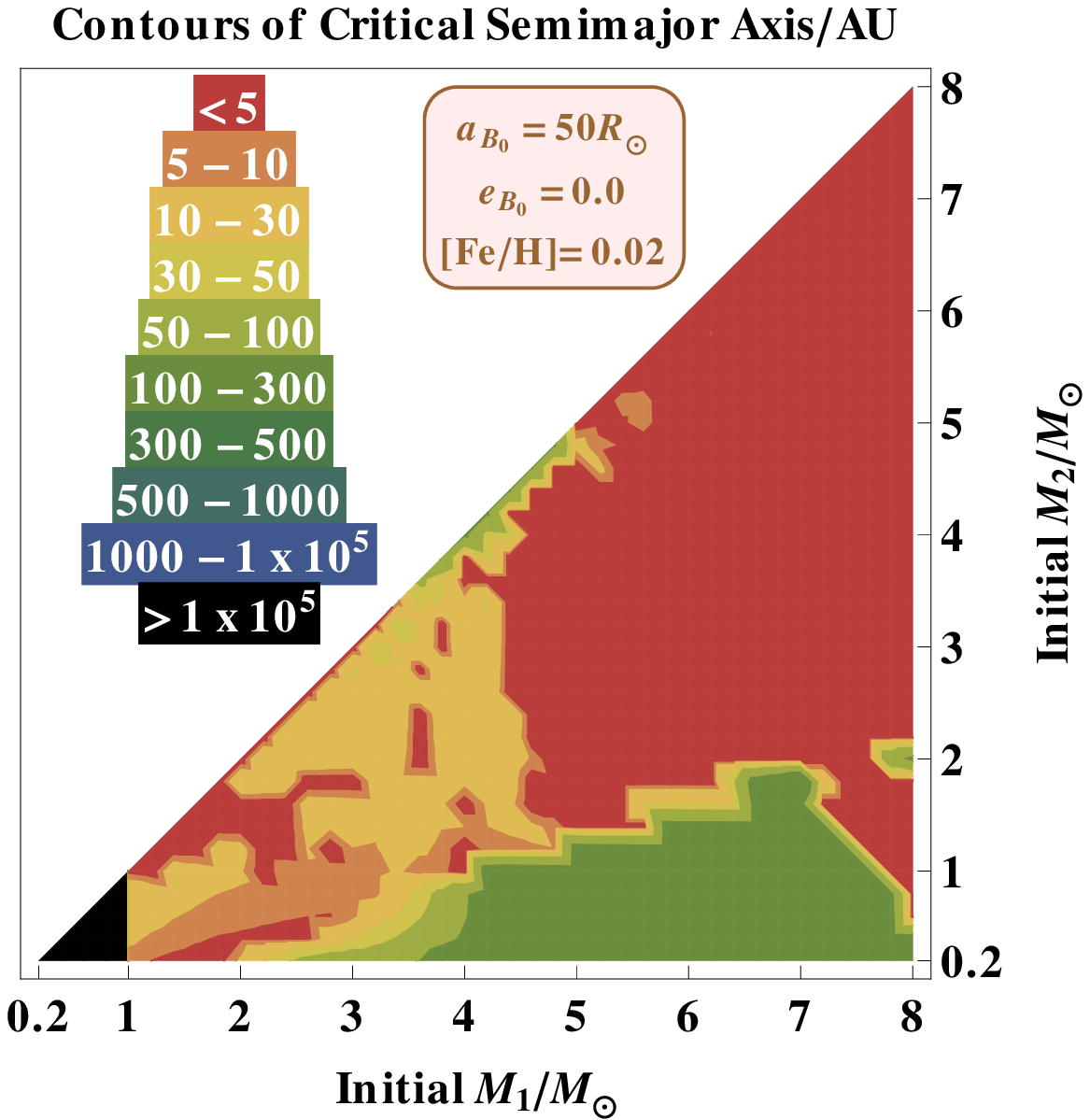,width=6cm}
\psfig{figure=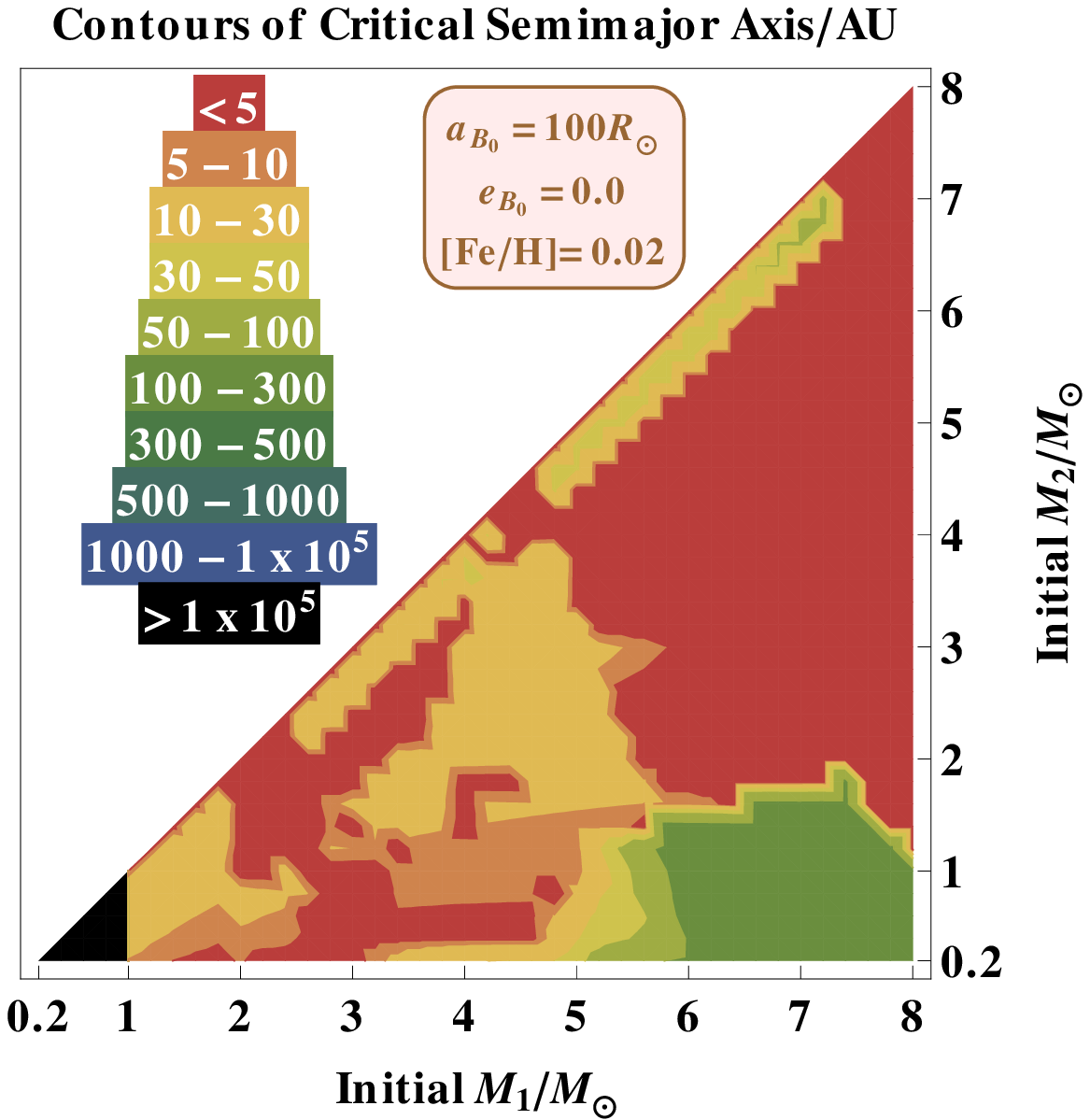,width=6cm}
}
\
\
\centerline{
\psfig{figure=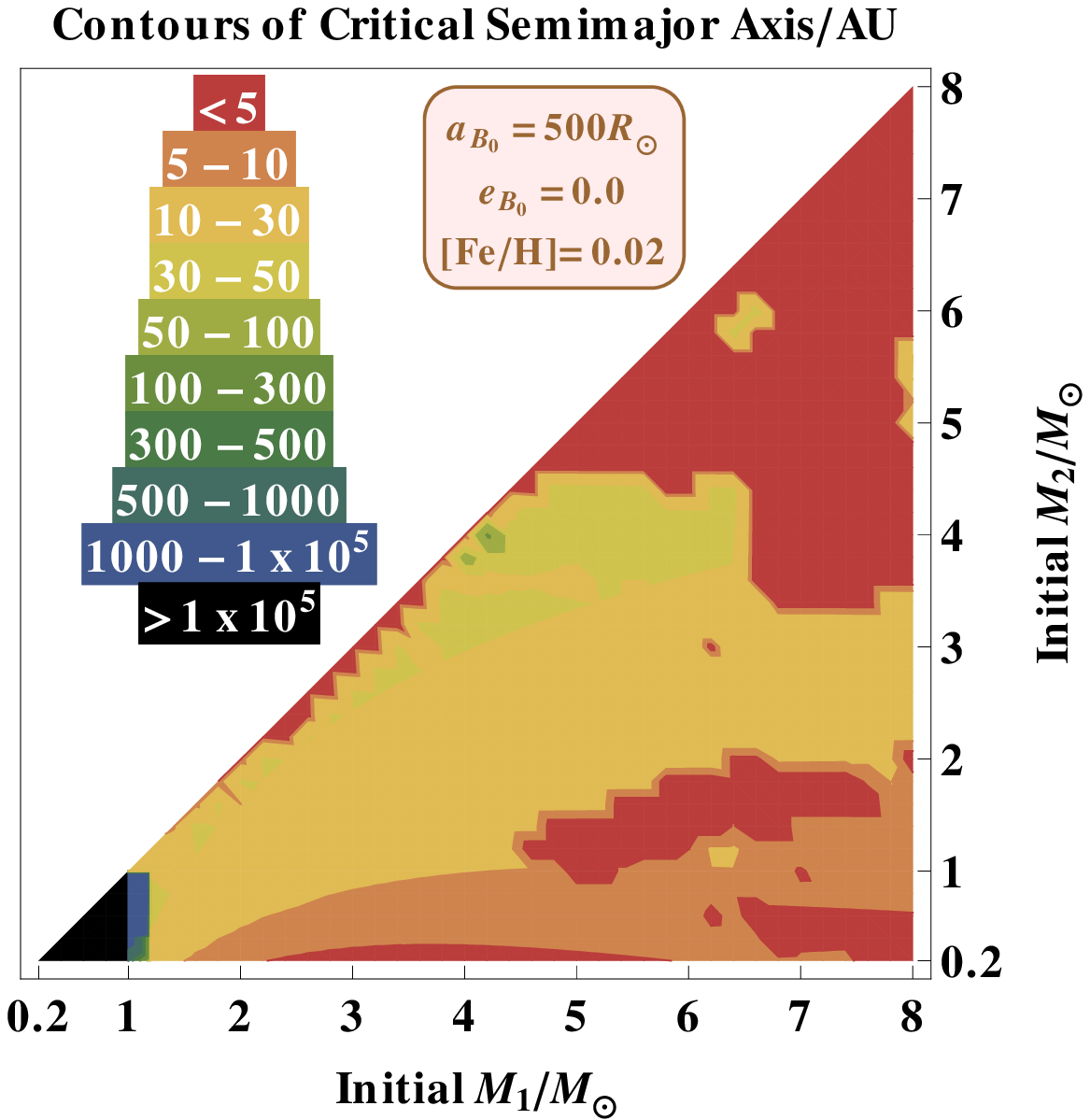,width=6cm} 
\psfig{figure=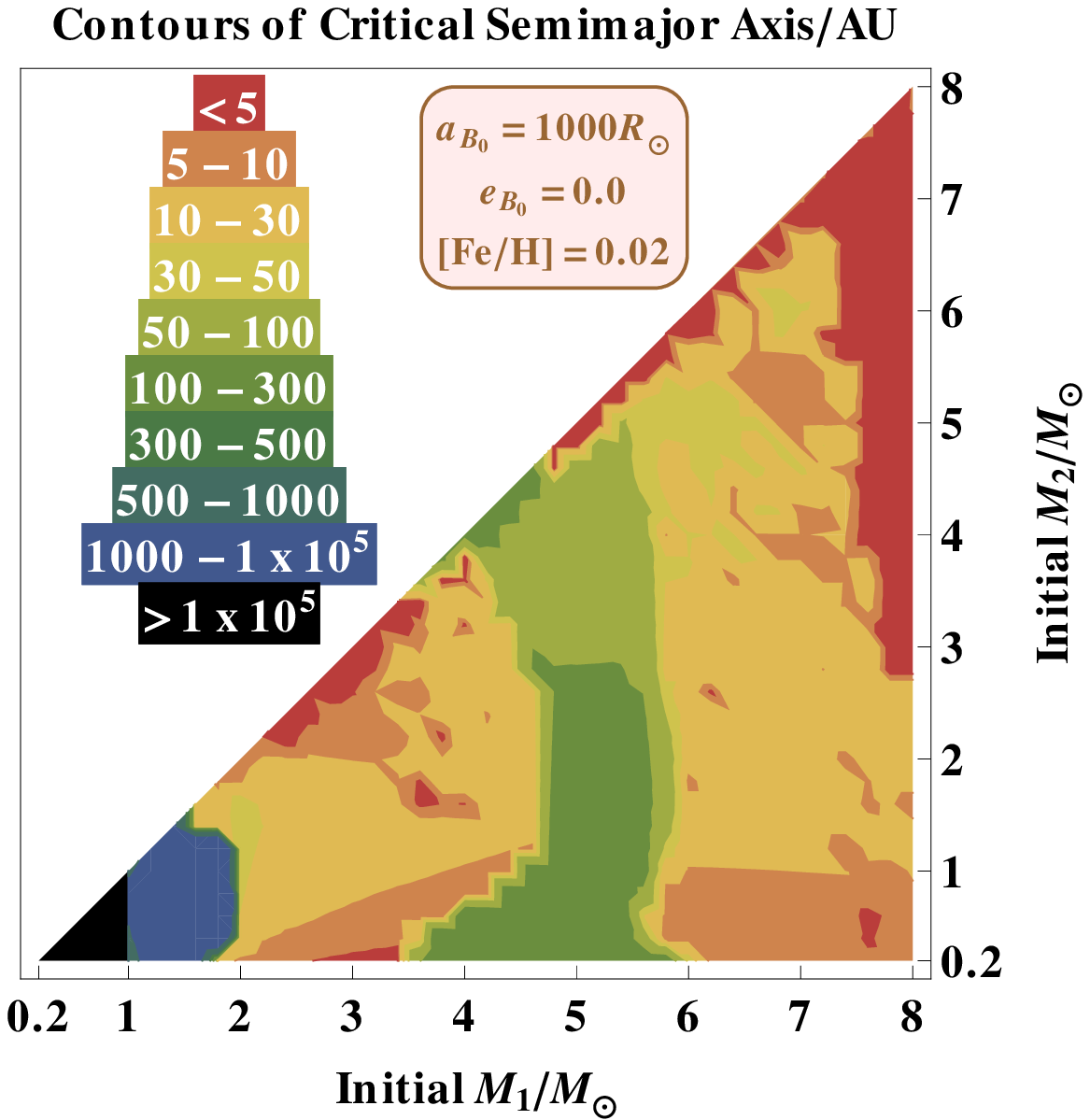,width=6cm} 
\psfig{figure=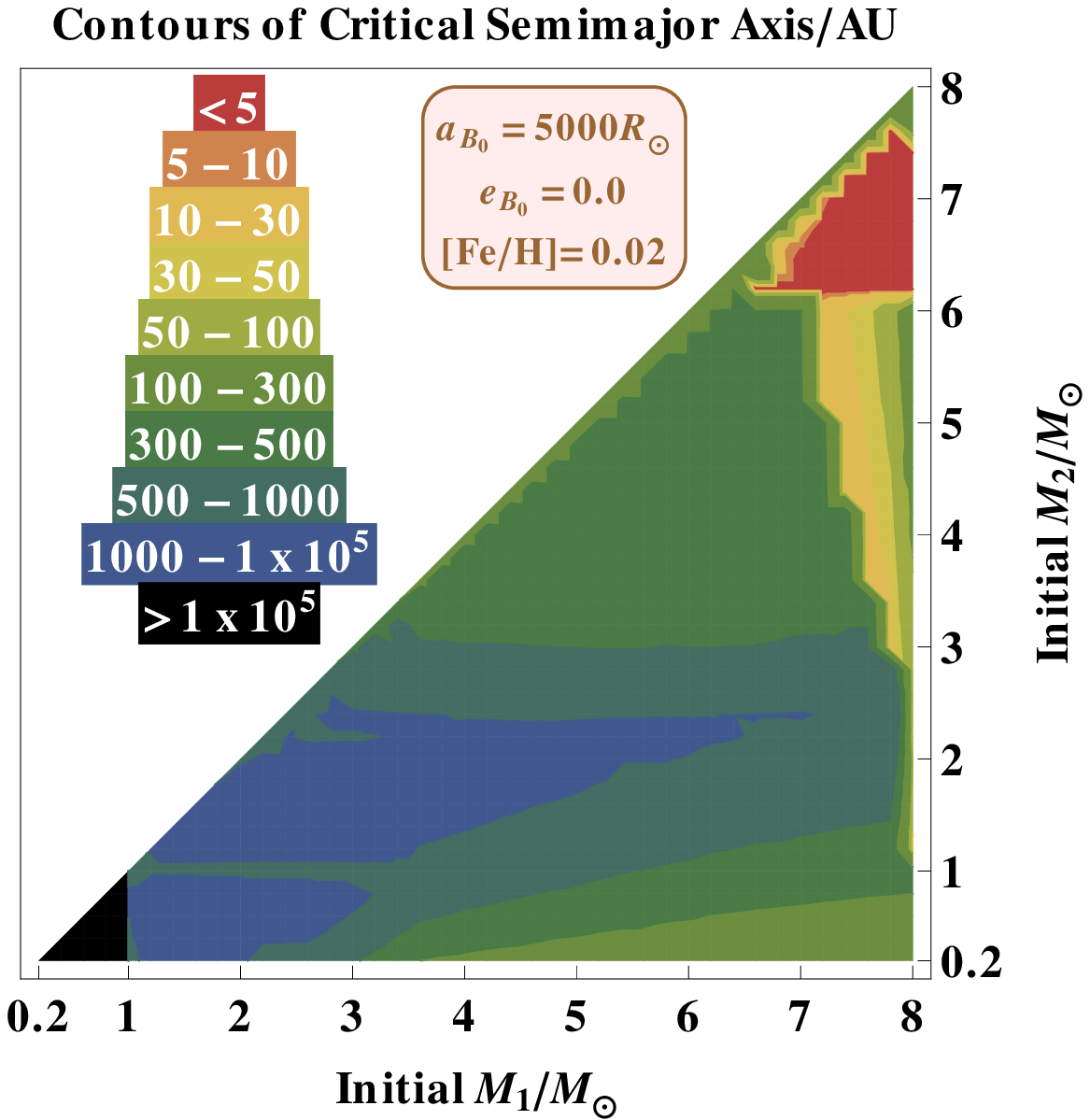,width=6cm} 
}
\
\
\centerline{
\psfig{figure=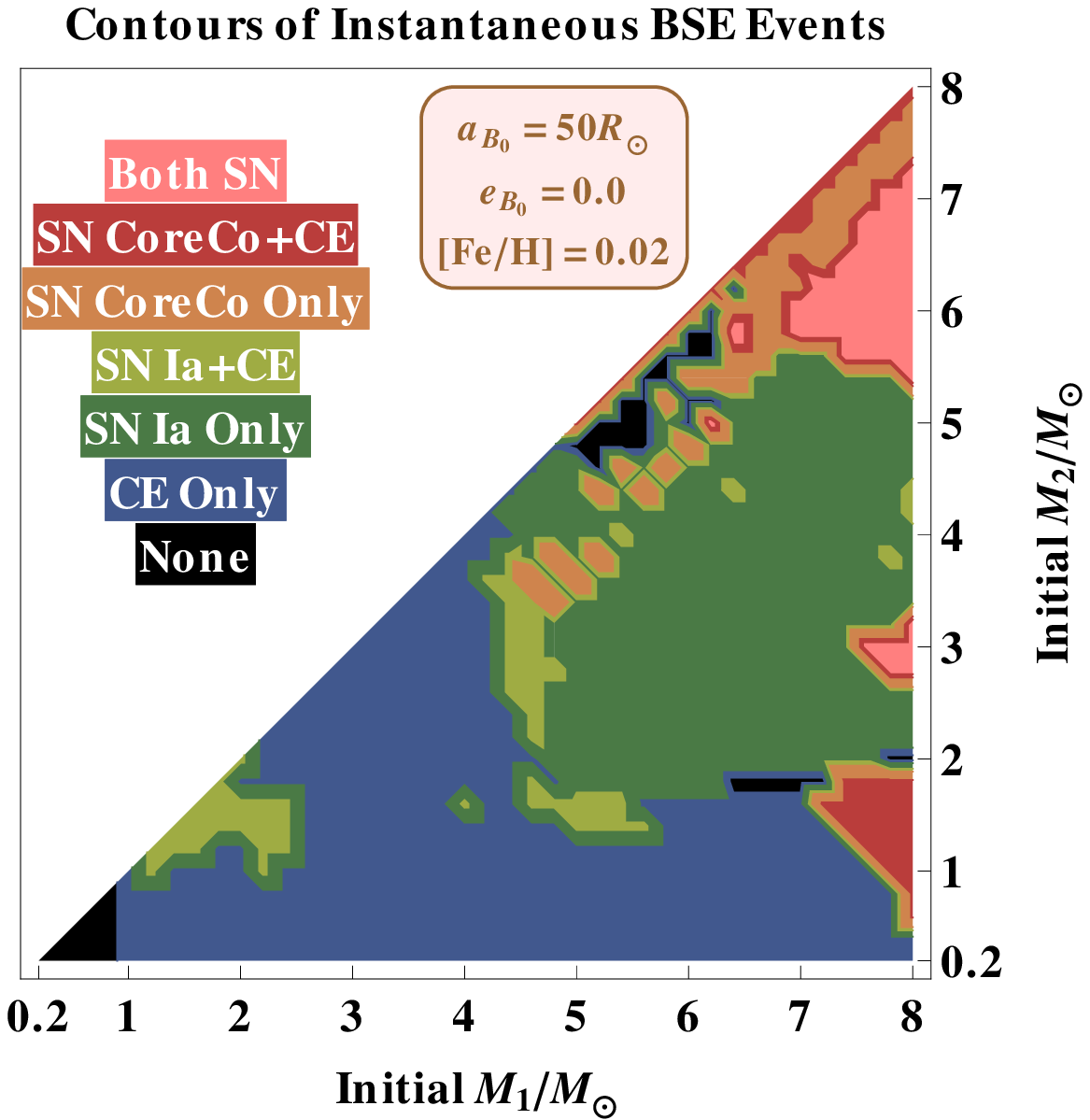,width=9cm} 
\psfig{figure=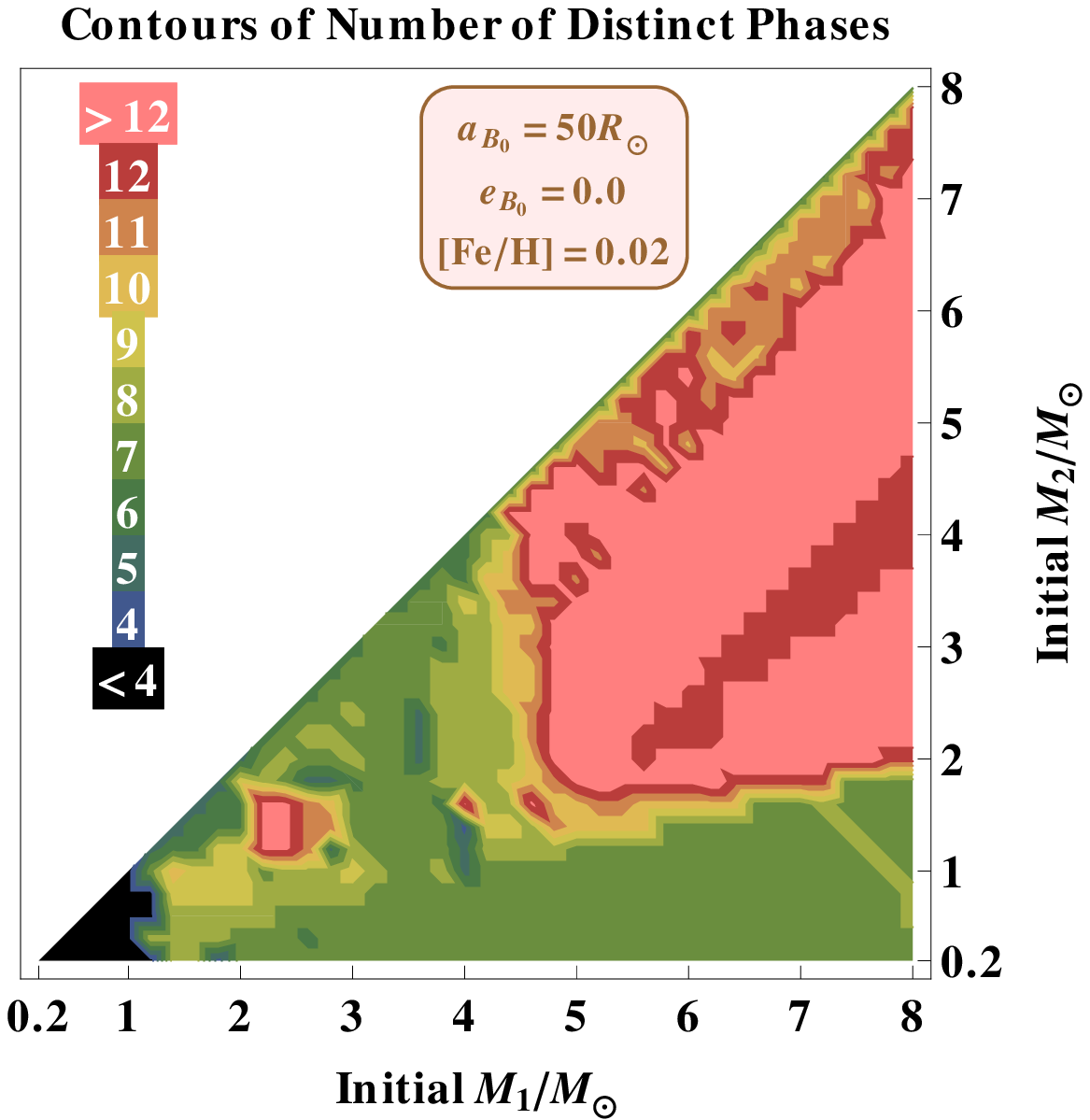,width=9cm} 
}
\caption{
Fiducial prospects for planetary retention.  The contours on the upper two panels
represent the critical semimajor axis at which a planet is guaranteed to remain
bound amidst binary stellar evolution as the
initial binary separation is increased from left to right.  Initially the binary has
a circular orbit and solar metallicity, and we adopt
$t_{\rm ce} = 10^3$ yr and $C_v \rightarrow \infty$.  The lower
panel provides qualitative detail about the evolution of the systems with $a_B = 50 R_{\odot}$.
The contour labels in the lower-left plot are CE Only- at least one common envelope phase
but no supernovae, SN Ia Only- thermonuclear supernova only, SN Ia+CE- thermonuclear supernova
plus at least one common envelope phase, SN CoreCo Only- core collapse supernovae only,
SN CoreCo+CE- core collapse supernovae plus at least one common envelope phase,
Both SN- both core collapse and thermonuclear supernovae.  The figure demonstrates 
that circumbinary planets are highly susceptible
to ejection.
}
\label{cont1}
\end{figure*}

\begin{figure*}
\centerline{
\psfig{figure=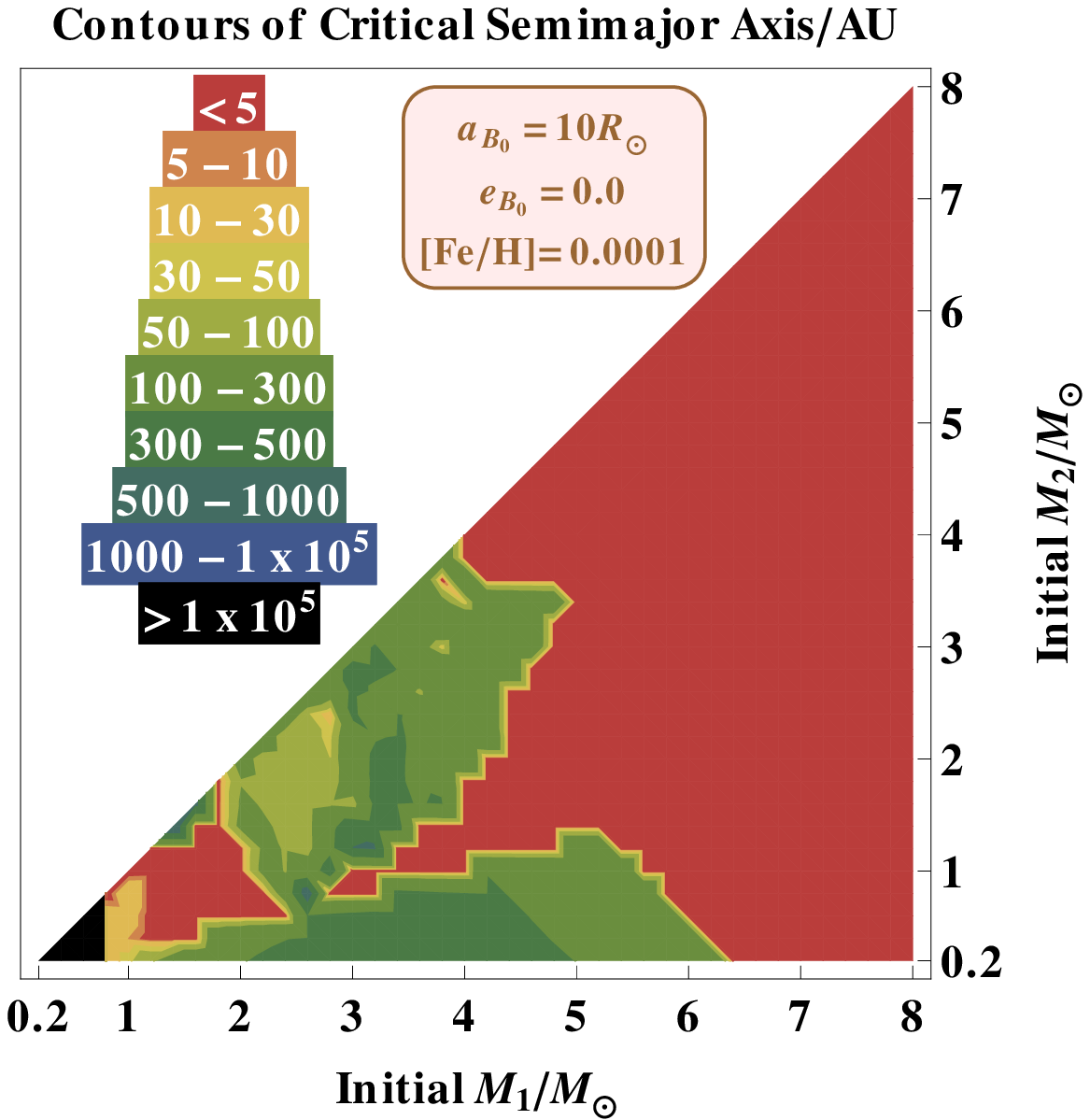,width=6cm} 
\psfig{figure=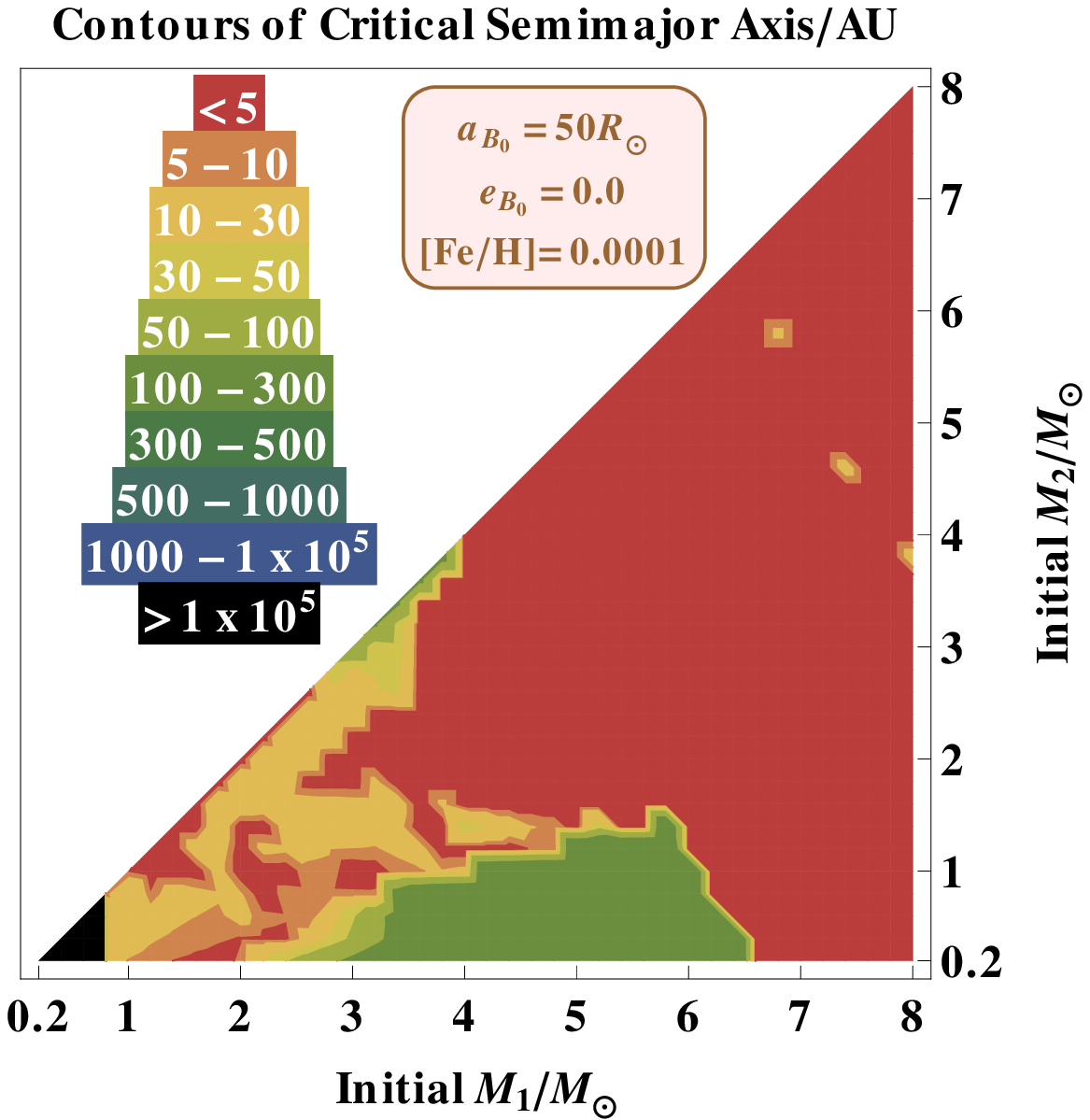,width=6cm}
\psfig{figure=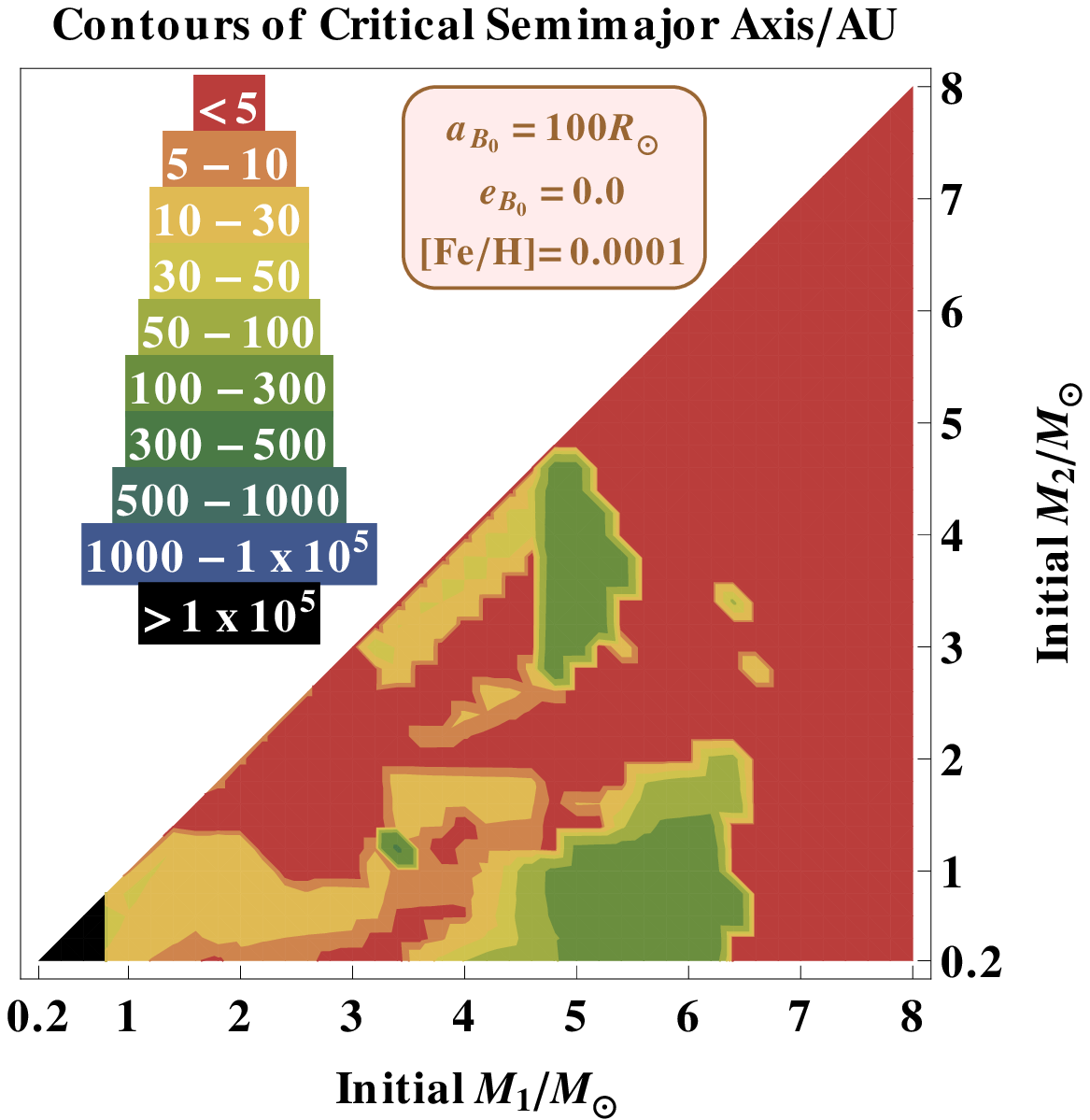,width=6cm}
}
\
\
\centerline{
\psfig{figure=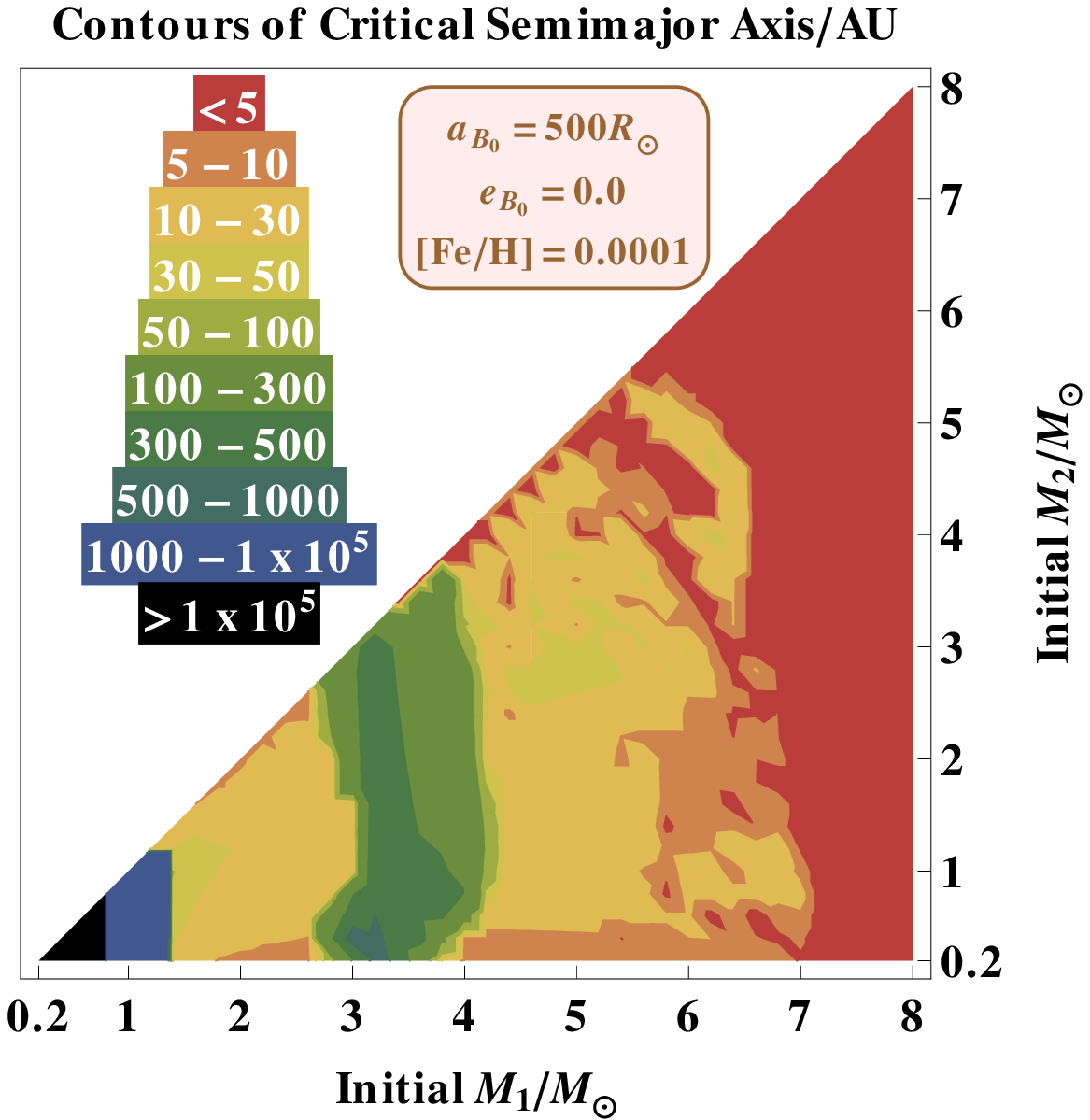,width=6cm} 
\psfig{figure=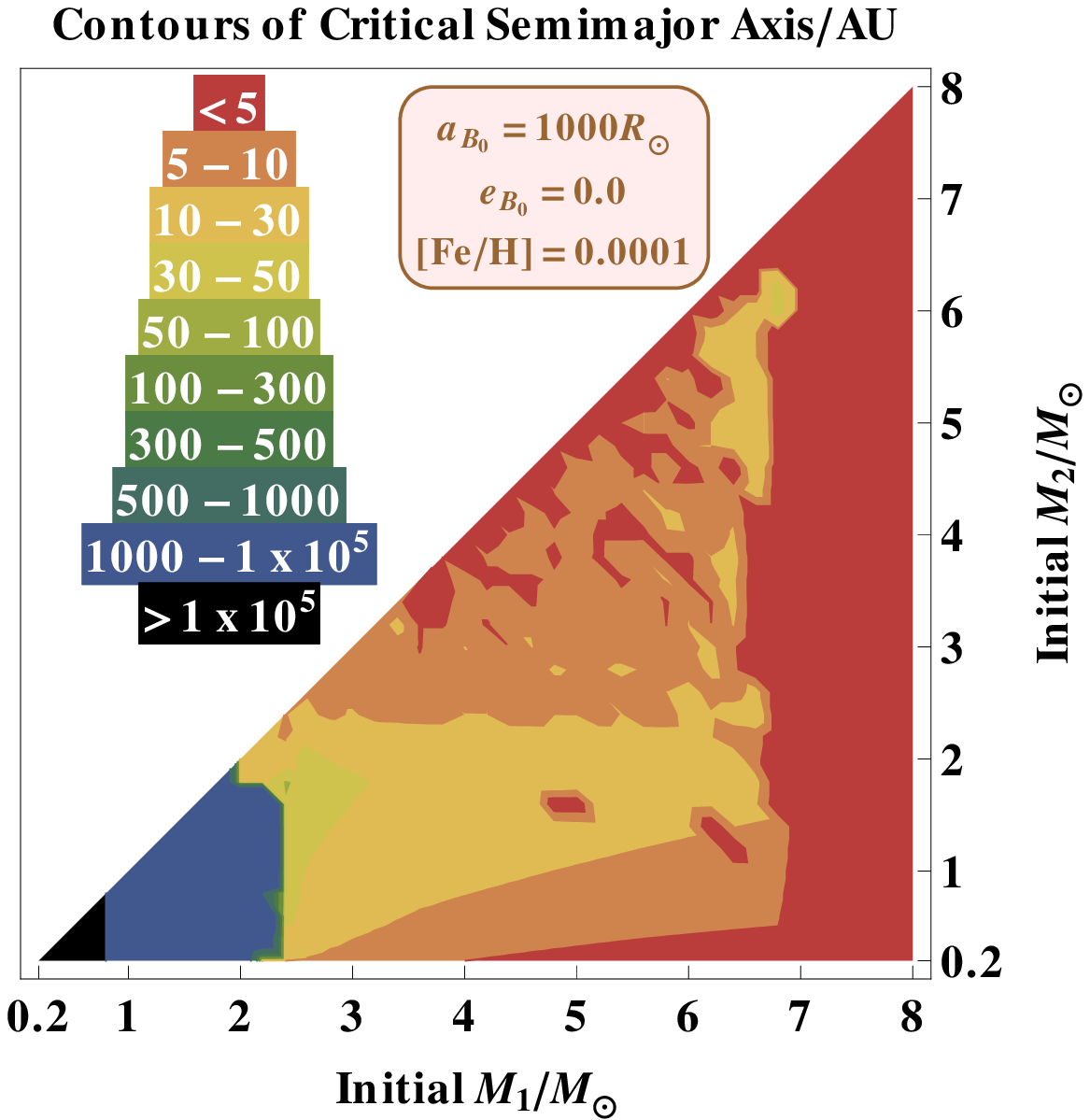,width=6cm} 
\psfig{figure=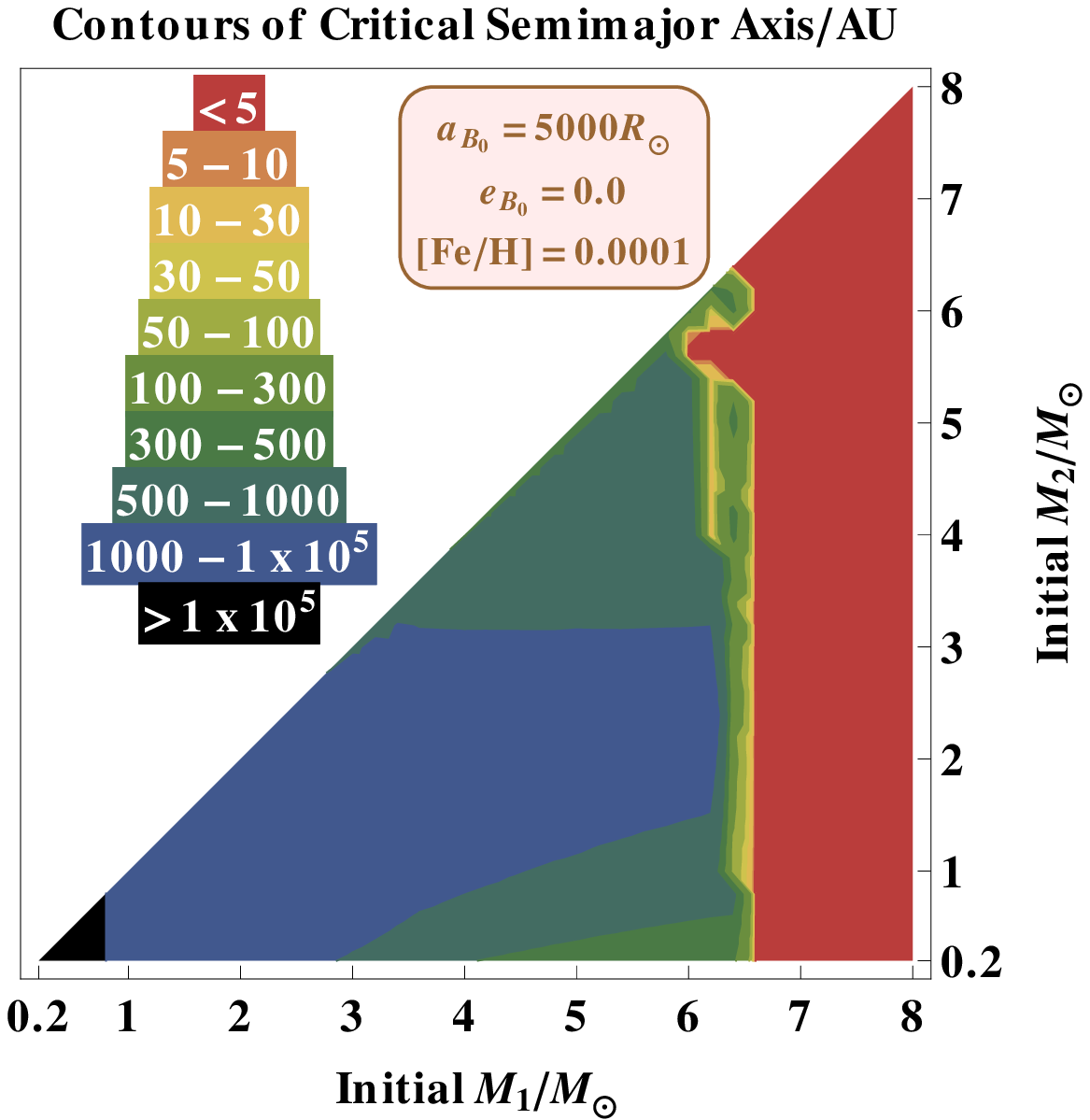,width=6cm} 
}
\centerline{
\psfig{figure=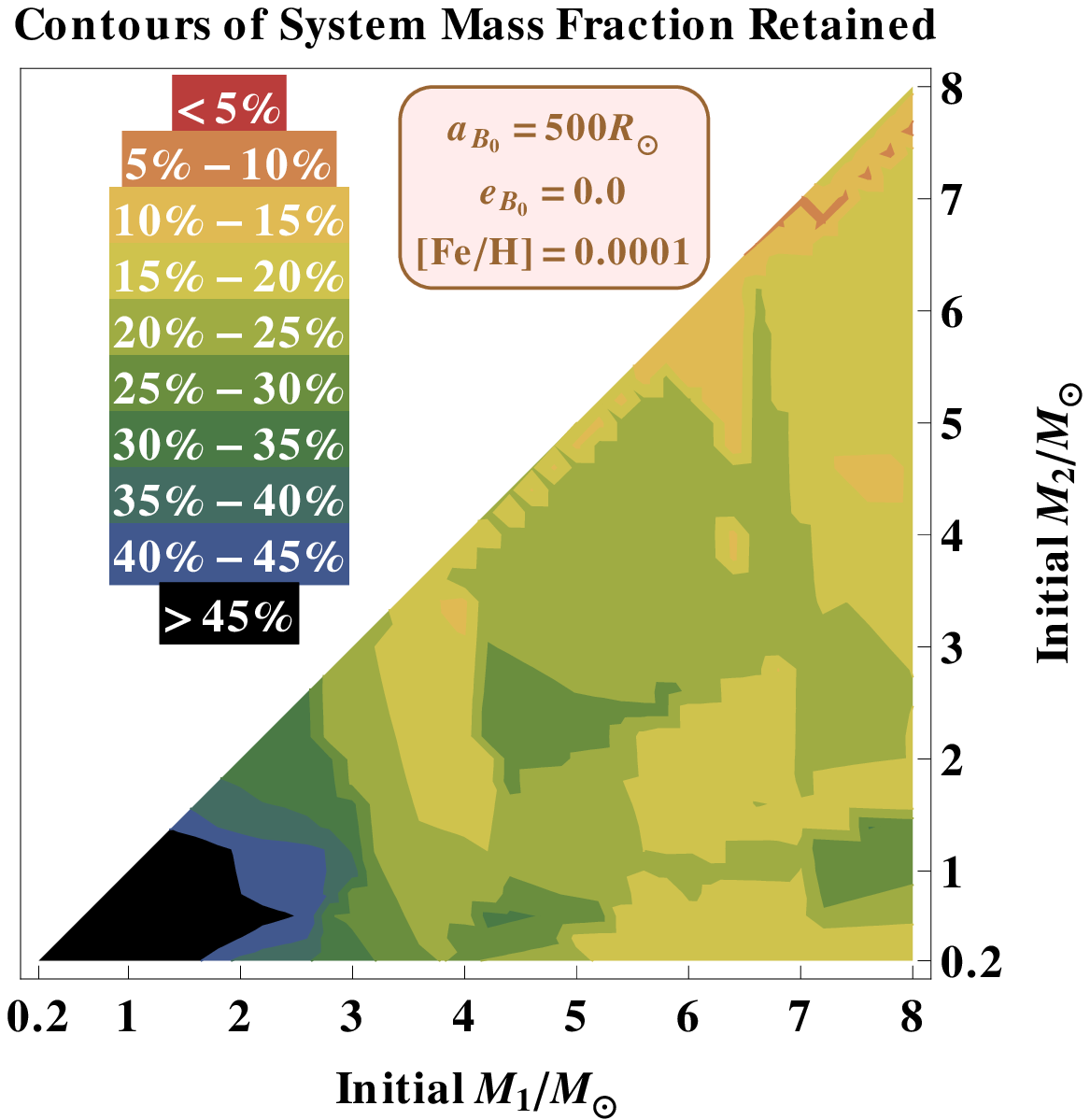,width=9cm} 
\psfig{figure=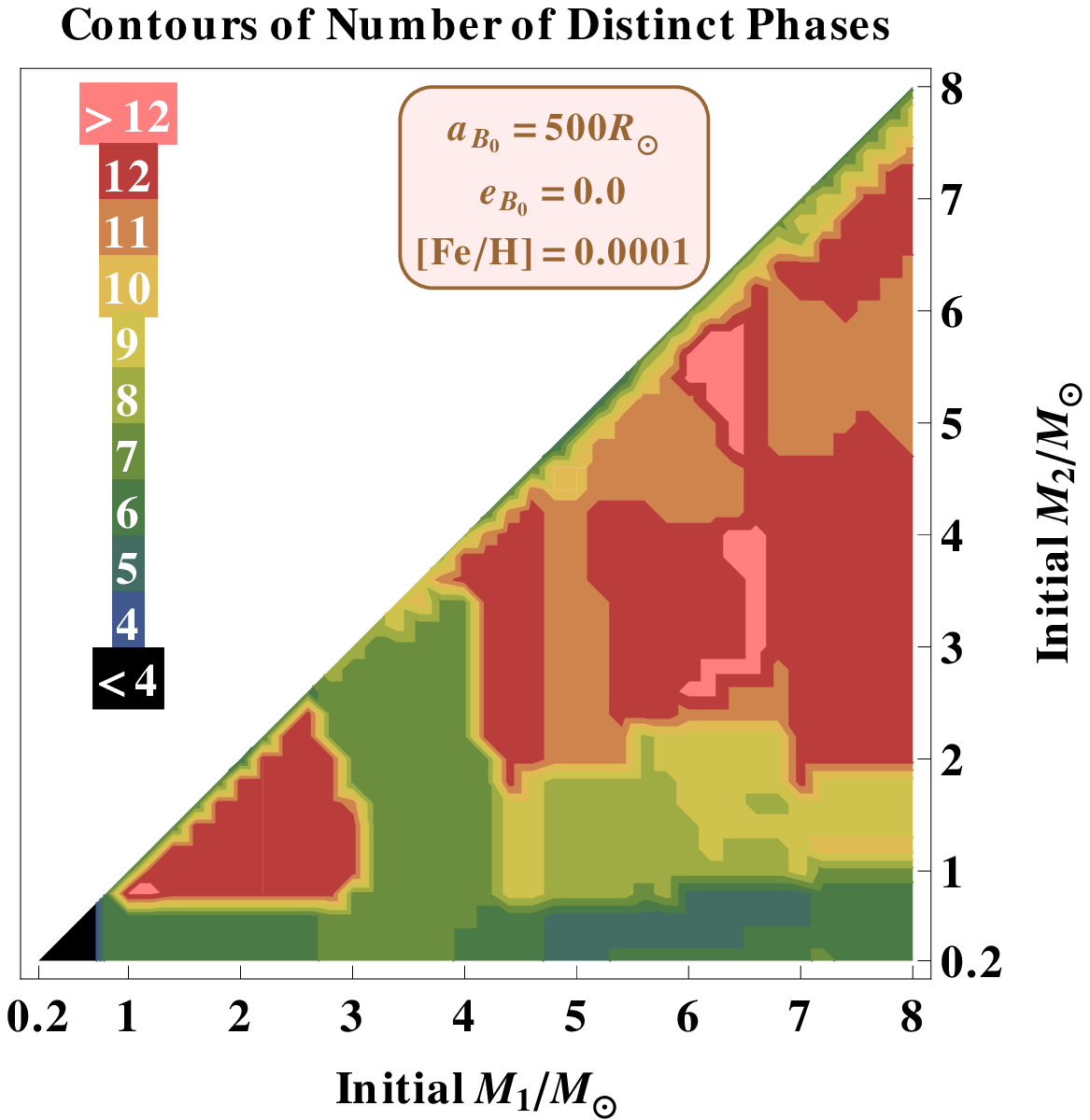,width=9cm} 
}
\caption{Low metallicity prospects for planetary retention.
The upper two panels are the same as Fig. \ref{cont1} but 
for [Fe/H] $= 0.0001$.  The lower panel provides more detail
on the the systems with $a_B = 500 R_{\odot}$.  The total
fraction of system mass loss over its lifetime is
a poor indicator of $a_{\rm crit}$, whereas
the total number of phases experienced by the binary
hints at the reason for the vertical green strip at
$3 M_{\odot} \le M_{1_0} \le 4 M_{\odot}$.  Although low-metallicity
progenitor masses are prone to core collapse supernova for $M \gtrsim 7 M_{\odot}$,
they afford circumbinary planets slightly more protection
than higher-metallicity systems for wide binary separations. 
}
\label{cont2}
\end{figure*}

\subsection{The Lower Mass Extreme}

We have argued that planetary retention cannot be guaranteed
if at least one progenitor star has $M \gtrsim 8 M_{\odot}$.
Now, we consider the lower mass extreme.

The lowest-mass stars ($M < 0.8 M_{\odot}$) have a lifetime which exceeds the
current age of the Universe \citep{paretal2011}.  Their mass-loss rates are
constant and negligible.  By themselves, their mass loss 
evolution does not perturb a planetary orbit.  However,
in a binary system, their interaction with a more massive stellar companion
could become a crucial aspect of their mutual evolution.  Therefore,
we include masses down to ($M = 0.2 M_{\odot}$) in our
simulations below.  For stars with masses $M \ll 0.1 M_{\odot}$,
planetary masses may no longer be comparatively negligible to
the stellar masses.
In that case, a less approximate treatment of the full
three-body problem with mass loss would be required.

\subsection{Simulation Parameter Choices}

Having restricted the mass phase space to 
$0.2 M_{\odot} \le M_1, M_2 \le 8.0 M_{\odot}$, reduced
the nonlinear fitting of mass loss to one parameter, $C_{v}^{\ast}$,
and established the variable $t_{\rm ce}$, we can now
consider the binary's initial orbit in more detail.
We treat both close and wide binaries in this study,
with binary separations of 
$a_B/R_{\odot} = 10,50,100,500,10^3, 5 \times 10^3, 10^4$ 
and $5 \times 10^4$, and binary eccentricities $e_B = 0.0, 0.5$ and $0.9$.  
Wide binaries which do not interact with each other 
typically have separations greater than $10^4 R_{\odot} \approx 47$ AU).
Therefore, in the extreme case of $a_B = 5 \times 10^4 R_{\odot}$,
each component of the binary evolves off of the main sequence
independently.  In all cases, the planet is considered
to be far enough away from the binary both to
not affect the orbit of the binary and to have an approximately
elliptical orbit around the binary.
Each stellar evolution track is simulated for 15 Gyr,
and we adopt $\kappa = 2\pi$ in all our simulations.

BSE contains several physical stellar parameters
which may be varied.  These include the accretion rate on to
the secondary and mass-loss prescriptions.  In order
to focus our study, we use the default values for all
physical stellar parameters except for metallicity.
For most of our simulations, we adopt either 
[Fe/H] $=$ [Fe/H]$_{\odot} \equiv 0.02$ 
or [Fe/H] $= 0.0001$, although we also sample 
metallicity values corresponding to
 [Fe/H] $= 0.0005, 0.001, 0.005$ and $0.01$.

Throughout each stellar evolution realization, the binary's orbit
evolves. The components may interact with each other in complex ways,
including merging through coalescence or collision, transferring mass
in dynamical, nuclear and thermal regimes, rejuvenating a companion
through this mass transfer, and inciting accretion-induced collapse to 
achieve, for example, a supernova-less $k = 12 \rightarrow 13$ transition.  
In all cases, the companions remain bound or merge.  Only
a supernova can unhinge the binary orbits.  Common envelope 
ejection occurs beyond the binary's orbit, and therefore should
not affect the binary orbit.

\subsection{Simulation Results}

We present our results through a series of descriptive contour plots.
On each, the $x$- and $y$-axes represent the initial mass of the primary
and secondary star, respectively.  We sampled each initial stellar mass at intervals
of $0.2 M_{\odot}$, such that $M_1 \ge M_2$.  Therefore, each plot
represents a grid of 800 points.  We can characterize many aspects
of the system evolution with these plots, such as evolutionary complexity,
timescales and mass loss factors.  Our primary interest, however,
is to determine the initial critical semimajor axis, $a_{\rm crit}$, at which
a planet is guaranteed to remain bound to the system.

\subsubsection{Dependence on Binary Separation}

The initial binary separation, $a_{B_0}$ largely
determines the sometimes complex manner in which
binary components interact with each other when
evolving off of the main sequence.  Because the planet
is considered to be far from the binary, 
if $a_{\rm crit} \lesssim a_{B}$ at any point,
then the planet is not guaranteed to remain bound.
The actual evolution of the planet in this case
is complex and would necessitate detailed modeling.

Figures \ref{cont1} and \ref{cont2} demonstrate how $a_{\rm crit}$ 
changes as $a_{B_0}$ increases from $10 R_{\odot}$ to $5 \times 10^3 R_{\odot}$ 
in both the [Fe/H] $=$ [Fe/H]$_{\odot} = 0.02$ and [Fe/H]$ = 0.0001$ cases.  The 
dearth of black and blue contours on the upper panels of these plots immediately 
demonstrates how prone planets are to escaping these systems.

First consider the black contours: all orbiting material is protected from 
ejection for $M_1, M_2 \le 0.8 M_{\odot}$ in all cases, and that this protected 
region is extended to $1.0 M_{\odot}$ for the Solar metallicity case.  
For these lowest-mass stars, their proximity to each other is unimportant.  
The lower panels of the figures demonstrate that in this regime, stellar 
evolution undergoes only a few phases; indeed, the stars never leave the 
main sequence.  Even after $15$ Gyr, the stars do not lose enough mass on the 
main sequence to have a noticeable effect on orbiting material.  This is 
the only evolutionary pathway that provides safety for Oort cloud comets with 
$a > 10^5$~AU, which are predicted to exist at galactocentric distances beyond
the Sun's \citep{braetal2010}.

Next consider the blue contours, found primarily in the widest initial binary 
separations, with $a_{B_0} = 5000 R_{\odot} \approx 23$ AU.  Comparing these
two plots in 
Figs \ref{cont1} and \ref{cont2} demonstrates that the 
higher metallicity case 
generally allows for greater planetary protection unless 
$M_{1_0} \gtrsim 6.6 M_{\odot}$, when a core-collapse supernova occurs.  A 
circumbinary planet with $a < 1000$ AU is guaranteed to remain bound if 
$M_{1_0}, M_{2_0} \le 2 M_{\odot}$, although in many cases the planet may still 
be safe further away.  This is the expected result from the single star case 
where mass loss is treated linearly in each phase (see Paper I), because at 
this separation the stars effectively evolve independently from each other.  
However, this treatment is conservative; later we demonstrate how one of these 
plots changes when $C_{v}^{\ast}$ is decreased and higher accuracy is achieved.  
For $a_{B_0} = 1000 R_{\odot} \approx 4.7$ AU, planets with $a > 1000$ AU may 
still survive if $M_{1_0}, M_{2_0} \le 2 M_{\odot}$, as long as $C_v \rightarrow \infty$.
The contour plots for $a_{B_0} = 1 \times 10^4 R_{\odot}$ and 
$a_{B_0} = 5 \times 10^4 R_{\odot}$ are nearly indistinguishable from the 
$a_{B_0} = 5000 R_{\odot}$ case and are not shown. 

For closer separations, the 
stars are more prone to interact with each other violently.
The lowest-right plots in both figures hint at the complexity of the stellar evolution 
for $a_{B_0} = 50 R_{\odot}$ and $a_{B_0} = 500 R_{\odot}$.  \cite{huretal2002} describe the 
complete details of the physical mechanisms behind the variations in these contour plots.
Here, we just point out some of the most important features.

When $a_{B_0} \le 1000 R_{\odot}$, prospects for planetary protection are primarily 
determined by common envelope and supernovae events.  The lowest-left plot of Fig. \ref{cont1} 
characterizes where in stellar mass phase space for $a_{B_0} = 50 R_{\odot}$ these events 
occur.  They fill almost the entire phase space.  Any contours on this plot that indicate 
a supernova has occurred automatically yield $a_{\rm crit} < 5$ AU (no protection)  in the 
corresponding upper panel figure, as expected.  Some contours which indicate that no supernova 
occurred {\it also} yield $a_{\rm crit} < 5$ AU.  In these cases, BSE computes a timescale for all 
phase transitions, and the mass-loss rate at some point during the evolution
is great enough to prevent planetary protection.

Now consider cases where $5$ AU $< a_{\rm crit} < 300$ AU for $a_{B_0} = 50 R_{\odot}$. 
Because we adopted $t_{\rm ce} = 1000$ yr for all these simulations, the variance
in $a_{\rm crit}$ is due to the amount of mass lost during the common envelope phase.
Later, we will explore how the results change when $t_{\rm ce}$ is altered, 
especially to less conservative values.  The green patch found at 
$\lbrace 4 M_{\odot} \lesssim M_{1_0} \lesssim 8 M_{\odot}$, $M_{2_0} \le 2 M_{\odot}\rbrace$ indicates that 
the fraction of system mass lost owing to common envelope evolution here is lowest.

A conspicuous feature of the $\lbrace a_{B_0} = 1000 R_{\odot}$, [Fe/H] = $0.02\rbrace$ and 
$\lbrace a_{B_0} = 500 R_{\odot}$, [Fe/H] = $0.0001\rbrace$ contour plots is a vertical green patch 
surrounded by yellow and orange contours.  This green patch arises because the common 
envelope in these systems is formed at a different phase of stellar evolution than 
the surrounding systems, at a phase where less of the system mass is lost.

Finally, the lower-left plot in Fig. \ref{cont2} displays contours of the total 
fraction of system mass lost after $15$ Gyr.  The plot demonstrates that there 
is almost no correlation, except in the lowest progenitor mass cases, to the 
corresponding middle panel $a_{\rm crit}$ plot.  This comparison illustrates how
the total system mass lost in a system is a poor proxy for $a_{\rm crit}$ (also compare
with Fig. 3 of \citealt*{verwya2012}).

\begin{figure*}
\centerline{
\psfig{figure=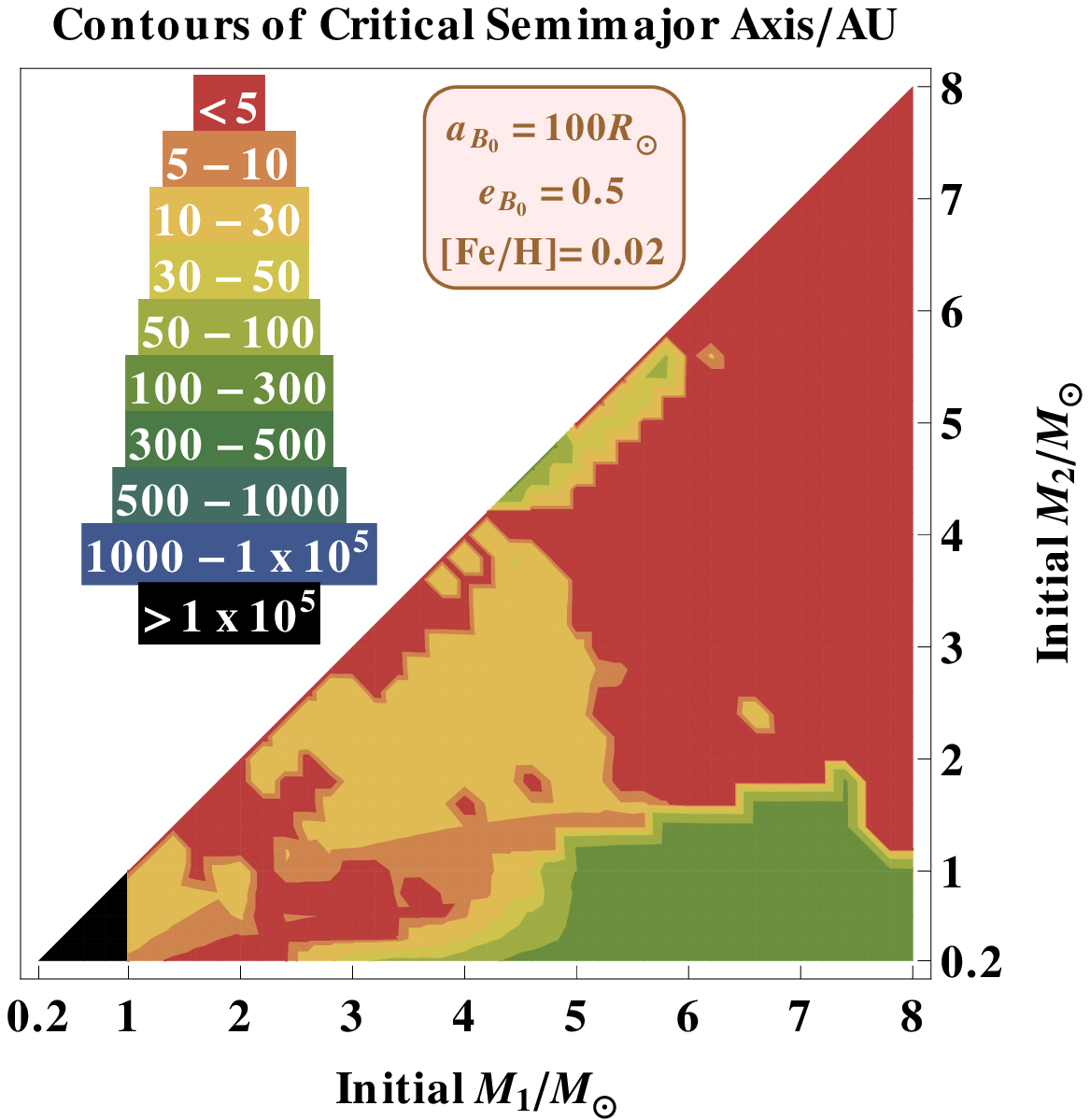,width=9cm}
\psfig{figure=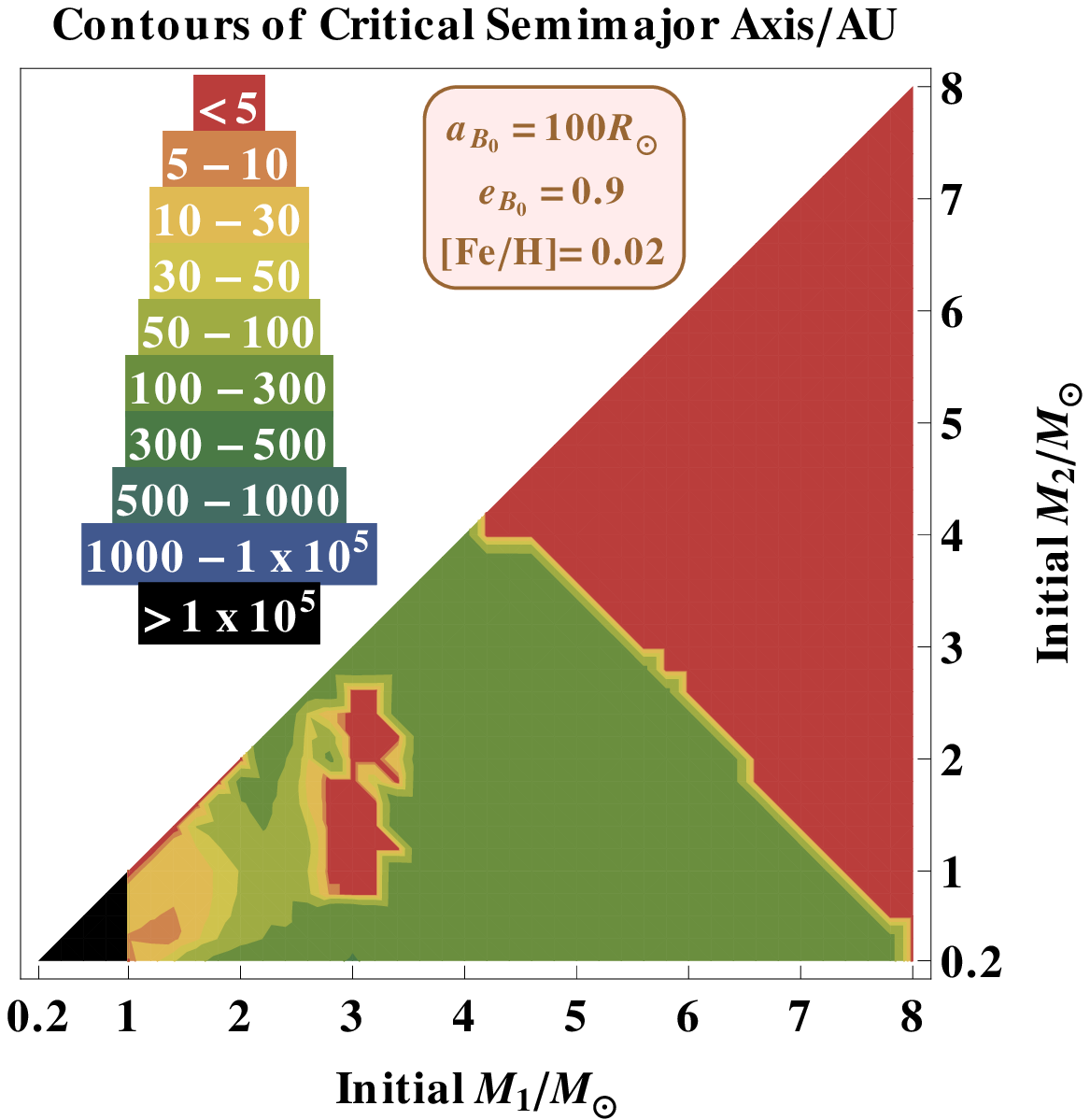,width=9cm}
}
\
\
\centerline{
\psfig{figure=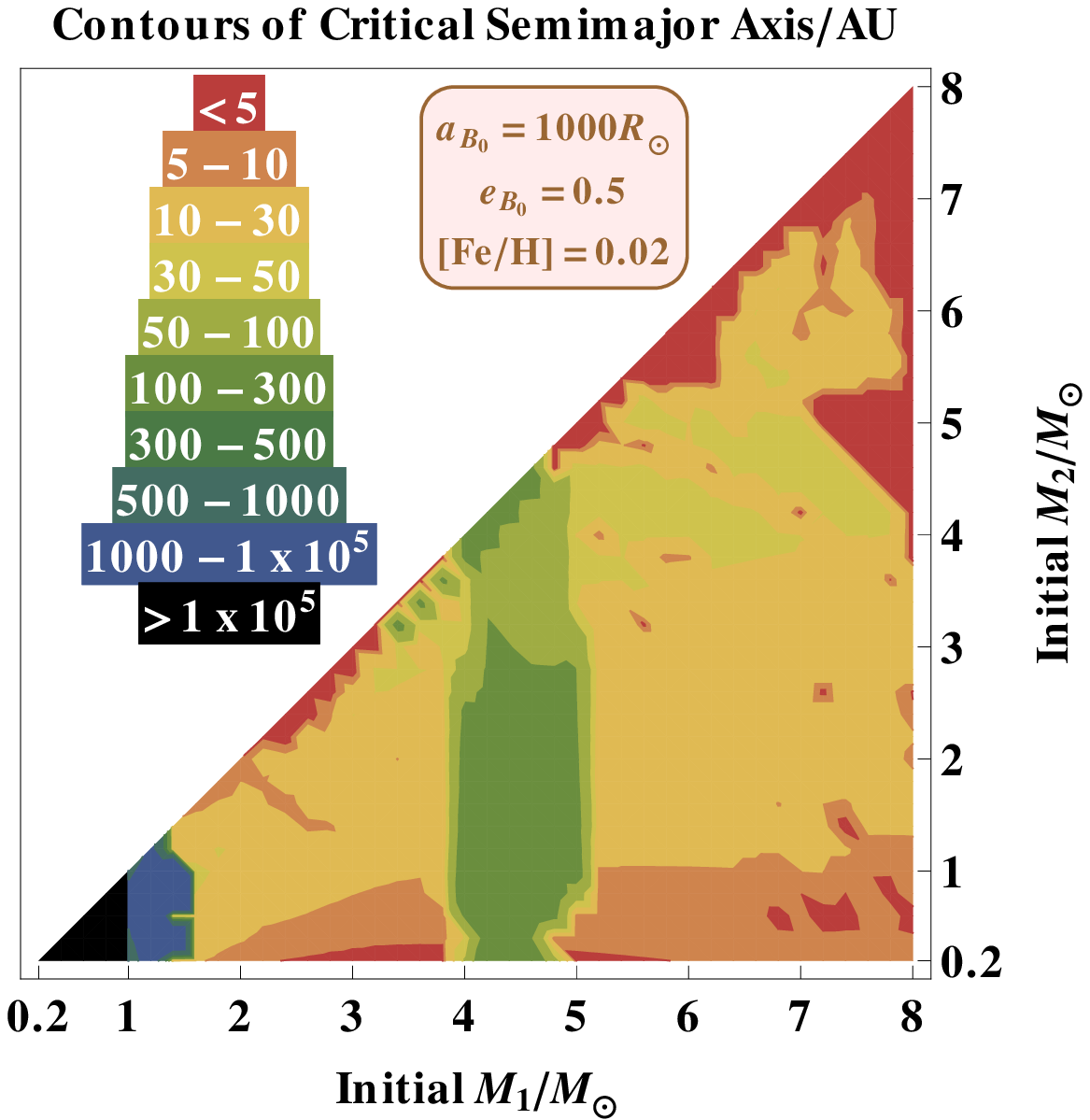,width=9cm}
\psfig{figure=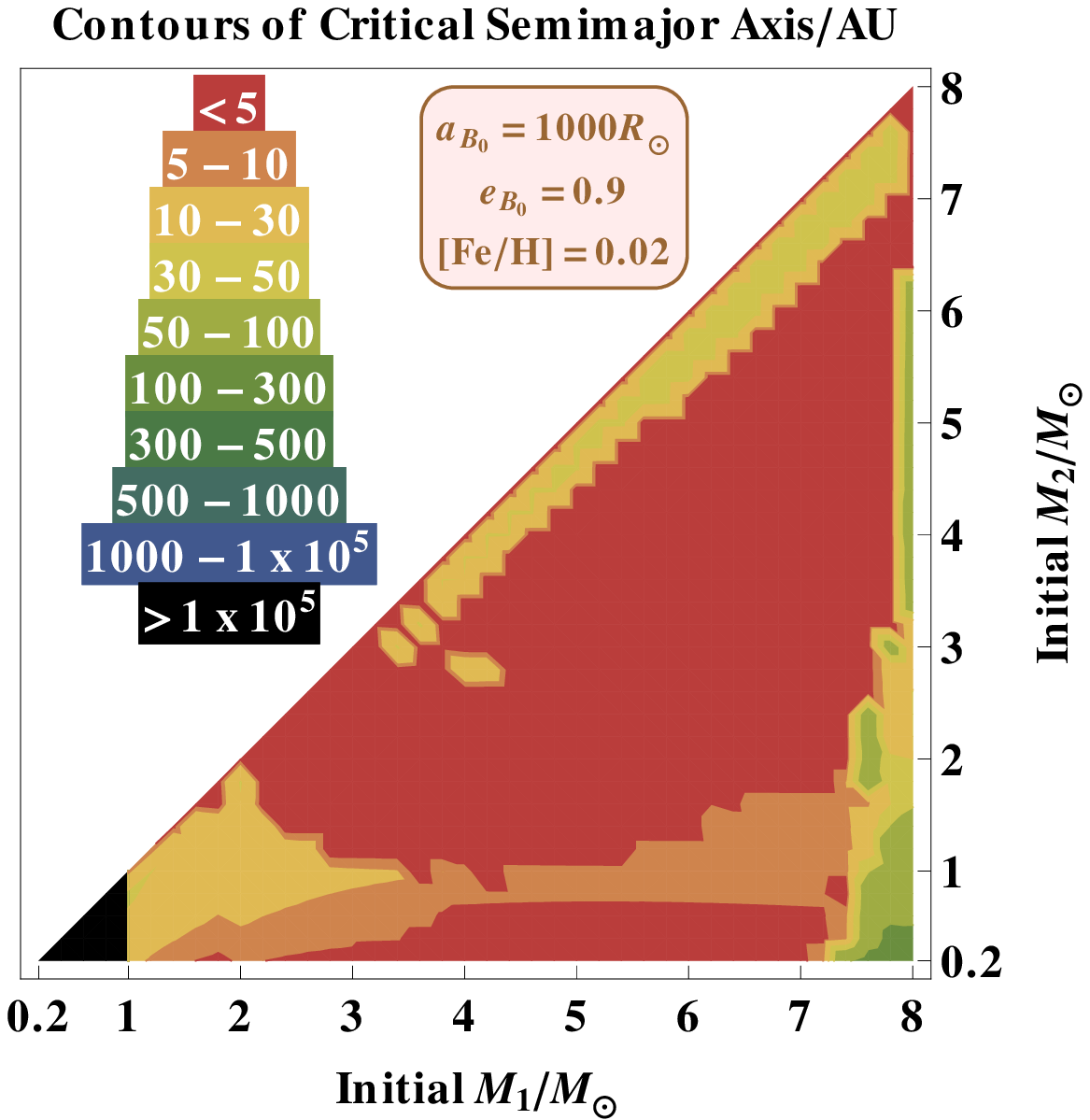,width=9cm}
}
\caption{
Critical semimajor axis for planet retention as a
function of initial binary eccentricity, $e_{B_0}$.
The upper panel and lower panel can, respectively, be directly
compared with the $a_B = 100 R_{\odot}$ and $a_B = 1000 R_{\odot}$
systems from Fig. \ref{cont1}.  Moderate initial binary eccentricities ($e = 0.5$)
have little effect on the systems.  However, high eccentricities ($e = 0.9$)
can cause close encounters which lead to phenomena that are
associated with smaller values of $a_{B}$.
}
\label{cont3}
\end{figure*}

\begin{figure*}
\centerline{
\psfig{figure=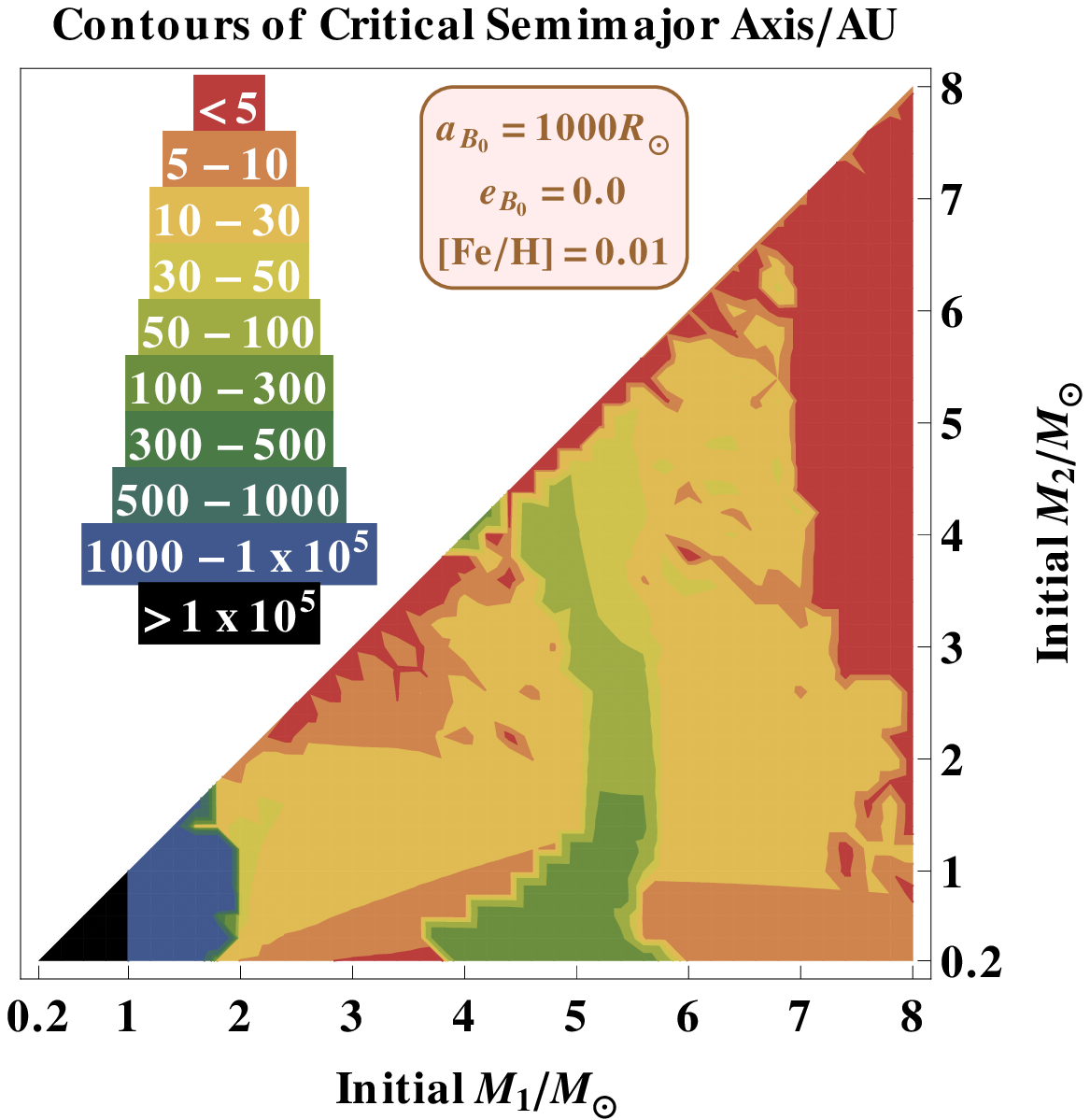,width=9cm}
\psfig{figure=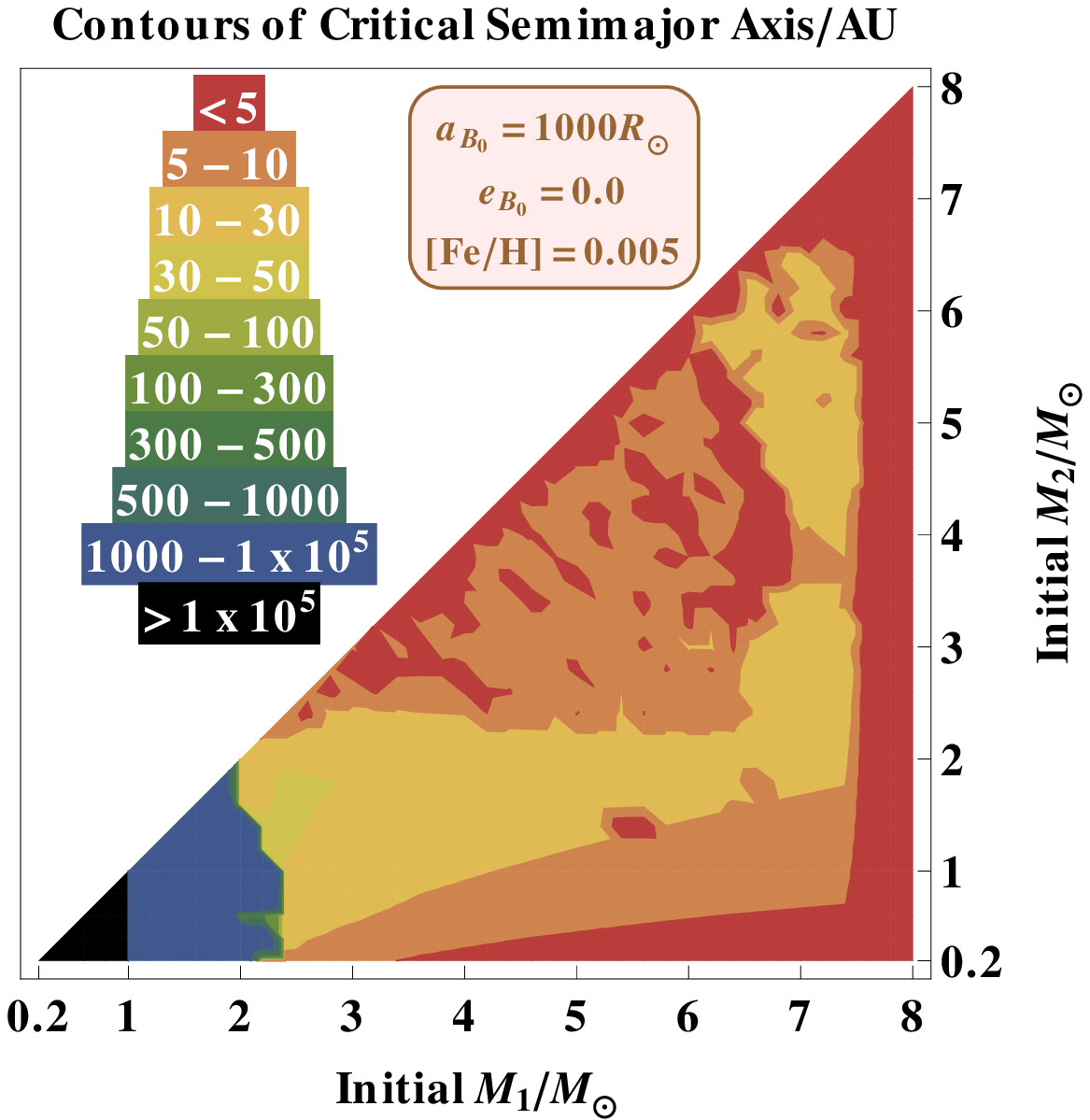,width=9cm}
}
\
\
\centerline{
\psfig{figure=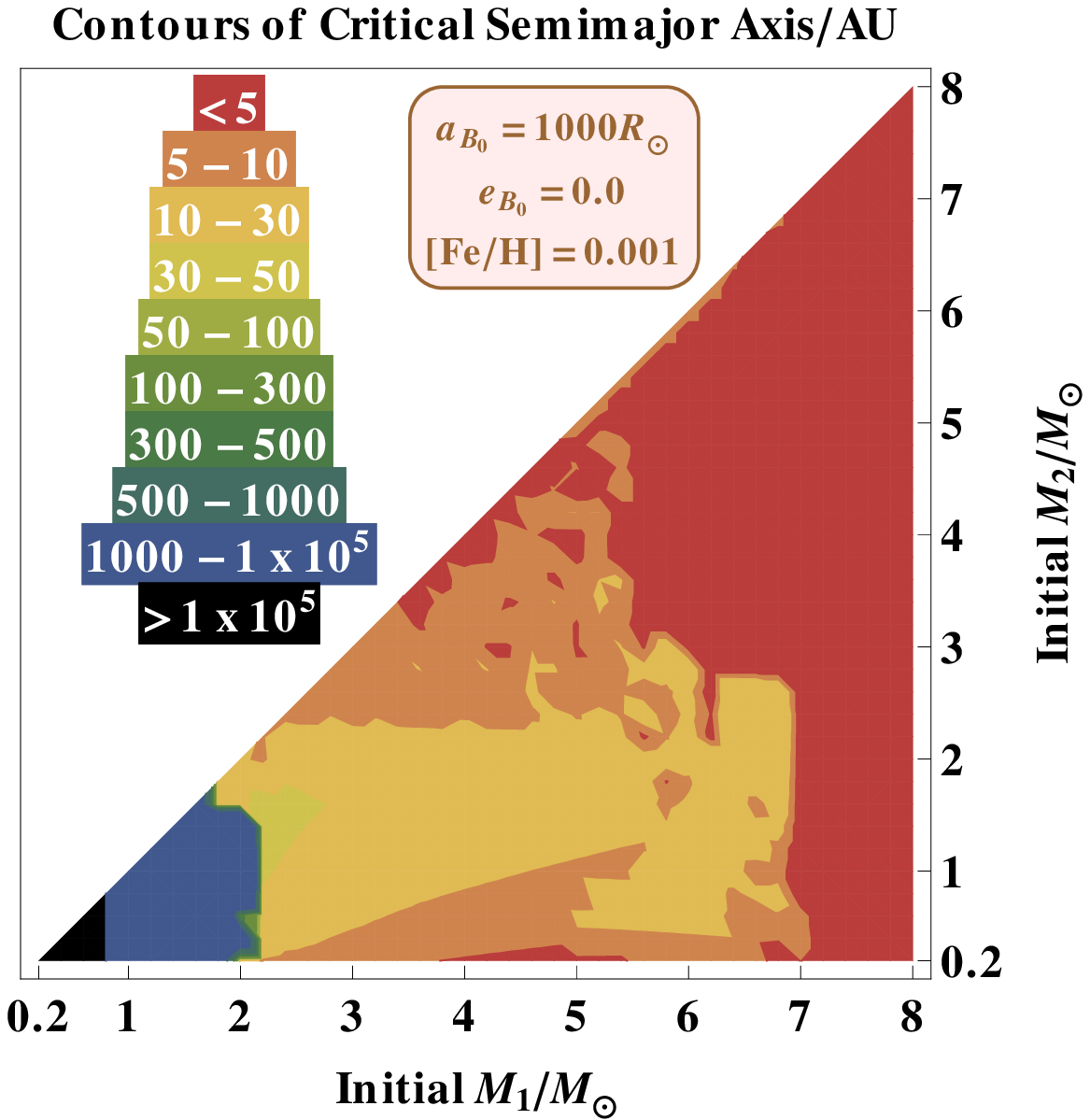,width=9cm}
\psfig{figure=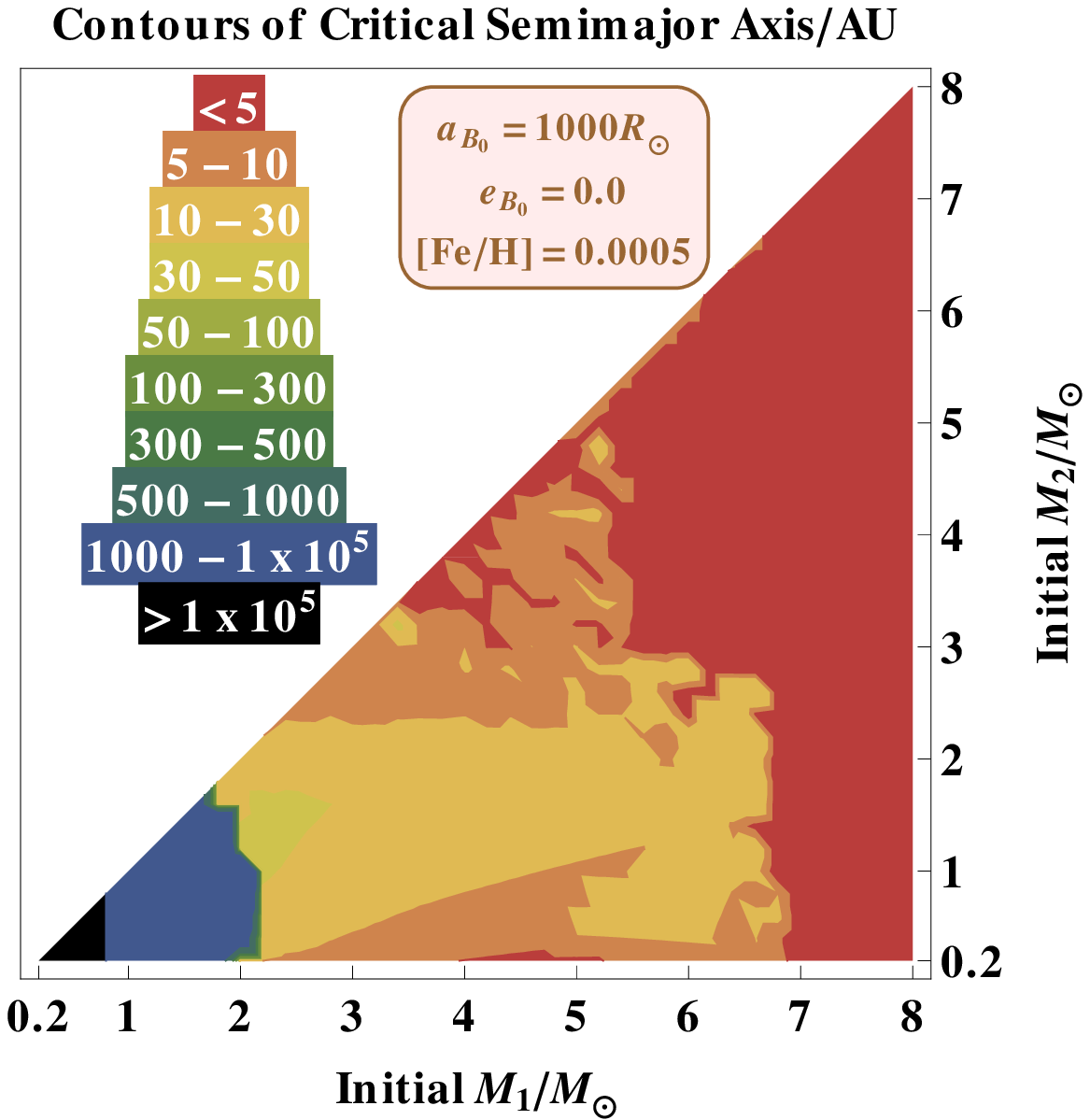,width=9cm}
}
\caption{
Critical semimajor axis for planet retention as a
function of stellar metallicity, [Fe/H].
These plots represent intermediate metallicity values
from the two extremes in Figs \ref{cont1} and \ref{cont2}
for $a_B = 1000 R_{\odot}$.
The critical semimajor axis is largely insensitive to
[Fe/H] values below about $0.001$.
}
\label{cont4}
\end{figure*}

\begin{figure*}
\centerline{
\psfig{figure=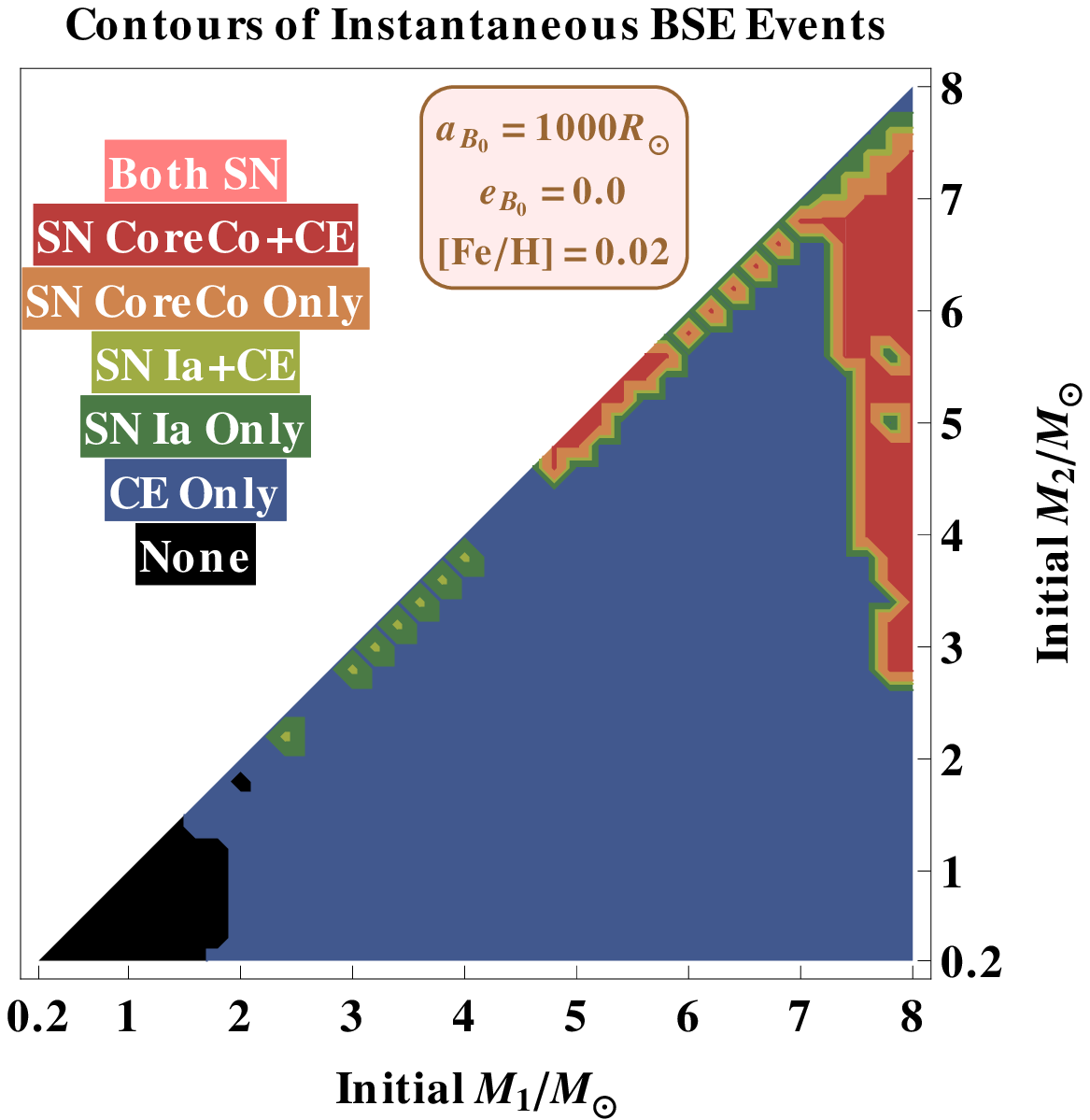,width=9cm}
\psfig{figure=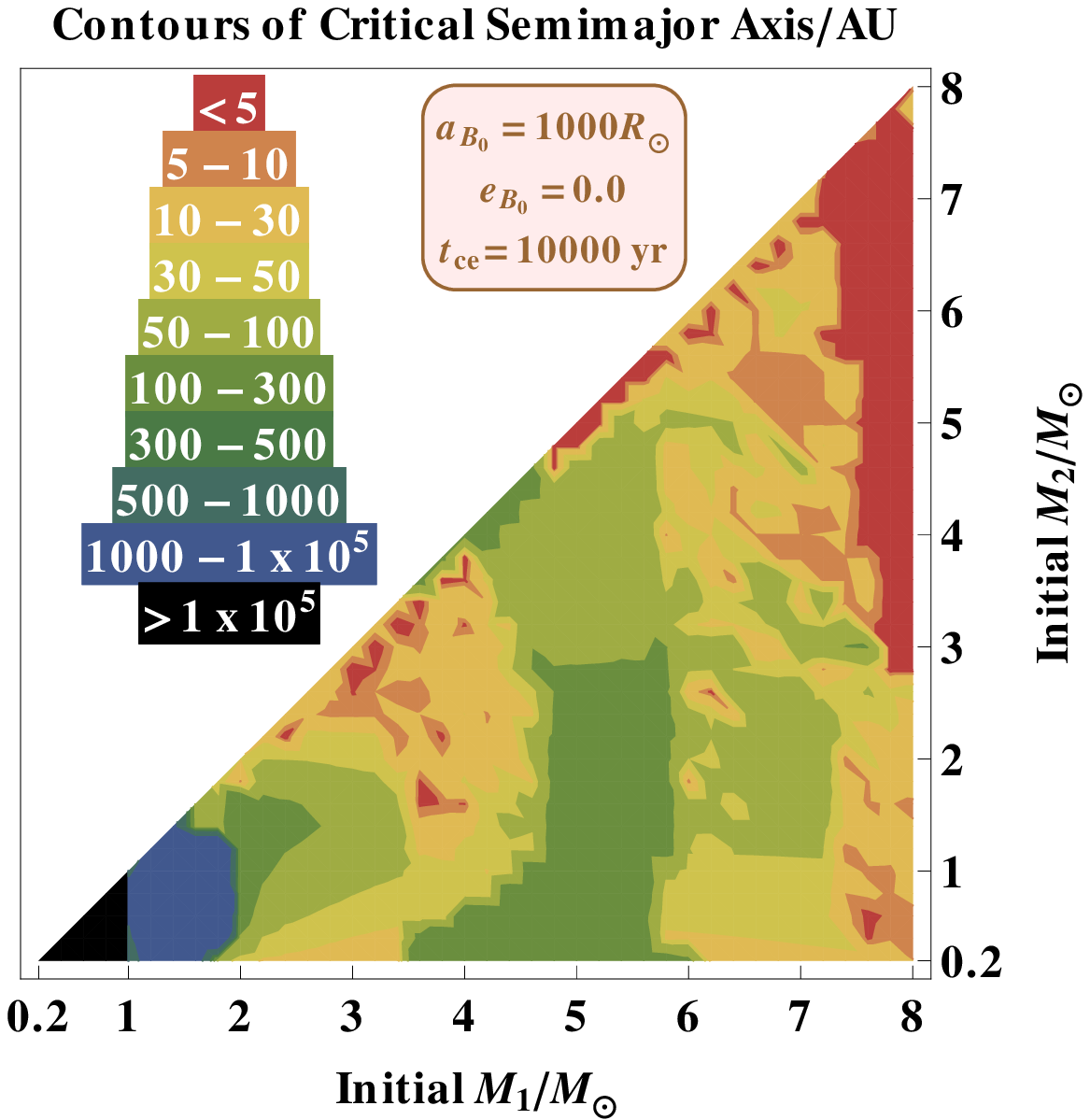,width=9cm}
}
\
\
\centerline{
\psfig{figure=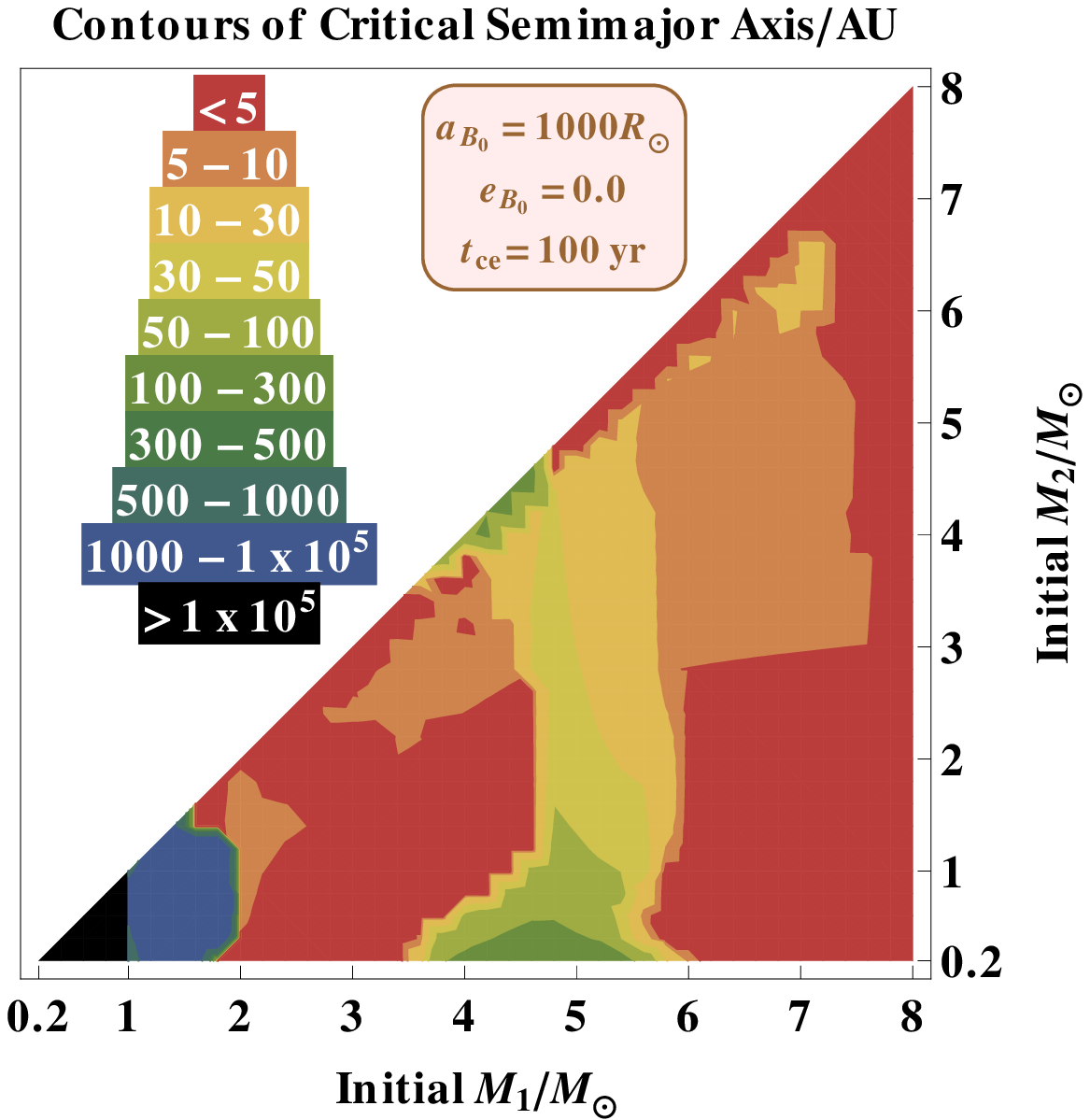,width=9cm}
\psfig{figure=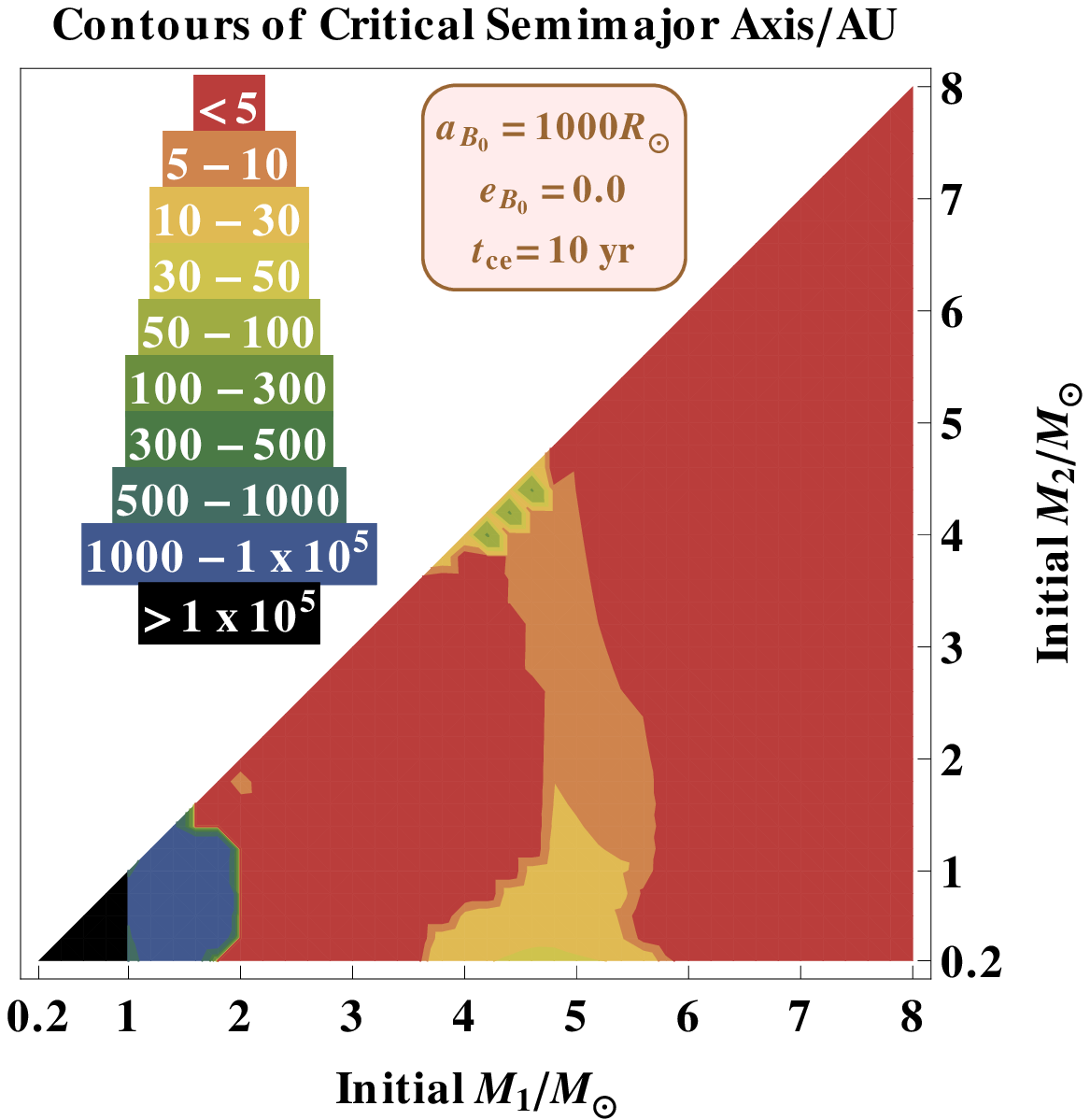,width=9cm}
}
\caption{
Critical semimajor axis for planet retention as a
function of common envelope evolution timescale, $t_{\rm ce}$.
These plots encompass the representative range
for the formation and ejection timescale of a common envelope 
($10^4$ yr in the upper right plot to $10^1$ yr in the lower right plot) 
and can be directly compared to the $a_B = 1000 R_{\odot}$
systems in Fig. \ref{cont1}, for which the fiducial
value of $t_{\rm ce} = 10^3$ yr is adopted.  The upper left
plot demonstrates our motivation for choosing this set of systems
for study: the common envelope phase dominates
the stellar evolution mass phase space for the initial binary 
orbit.  The contour labels in the upper-left plot are described 
in the Fig. \ref{cont1} caption.}
\label{cont5}
\end{figure*}

\begin{figure*}
\centerline{
\psfig{figure=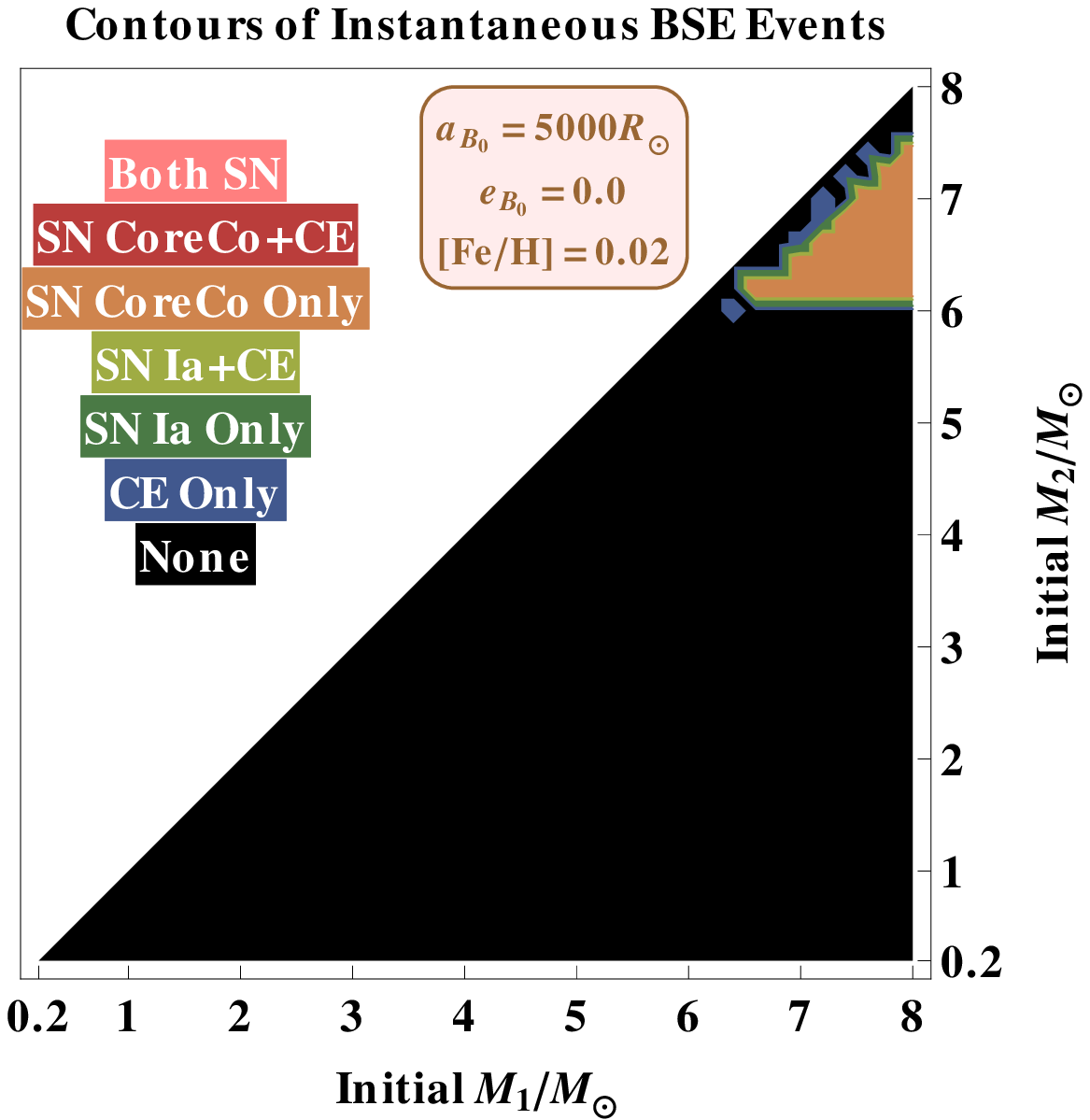,width=9cm}
\psfig{figure=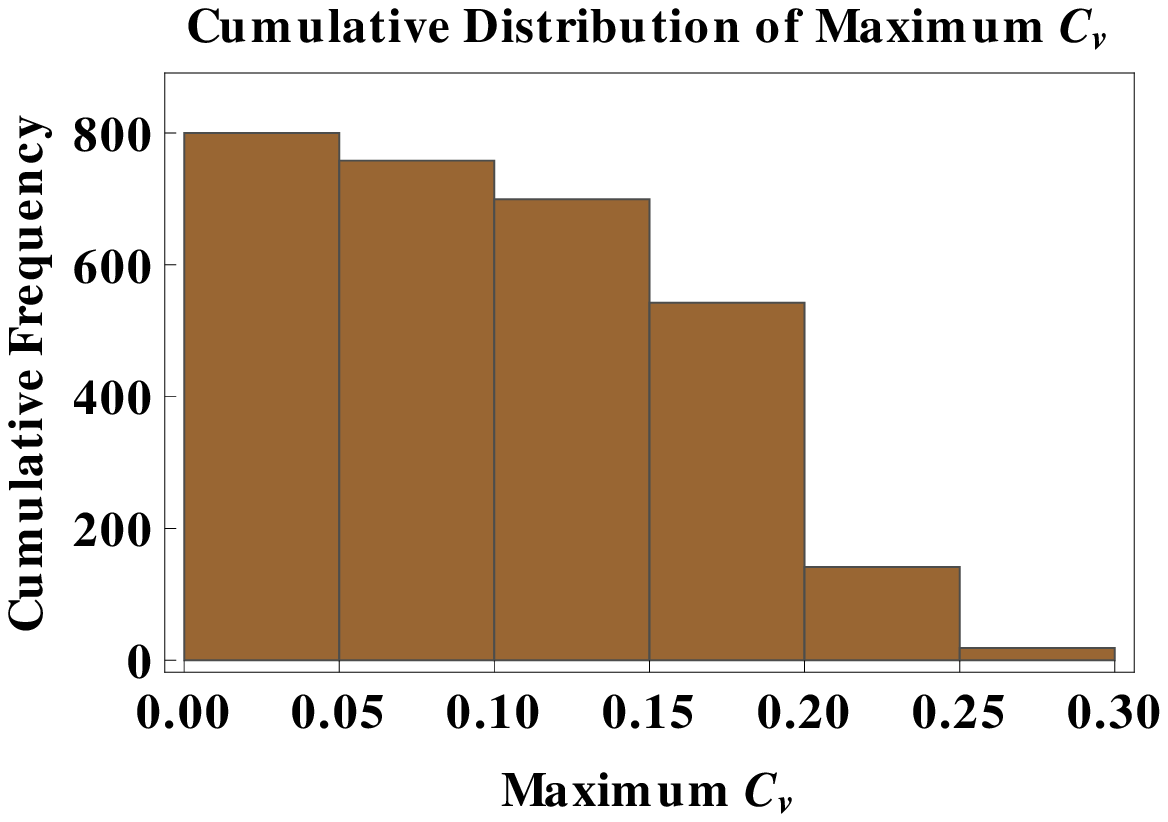,width=9cm,height=9cm}
}
\
\
\centerline{
\psfig{figure=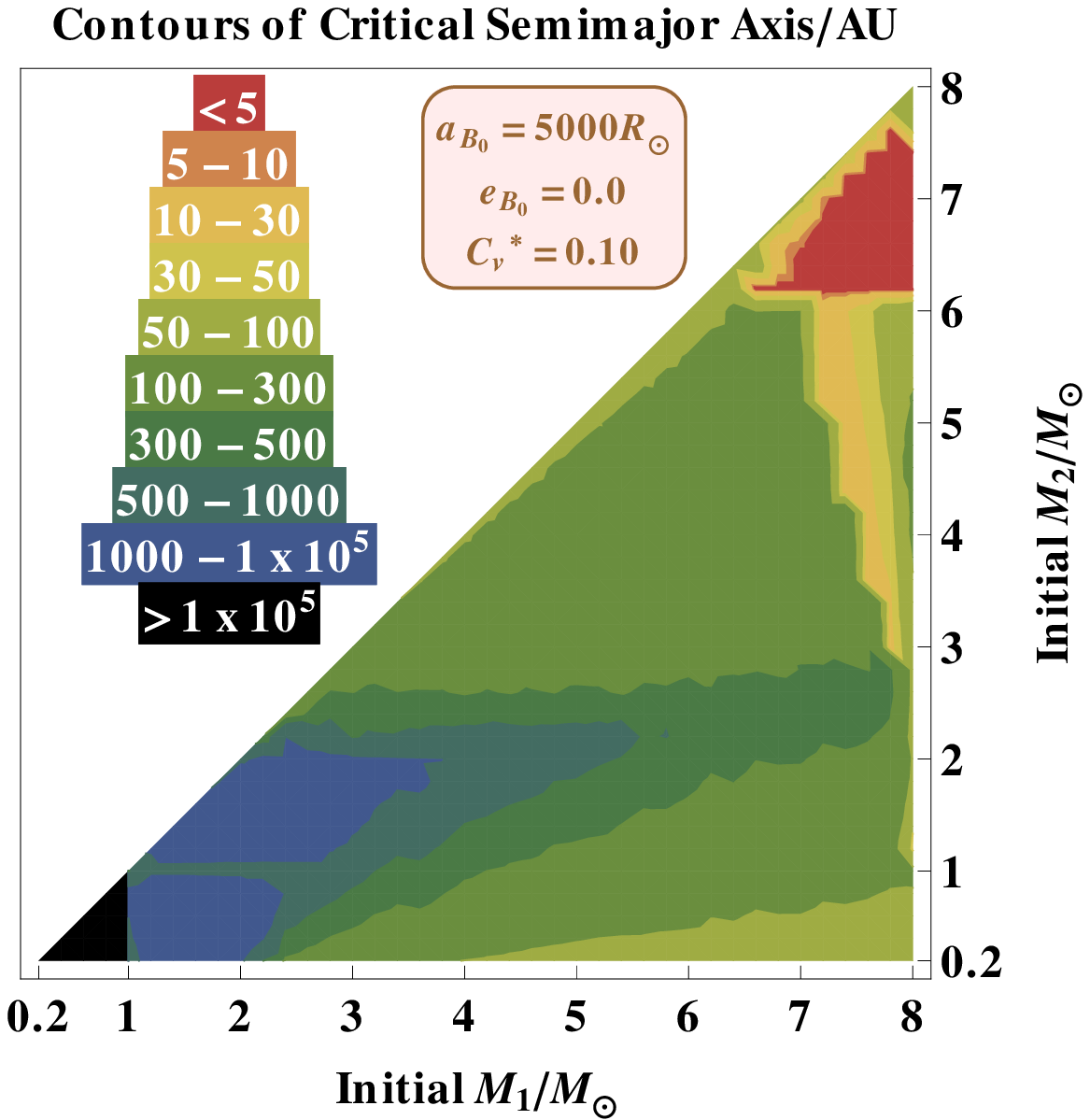,width=9cm}
\psfig{figure=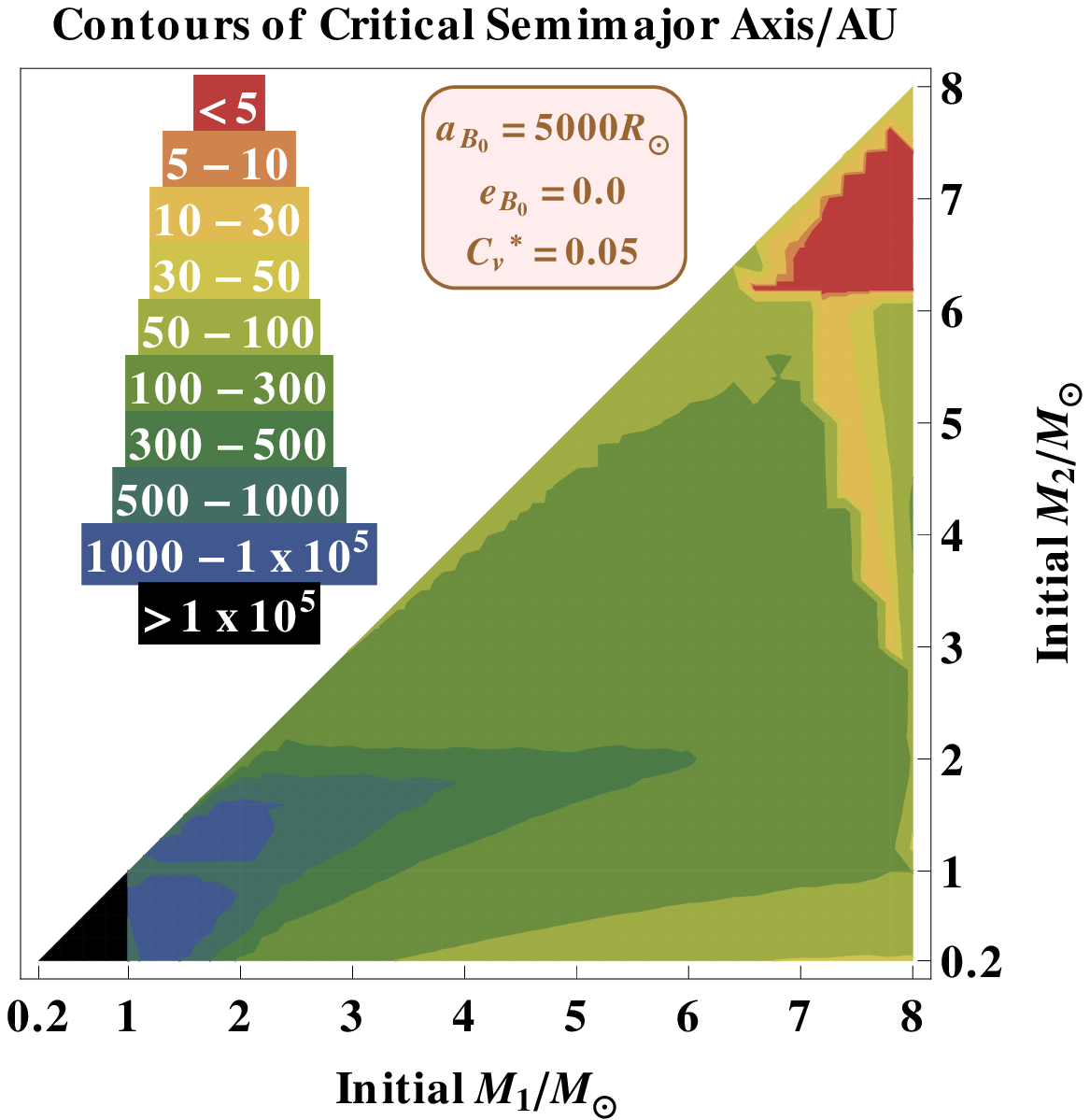,width=9cm}
}
\caption{
Critical semimajor axis for planet retention as a
function of $C_{v}$.
The upper left plot, composed of 800 points, demonstrates 
our motivation for choosing 
a set of systems with $a_B = 5000 R_{\odot}$:
the vast majority of stellar evolution realizations featured
in this set of systems feature well-defined and nonviolent
phase transitions.  Therefore, the critical semimajor axis
is primarily determined by mass loss within phases rather
than in between phases.  The contour labels in the upper-left plot are described 
in the Fig. \ref{cont1} caption.
The upper-right plot demonstrates that
our choices of $C_{v}^{\ast} = 0.10$ and $C_{v}^{\ast} = 0.05$
affect the vast majority of the systems sampled.  
The $C_{v}^{\ast} \rightarrow \infty$ case is plotted in
Fig. \ref{cont1}.  Lower values of
$C_v$ provide slightly more accurate but significantly more 
computationally expensive results.
}
\label{cont6}
\end{figure*}

\subsubsection{Dependence on Binary Eccentricity}

Related to initial binary separation is initial binary eccentricity. 
The pericentre of the orbit helps determine how close to each other 
the stars orbit.  Fig. \ref{cont3} explores how the 
$a_{B_0} = 100 R_{\odot}$ and $a_{B_0} = 1000 R_{\odot}$ $a_{\rm crit}$ 
contour plots from Fig. \ref{cont1} change when the binary's 
initial eccentricity is moderate ($e_{B_0} = 0.5$) 
or high ($e_{B_0} = 0.9$).  The plots reveal 
that $a_{\rm crit}$ is relatively insensitive to 
nonzero binary eccentricities unless the eccentricity is high.  
In the high-eccentricity case, the plots resemble those obtained from smaller 
initial binary separations and circular orbits.  This is consistent
with the finding of \cite{huretal2002} that it is the initial
semi-latus rectum, or angular orbital momentum, rather than 
semimajor axis and eccentricity which determine how a system evolves.

However, there are some differences.  For $a_{B_0} = 100 R_{\odot}$, 
with the initial pericentre at 
$10 R_{\odot}$, a bifurcation in the plot occurs where 
$M_{1_0} + M_{2_0} \approx 8 M_{\odot}$, resembling the circular 
$10 R_{\odot}$ case.  However, the behaviour at 
$\lbrace$$2 M_{\odot} \le M_{1_0} \le 4 M_{\odot}$~, 
$1 M_{\odot} \le M_{2_0} \le 3 M_{\odot}\rbrace$, is markedly 
different.  The red patch bounded by this region in the
circular case is caused by thermonuclear supernovae.
However, in the highly eccentric case, the conditions favorable
to producing this phenomena vanish.  The resulting
value of $a_{\rm crit}$ is then dominated by common envelope evolution.

For $a_{B_0} = 1000 R_{\odot}$, with 
an initial pericenter at $100 R_{\odot}$, the $a_{\rm crit}$ 
distribution is lower than in the circular $a_{B_0} = 100 R_{\odot}$ case.
The explanation again relates to thermonuclear supernova.  However,
here the high eccentricity of the binary {\it promotes} the conditions
necessary for this explosion, thereby suggesting that for high values
of $e_{B_0}$, the dependence of $a_{\rm crit}$ on $e_{B_0}$ may be complex
and best treated on an individual basis.

\subsubsection{Dependence on Metallicity}

In Fig. \ref{cont4}, we explore how $a_{\rm crit}$ changes during the transition 
between [Fe/H] = $0.02$ (Fig. \ref{cont1}) and 
[Fe/H] = $0.0001$ (Fig. \ref{cont2}) through the 
$a_{B_0} = 1000 R_{\odot}$ case.  As [Fe/H] decreases,
primaries with $M_{1_0} \gtrsim 7 M_{\odot}$ are more
likely to experience core collapse supernovae.
However, for $M_{1_0} \lesssim 7 M_{\odot}$ and [Fe/H] $\le 0.005$, 
the plots are nearly identical.  The primary differences 
occur for $4.0 M_{\odot} \lesssim M_{1_0} \lesssim 7.0 M_{\odot}$ 
and $M_{2_0} \le 0.6  M_{\odot}$.  In this regime, no supernovae
occur, and $a_{\rm crit}$ is dominated by common envelope evolution.
The amount of mass loss during this evolution is greatest
for the higher metallicity stars.  This is the reason for the
red strip on the bottom of the upper-right plot in the figure.

\subsubsection{Dependence on Common Envelope Evolution}

In order to demonstrate the dependence of common 
envelope timescale on $a_{\rm crit}$, we select a set of systems for study whose stellar 
evolution is dominated by a common envelope phase that BSE treats as instantaneous.
The upper left plot in Fig. \ref{cont5} presents our selection.  The other 
plots in the figure demonstrate how $a_{\rm crit}$ changes when $t_{\rm ce}$ is 
varied from $10$ yr to $10^4$ yr (the $10^3$ yr case is shown in Fig. \ref{cont1}).

Fig. \ref{cont5} demonstrates how conservative our fiducial choice of $t_{\rm ce} = 10^3$ yr is.  
For most of the stellar mass phase space on these plots, 
the initial orbital period of the binary is less than $10$ yr.  Therefore, if common 
envelope formation and destruction proceeds on the order of a few dynamical 
timescales, few planets would be guaranteed to survive.  Those that do 
would have $a_{\rm crit} \le a_{B_0}$, which destroys our assumptions and likely 
would destroy the planet as well.  Additionally, note the sharp transition between 
the blue and red contours in the $t_{ce} = 10$ yr case, indicating the importance 
of the onset of common envelope evolution.

\subsubsection{Dependence on Nonlinear Mass Loss}

Here we relax the conservative assumption $C_{v}^{\ast} \rightarrow \infty$ for 
a subset of cases.  Doing so for all stellar evolution realizations would be 
ideal but is too computationally expensive.  Further, because $a_{\rm crit}$ 
is determined by phase transitions for the vast majority of phase space, relaxing 
the assumption $C_{v}^{\ast} \rightarrow \infty$ is largely unnecessary.  However, 
for low progenitor masses and large separations, the choice of $C_{v}^{\ast}$ 
affects $a_{\rm crit}$\footnote{Unlike for multiple star systems,
in single star systems with $M < 8 M_{\odot}$, $a_{\rm crit}$ is dominated by 
the choice of $C_{v}^{\ast}$ for the vast majority of the stellar mass phase space.}.

In Fig. \ref{cont6}, we choose a set of systems where $a_{\rm crit}$ is 
dominated by $C_{v}^{\ast}$ (upper-left plot).  This set features widely
separated stars which evolve physically nearly independently of one another. 
We then computed the maximum value of $C_v$ achieved throughout each of the 
800 stellar evolution realizations (upper-right plot) and used those to 
sample values of $C_{v}^{\ast}$ that would ensure splitting the phases with the most nonlinear 
stellar mass-loss rates into multiple linear stages for the vast majority of simulations.
The $C_{v}^{\ast} \rightarrow \infty$ case is plotted in Fig. \ref{cont1}.

Setting $C_{v}^{\ast}=0.05$ (lower-right plot) eliminates the dark blue contour from Fig.~\ref{cont1} 
for $M_{1_0} \gtrsim 2 M_{\odot}$ and reduces $a_{\rm crit}$ by a factor of 
at least a few across the entire stellar mass phase space.  This result reinforces
the need for detailed modeling of individual systems that feature highly nonlinear
mass loss during, for example, the last epochs of giant branch evolution.

\section{Discussion}

Here we discuss four relevant related topics to this work.  The first 
compares planetary ejection rates for single and binary stars, which 
helps inform their relative contributions to the free-floating planet 
population.  The second
importantly considers how the violent behavior which often accompanies P-type 
planetary orbits extends to other configurations and higher multiplicities.
The third discusses the limitation of our knowledge of some aspects of stellar
evolution.  In the fourth, we consider partitions of the runaway regime into 
multiple stages, and show that in some cases there exist analogues to 
Eqs (\ref{aadia}) and (\ref{eadia}).

\subsection{Escape Rate Comparison With Single Stars}

Observations suggest that nearly two free-floating planets exist 
per main sequence star \citep{sumetal2011}.  This vast population 
cannot be explained by instability-induced planet-planet scattering 
alone \citep{verray2012}.  Other potential sources of free-floating 
planets arise from escape in evolved single star and binary star 
systems.  Although a detailed ejection rate computation based on initial 
mass functions, binary fractions, distributions of binary 
separations, and semimajor axis-based planetary distributions 
is beyond the scope of this study, we can provide some 
qualitative estimates here.

The critical semimajor axis is given by Eq. (\ref{acrit}), which is 
applicable for any stellar multiplicity and a planetary orbit that 
surrounds and is far from the central star or stars.  This equation 
contains the total initial system mass $\mu_0$.  On the critical 
semimajor axes contour plots in this paper (Figs. \ref{cont1}-\ref{cont6}), 
lines of constant $\mu_0$ would be approximately diagonal from the 
lower right to upper left.  Therefore, the tight binaries in 
Figs. \ref{cont1}-\ref{cont3} demonstrate that for 
$\mu_0 \ge 2 M_{\odot}$, a planet at several tens of AU away from 
the binary stars is typically prone to escape, and a planet 
at several hundred AU away is prone to escape in nearly all 
cases.  In contrast, direct 2-body numerical simulations from 
Paper I illustrate that for single stars, for no value of 
$\mu_0$ does a planet with a semimajor axis under 100 AU escape.  
Further, at 500 AU, planets begin to escape at 
$\mu_0\ge 3 M_{\odot}$.  Therefore, for a given system mass, 
planets are more prone to escape in a binary system than 
in a single star system.

This conclusion is critically supported by the lack of 
thermonuclear supernovae (see Section 3.2.3) and common 
envelopes (see Section 3.2.4) in single star systems.  The bottom 
left contour plot of Fig. \ref{cont1} illustrates that either 
or both of these violent phenomena occur in the vast majority 
of the stellar mass phase space for that set of tight 
($50 R_{\odot}$), Solar-metallicity binary systems.  In 
particular, the common envelope which typically forms 
for $2 \le \mu_0/M_{\odot} \le 6$ values is not present 
in single star systems and promotes escape for all 
reasonable blow-off timescales ($10$ yr - $10^4$ yr).
Even in moderately wider binaries, with binary separations 
of several AU, the stellar mass phase space is dominated 
by common envelope formation and ejection (see the upper left plot of 
Fig. \ref{cont5}).  If the common envelope blow-off timescale 
is as short as 10 yr, then the bottom right plot of 
Fig. \ref{cont5} demonstrates that nearly all planets, 
regardless of their semimajor axes, are prone to escape 
for $\mu_0 \ge 4 M_{\odot}$.  In contrast, in the 
single star case, no planet under 100 AU will escape 
for $\mu_0 \ge 4 M_{\odot}$ (right panel of Fig. 14 in Paper I).

For the lowest mass systems, with total masses under one 
Solar mass, we can crudely estimate that planetary escape for 
single stars is negligible based on arguments from Paper I.  
Similarly, Figs. \ref{cont1}-\ref{cont6} show that 
$\mu_0 \le 1 M_{\odot}$ binary systems protect planets within 
$10^5$ AU.  These low-mass systems represent the majority of 
stars in the Milky Way, based on a wide selection of initial
mass functions from \cite{paretal2011}.  Therefore, we 
conclude that these systems do not contribute to the 
free-floating planet population, regardless of multiplicity.  
For higher-mass systems, as summarized by 
\cite{basetal2010}, the binary frequency of Solar-type stars 
is approximately $60\%$; this value increases to nearly $100\%$ 
for O-type stars.  Therefore, more stars appear in binaries 
than not for $\mu_0 \ge 2 M_{\odot}$.  Further, as argued 
above, planets are more susceptible to escape from binaries 
than from single stars for $\mu_0 \ge 2 M_{\odot}$.
Therefore, assuming that the number of planets which 
survive main sequence evolution is approximately equal
in systems of all stellar multiplicities, the contribution 
to the free-floating planet 
population is greater in multiple star systems than in 
single star systems for $\mu_0 \ge 2 M_{\odot}$.  For the 
intermediate range $1 \le \mu_0/M_{\odot} \le 2$, the 
comparison of ejection contributions is less clear, because the 
binary fraction might be less than half \citep{basetal2010}.

\subsection{Higher Multiplicities and S-Type Orbits}

A detailed extension of these results to planets
on S-type orbits or to stellar
systems with more than 2 stars is 
nontrivial\footnote{Kratter \& Perets (2012) consider
a planet which can safely ``hop'' from one
star to another during post-main-sequence
evolution of an S-type binary and remain in a long-term stable
orbit after hopping.  If, however, both 
of these stars are evolving on the main-sequence,
then a planet may ``bounce'' between the stars
but fail to achieve a long-term stable orbit 
around either \citep{moever2012}.}.
However, on qualitative grounds we claim that
$a_{\rm crit}$ for both of those system types
is generally lower than for the binary 
systems studied here.

A planet orbiting one star in a close binary is likely
to be destroyed before entering a regime where it is prone
to escape.  Although a planet may briefly survive engulfment by 
a stellar envelope of a single star \citep{beaetal2011},
a planet residing in or near a binary common envelope would be subject
to 3-body forces that would likely cause a collision.  Similarly,
a supernova by either star might incite a 3-body collision 
or escape if
not direct destruction of the planet.  A 
planet's orbit that is entangled in
a mass transfer stream between the two stars would likely harbor
a quite interesting but destructive orbit.

If the planet orbits one star in a wide binary, then stellar evolution
of the parent star would extend the planet's orbit, possibly beyond
a region of stability in the three-body 
problem \citep[e.g.][]{holwie1999,donnison2009}.  
If the parent star explodes, the planet is ejected, regardless 
of the mass of the binary companion.  If instead the binary 
companion explodes, then the planet and its parent star 
together will become unbound 
from the companion's remnant but could remain bound to one another.
The planet's fate then depends on single star evolution.

Now consider P-type orbits in systems with higher multiplicities.
Examples include combinations of stars and possibly other planets
which are all orbited by a distant planet in an approximately elliptical
orbit.  Formation scenarios in which a planet
can form from core accretion in a, for example, circumternary
disc, have yet to be explored fully.  However, second-generation
planets may form in a variety of exotic systems \citep{perets2010}
and if the claim by \cite{sumetal2011} that free-floating bodies 
are more abundant than main-sequence stars is true,
these planets may be captured by multiple star
systems.  Although additional stars add to the total mass of a
system, thereby inhibiting escape according to Eq. (\ref{inhibit}), 
this effect is overshadowed
by the increased likelihood of repeated, violent mass loss events.
Further, binary evolution represents the simplest-possible outcomes
for closely interacting stars, and violent phenomena that have not
yet been characterized may exist for systems of higher multiplicity.
Many multiple systems contain binary components and thus suggests that
as a minimum an orbiting planet is subject to the restrictions
suggested in this work.

\subsection{Improved Physics}

As researchers gain a better understanding
of the underlying physical processes featured by BSE,
the quantitative results of this work are likely to be
modified.  Particular events, such as common envelope
evolution, are largely unexplored. Only recently
has the dynamics of mass transfer in binaries 
on non-circular orbits been investigated.
\cite{sepetal2007,sepetal2009,sepetal2010} developed a 
formalism for mass transfer in eccentric
binaries that represents a step forward in 
understanding this physical process.
\cite{vanetal2008,vanetal2010} find that the mass-transfer rate for 
low-mass binaries is never large enough to allow for mass loss
from the system.  They claim that matter can escape the binary
if the kinetic energy from fast rotation plus the radiative
energy of the hot spot exceeds the binding energy of the system.
Mass transfer in cataclysmic variables might 
repeatedly vary by over an order of magnitude owing to nova outbursts
\citep{koletal2001}.  \cite{menetal2008} considers the amount 
of mass lost for the highest metallicity stars 
([Fe/H] $ = 0.04-0.1$), a regime not treated by BSE.
Recent hydrodynamic models of mass transfer
between binaries \citep{lajsil2011a,lajsil2011b}
attempt to circumvent the limitations of the
Roche lobe formalism, which was developed for
restricted cases.  Although these simulations
are just beginning to tackle the complexities
of mass loss in eccentric binaries, they
demonstrate that some transferred material
is ejected, while other becomes loosely
bound.

\subsection{Multi-phasic Runaway Regime Evolution}

Investigation of the runaway regime might help
assess the likelihood of planetary escape
when $\Psi \gg 1$.  Nonlinear mass loss in this regime may
be treated in a similar manner to the adiabatic regime
in specific cases.  By assuming that mass loss occurs in a 
series of consecutive linear stages, we may
derive equations analogous to 
Eqs. (\ref{aadia}) and (\ref{eadia}).  However,
the evolution of $a$ and $e$ in the runaway
regime is more complex (see Paper I) and 
we have derived analytical evolution equations
only when the planet is assumed to begin and
hence remain at pericentre or apocentre.

In the first case, for $f = f(t) = 0^{\circ}$, despite the added 
complexity of the relations, they couple together to
cancel out all intermediate stage terms so that
$a_i$ and $e_i$ may be expressed in terms of $a_0$,
$e_0$ and $\beta_i$ only so that:

\begin{equation}
a_i =
\frac{a_0 \left(1 - e_0 \right)}
{2 - \beta_{i}^{-1} \left( 1 + e_0 \right)}
\end{equation}

\noindent{and}

\begin{equation}
e_i = 
\beta_{i}^{-1} \left( 1 + e_0 \right) - 1,
\label{eperi}
\end{equation}

\noindent{such} that $a_0$ and $e_0$ are the initial
values at the beginning of the runaway stage.
Therefore, in this case, the prospects for planetary
ejection are {\it independent} of both $\alpha_i$ and
the intermediate stages of runaway evolution, as long
as $\alpha_i$ is high enough to ensure $\Psi \gtrsim 1$.
Further, as in the single stage case, Eq. (\ref{eperi})
demonstrates that the fraction of mass remaining in the
star at the moment of ejection is equal to $(1+e_0)/2$,
independent of the details of the intermediate stages.

If $f_0 = 180^{\circ}$, then the planet's orbit 
first circularizes before possibly expanding and causing
ejection.  The equations leading to this
circularization feature the same cancellations, and

\begin{equation}
a_i = 
\frac{a_0 \left(1 + e_0 \right)}
{2 - \beta_{i}^{-1} \left( 1 - e_0 \right)}
\end{equation}

\noindent{and}

\begin{equation}
e_i = 
1 - \beta_{i}^{-1} \left( 1 - e_0 \right)
.
\label{eapo}
\end{equation}

\noindent{Hence,} prospects for planetary orbit circularization
in this case are also independent of both $\alpha_i$ and
the intermediate stages of runaway evolution, as long
as $\alpha_i$ is high enough to ensure $\Psi \gtrsim 1$.
The planet becomes circularized when the fraction of mass
remaining is equal to $1-e_0$, the same result from Paper I.

If we consider the mass loss to be instantaneous
and $f_0$ is known then multi-phasic runaway evolution 
may be treated as a consecutive series of impulsive 
approximations (see Section 2.7 of Paper I).  We can
then create relations linking $a_i$, $e_i$ and $f_i$ to 
the same variables from all previous runaway stages before ejection.
However, because the transition between the adiabatic and runaway
regimes is not sharp (as partially 
evidenced by the presence of $\kappa$),
attempting to link the evolution in both with the above methods 
would likely lead to an unphysical result unless the transitional
regime is bypassed.  

\section{Conclusion}

Multiple stars violently interact in ways that a single
evolving star cannot and the effect on orbiting material 
is hence greater than in the single star case.
Conservatively, planetary material residing beyond a few hundred AU 
orbiting multiple stars each more massive than the Sun
and whose minimum pairwise separation is less than $100 R_{\odot}$ is unlikely to remain bound
during post-main sequence evolution.  All Oort cloud analogs
in post-main-sequence multiple star systems would be disrupted and feature escape,
independent of stellar separations.  Planets residing at just a few tens of AU
from a central concentration of stars may be subject to escape in a wide 
range of multiple star systems.  These systems may provide a 
significant contribution to the free-floating
planet population.

The techniques we utilized in order to obtain these results may aid in future studies
of individual cases \citep[e.g.][]{verwya2012}.  We have shown that the prospects for planetary escape
are determined entirely by the evolutionary stage with the greatest 
mass-loss rate relative to the remaining system mass.  Nonlinear mass-loss rate profiles within a phase can be 
analytically treated by a partition into linear segments, where the 
extent of the partition determines the accuracy of the model.  For single
stars which become white dwarfs, specific constraints may be placed on
the maximum stellar mass-loss rate given assumptions about an orbiting planet.

\section*{Acknowledgments}

We thank the referee, Valeri Makarov, for useful comments, 
John Eldridge for providing the updated reference for the
metallicity-dependent mass loss prescription for naked helium
stars, and Steinn Sigurdsson, John Debes and Richard Wade for 
helpful discussions.  CAT thanks Churchill College for his fellowship.

\label{lastpage}


\begin{thebibliography}{99}

\bibitem[Bakos et al.(2006)]{baketal2006} Bakos, G.~{\'A}., 
P{\'a}l, A., Latham, D.~W., Noyes, R.~W., 
\& Stefanik, R.~P.\ 2006, ApJL, 641, L57 

\bibitem[Bastian et al.(2010)]{basetal2010} Bastian, N., 
Covey, K.~R., \& Meyer, M.~R.\ 2010, ARA\&A, 48, 339 

\bibitem[Bear et al.(2011)]{beaetal2011} Bear, E., Soker, N., 
\& Harpaz, A.\ 2011, ApJL, 733, L44 

\bibitem[Belczynski et al.(2011)]{beletal2011} 
Belczynski, K., Wiktorowicz, G., Fryer, C., Holz, D., 
\& Kalogera, V.\ 2011, arXiv:1110.1635 

\bibitem[Beuermann et al.(2010)]{beuetal2010} Beuermann, K., 
Hessman, F.~V., Dreizler, S., et al.\ 2010, A\&A, 521, L60

\bibitem[Beuermann et al.(2011)]{beuetal2011} 
Beuermann, K., Buhlmann, J., Diese, J., et al.\ 2011, A\&A, 526, A53 

\bibitem[Bihain et al.(2009)]{bihetal2009} Bihain, G., 
Rebolo, R., Zapatero Osorio, M.~R., et al.\ 2009, A\&A, 506, 1169 

\bibitem[Brasser et al.(2010)]{braetal2010} Brasser, R., 
Higuchi, A., \& Kaib, N.\ 2010, A\&A, 516, A72 


\bibitem[Cochran et al.(1997)]{cocetal1997} Cochran, W.~D., 
Hatzes, A.~P., Butler, R.~P., \& Marcy, G.~W.\ 1997, ApJ, 483, 457 

\bibitem[Correia et al.(2008)]{coretal2008} Correia, A.~C.~M., 
Udry, S., Mayor, M., et al.\ 2008, A\&A, 479, 271 

\bibitem[Crowther et al.(2010)]{croetal2010} Crowther, P.~A., 
Schnurr, O., Hirschi, R., et al.\ 2010, MNRAS, 408, 731 

\bibitem[Desidera et al.(2011)]{desetal2011} 
Desidera, S., Carolo, E., Gratton, R., et al.\ 2011, A\&A, 533, A90 


\bibitem[Donnison(2009)]{donnison2009} 
Donnison, J.~R.\ 2009, PlanSS, 57, 771 

\bibitem[Doyle et al.(2011)]{doyetal2011} Doyle, L.~R., Carter, 
J.~A., Fabrycky, D.~C., et al. 2011, Science, 333, 1602 

\bibitem[Duquennoy \& Mayor(1991)]{duqmay1991} 
Duquennoy, A., \& Mayor, M.\ 1991, A\&A, 248, 485 

\bibitem[Eggenberger et al.(2006)]{eggetal2006} 
Eggenberger, A., Mayor, M., Naef, D., et al.\ 2006, A\&A, 447, 1159 

\bibitem[Fesen et al.(2007)]{fesetal2007} Fesen, R.~A., 
H{\"o}flich, P.~A., Hamilton, A.~J.~S., et al.\ 2007, ApJ, 658, 396 

\bibitem[Guenther et al.(2009)]{gueetal2009} 
Guenther, E.~W., Hartmann, M., Esposito, M., et al.\ 2009, 
A\&A, 507, 1659 

\bibitem[Hamuy \& Pinto(2002)]{hampin2002} 
Hamuy, M., \& Pinto, P.~A.\ 2002, ApJL, 566, L63 

\bibitem[Holman \& Wiegert(1999)]{holwie1999} 
Holman, M.~J., \& Wiegert, P.~A.\ 1999, AJ, 117, 621 

\bibitem[Hurley et al.(2000)]{huretal2000} Hurley, J.~R., Pols, 
O.~R., \& Tout, C.~A.\ 2000, MNRAS, 315, 543 

\bibitem[Hurley et al.(2002)]{huretal2002} Hurley, J.~R., Tout, 
C.~A., \& Pols, O.~R.\ 2002, MNRAS, 329, 897 

\bibitem[Ivanova(2011)]{ivanova2011} Ivanova, N.\ 2011, 
arXiv:1108.1226 

\bibitem[Kolb et al.(2001)]{koletal2001} 
Kolb, U., Rappaport, S., Schenker, K., \& 
Howell, S.\ 2001, ApJ, 563, 958 

\bibitem[Kratter \& Perets(2012)]{kraper2012}
Kratter, K.~M., Perets, H.~B.\ 2012, In Prep 

\bibitem[Kuzuhara et al.(2011)]{kuzetal2011} Kuzuhara, M., 
Tamura, M., Ishii, M., et al.\ 2011, AJ, 141, 119 

\bibitem[Lada(2006)]{lada2006} 
Lada, C.~J.\ 2006, ApJL, 640, L63 

\bibitem[Lajoie \& Sills(2011a)]{lajsil2011a} 
Lajoie, C.-P., \& Sills, A.\ 2011a, ApJ, 726, 66 

\bibitem[Lajoie \& Sills(2011b)]{lajsil2011b} 
Lajoie, C.-P., \& Sills, A.\ 2011b, ApJ, 726, 67

\bibitem[Lee et al.(2009)]{leeetal2009} Lee, J.~W., Kim, S.-L., 
Kim, C.-H., et al.\ 2009, AJ, 137, 3181 

\bibitem[Lowrance et al.(2002)]{lowetal2002} Lowrance, P.~J., 
Kirkpatrick, J.~D., \& Beichman, C.~A.\ 2002, ApJL, 572, L79

\bibitem[Lucas \& Roche(2000)]{lucroc2000} 
Lucas, P.~W., \& Roche, P.~F.\ 2000, MNRAS, 314, 858 


\bibitem[Mazzali et al.(2007)]{mazetal2007} Mazzali, P.~A., 
R{\"o}pke, F.~K., Benetti, S., \& Hillebrandt, W.\ 2007, Science, 315, 825 

\bibitem[Meng et al.(2008)]{menetal2008} 
Meng, X., Chen, X., \& Han, Z.\ 2008, A\&A, 487, 625 

\bibitem[Moeckel \& Veras(2012)]{moever2012}
Moeckel, N., Veras, D.\ 2012, MNRAS, In Press

\bibitem[Mugrauer \& Neuh{\"a}user(2009)]{mugneu2009} 
Mugrauer, M., \& Neuh{\"a}user, R.\ 2009, A\&A, 494, 373 

\bibitem[Nugis \& Lamers(2000)]{nuglam2000} 
Nugis, T., \& Lamers, H.~J.~G.~L.~M.\ 2000, A\&A, 360, 227 

\bibitem[O'Connor \& Ott(2011)]{ocoott2011} 
O'Connor, E., \& Ott, C.~D.\ 2011, ApJ, 730, 70 

\bibitem[Pakmor et al.(2010)]{paketal2010} 
Pakmor, R., Kromer, M., R{\"o}pke, F.~K., et al.\ 2010, Nature, 463, 61 

\bibitem[Parravano et al.(2011)]{paretal2011} Parravano, A., McKee, 
C.~F., \& Hollenbach, D.~J.\ 2011, ApJ, 726, 27 

\bibitem[Passy et al.(2011)]{pasetal2011} Passy, J.-C., 
De Marco, O., Fryer, C.~L., et al.\ 2011, arXiv:1107.5072 

\bibitem[Perets(2010)]{perets2010} Perets, H.~B.\ 2010, 
arXiv:1001.0581 

\bibitem[Potter et al.(2011)]{potetal2011} Potter, S.~B., 
Romero-Colmenero, E., Ramsay, G., et al.\ 2011, MNRAS, 416, 2202 

\bibitem[Qian et al.(2011)]{qiaetal2011} Qian, S.-B., Liu, L., 
Liao, W.-P., et al.\ 2011, MNRAS, 414, L16 

\bibitem[Qian et al.(2012)]{qiaetal2012} 
Qian, S.-B., Zhu, L.-Y., Dai, Z.-B., et al.\ 2012, ApJL, 745, L23 

\bibitem[Raghavan et al.(2006)]{ragetal2006} Raghavan, D., Henry, 
T.~J., Mason, B.~D., et al.\ 2006, ApJ, 646, 523 

\bibitem[Sepinsky et al.(2007)]{sepetal2007} 
Sepinsky, J.~F., Willems, B., Kalogera, V., \& 
Rasio, F.~A.\ 2007, ApJ, 667, 1170 

\bibitem[Sepinsky et al.(2009)]{sepetal2009} 
Sepinsky, J.~F., Willems, B., Kalogera, V., \& 
Rasio, F.~A.\ 2009, ApJ, 702, 1387 

\bibitem[Sepinsky et al.(2010)]{sepetal2010} 
Sepinsky, J.~F., Willems, B., Kalogera, V., \& 
Rasio, F.~A.\ 2010, ApJ, 724, 546

\bibitem[Sigurdsson et al.(2003)]{sigetal2003} 
Sigurdsson, S., Richer, H.~B., Hansen, B.~M., Stairs, I.~H., 
\& Thorsett, S.~E.\ 2003, Science, 301, 193 

\bibitem[Sumi et al.(2011)]{sumetal2011} Sumi, T., Kamiya, K., 
Bennett, D.~P., et al.\ 2011, Nature, 473, 349 

\bibitem[Taam \& Ricker(2010)]{taaric2010} 
Taam, R.~E., \& Ricker, P.~M.\ 2010, New Astronomy Review, 54, 65 

\bibitem[van Rensbergen et al.(2008)]{vanetal2008} 
van Rensbergen, W., De Greve, J.~P., De Loore, C., 
\& Mennekens, N.\ 2008, A\&A, 487, 1129

\bibitem[van Rensbergen et al.(2010)]{vanetal2010} 
van Rensbergen, W., De Greve, J.~P., Mennekens, N., 
Jansen, K., \& De Loore, C.\ 2010, A\&A, 510, A13

\bibitem[Veras et al.(2011)]{veretal2011} Veras, D., Wyatt, M.~C., 
Mustill, A.~J., Bonsor, A., \& Eldridge, J.~J.\ 2011, MNRAS, 417, 2104 

\bibitem[Veras \& Raymond(2012)]{verray2012} Veras, D., Raymond, S.~N.,
MNRAS Letters, In Press

\bibitem[Veras \& Wyatt(2012)]{verwya2012} Veras, D., Wyatt, M.~C.,
MNRAS, In Press

\bibitem[Welsh et al.(2012)]{weletal2012} Welsh, W., et al.\ 2012,
Nature, 481, 475

\bibitem[Zapatero Osorio et al.(2000)]{zapetal2000} Zapatero 
Osorio, M.~R., B{\'e}jar, V.~J.~S., Mart{\'{\i}}n, E.~L., et al.\ 2000, 
Science, 290, 103 

\bibitem[Zapatero Osorio et al.(2002)]{zapetal2002} Zapatero 
Osorio, M.~R., B{\'e}jar, V.~J.~S., Mart{\'{\i}}n, E.~L., et al.\ 2002, 
ApJ, 578, 536 





\end{thebibliography}
\end{document}